\documentclass[fleqn,usenatbib]{mnras}

\usepackage{newtxtext,newtxmath}

\usepackage[T1]{fontenc}

\DeclareRobustCommand{\VAN}[3]{#2}
\let\VANthebibliography\thebibliography
\def\thebibliography{\DeclareRobustCommand{\VAN}[3]{##3}\VANthebibliography}

\usepackage{graphicx}
\usepackage{amsmath}	

\usepackage{amssymb}	
\usepackage{mathtools}
\usepackage{subfigure}
\usepackage{cleveref}
\crefname{figure}{}{}
\Crefname{figure}{}{}
\crefrangelabelformat{figure}{#3#1--#2#4}

\usepackage{color}
\usepackage[normalem]{ulem}  
\usepackage{xspace}
\usepackage{xcolor}
\definecolor{green2}{rgb}{0,0.5,0}
\definecolor{green2}{rgb}{0.2,0.6,0.1}
\definecolor{orange}{rgb}{1,0.5,0}

\title[End-to-End Kilonova Modeling]{2D End-to-End Modeling of Kilonovae from Binary Neutron-Star Merger Remnants}

\author[L. Sippens Groenewegen et al.]{Lieke Sippens Groenewegen$^{1}$\thanks{E-mail: lieke@freedom.nl}, Sanjana Curtis$^{2, 6}$\thanks{E-mail: sanjana.curtis@oregonstate.edu},
Philipp M\"osta$^{3}$,
Daniel Kasen$^{4,5,6}$, Daniel Brethauer$^{6}$
\\
$^{1}$Institute of Physics, University of Amsterdam, Science Park 904, 1098 XH Amsterdam, The Netherlands\\
$^{2}$Department of Physics, Oregon State University, Corvallis, OR 97331, USA \\
$^{3}$GRAPPA, Anton Pannekoek Institute for Astronomy and Institute of
High-Energy Physics, University of Amsterdam,\\
Science Park 904, 1098 XH Amsterdam, The Netherlands\\
$^{4}$Department of Physics, University of California, Berkeley, CA 94720, USA\\
$^{5}$Nuclear Science Division, Lawrence Berkeley National Laboratory, Berkeley, CA 94720, USA\\
$^{6}$Department of Astronomy and Theoretical Astrophysics Center, University of California, Berkeley, CA 94720, USA\\
}
\graphicspath{{./}{figures_submitted/}}
\date{Accepted XXX. Received YYY; in original form ZZZ}

\pubyear{2025}

\begin{document}
\label{firstpage}
\pagerange{\pageref{firstpage}--\pageref{lastpage}}
\maketitle

\begin{abstract}
We investigate the kilonova emission resulting from outflows produced in a three-dimensional (3D) general-relativistic magnetohydrodynamic (GRMHD) simulation of a hypermassive neutron star (HMNS) remnant. We map the outflows into the \texttt{FLASH} hydrodynamics code to model their expansion in axisymmetry, and study the effects of employing different $r$-process heating rates. Except for the highest heating rate prescription, we find no significant differences with respect to overall ejecta dynamics and morphology compared to the simulation without heating. Once homologous expansion is attained, typically after $\sim$ 2s for these ejecta, we map the outflows to the \texttt{Sedona} radiative transfer code and compute the spectral evolution of the kilonova and broadband light curves in various Legacy Survey of Space and Time (LSST) bands. The kilonova properties depend on the remnant lifetime, with peak luminosities and peak timescales increasing for longer-lived remnants that produce more massive ejecta. For all models, there is a strong dependence of both the bolometric and broadband light curves on the viewing angle. While the short-lived (12ms) remnant produces higher luminosities when viewed from angles closer to the pole, longer-lived remnants (240ms and 2.5s) are more luminous when viewed from angles closer to the equator. Our results highlight the importance of self-consistent, long-term modeling of merger ejecta, and taking viewing-angle dependence into account when interpreting observed kilonova light curves. We find that magnetized outflows from a HMNS---if it survives long enough---could explain blue kilonovae, such as the blue emission seen in AT2017gfo. 

\end{abstract}




\section{Introduction}

Binary neutron-star (BNS) mergers are among the most energetic events in the Universe. Their violent collisions expel neutron-rich matter at extreme temperatures and densities, and produce gravitational waves, short gamma-ray bursts, and an electromagnetic transient
known as a kilonova \citep{Metzger_2019}. These multi-messenger events offer a rare opportunity to probe fundamental physics, including the behavior of matter at extreme densities and the still unknown dense-matter equation of state (EOS). These mergers are also a primary astrophysical site for the rapid neutron-capture process ($r$-process), responsible for the formation of the heaviest elements in the Universe (\citealt{Lattimer_1977,Eichler_1989,Thielemann_2017,Cowan_2021}).

On August 17, 2017, the Laser Interferometer Gravitational-Wave Observatory (LIGO)-Virgo detector network recorded GW170817, the first gravitational wave signal from a BNS merger (\citealt{Abbott_2017}). A corresponding electromagnetic counterpart, AT2017gfo, was detected by several telescopes eleven hours later in the galaxy NGC 4993 at a distance of $\sim40$ Mpc. The observed light curve and spectra (e.g., \citealt{Chornock_2017, Nicholl_2017,Cowperthwaite2017,tanaka2017kilonova}) closely matched theoretical predictions for $r$-process powered kilonovae, providing the first direct evidence that BNS mergers are indeed sites of heavy-element nucleosynthesis (e.g., \citealt{Smartt2017, Drout2017, Kasen_2017, Rosswog_2018}). 

\begin{figure}
    \centering
    \includegraphics[width=0.95\columnwidth]{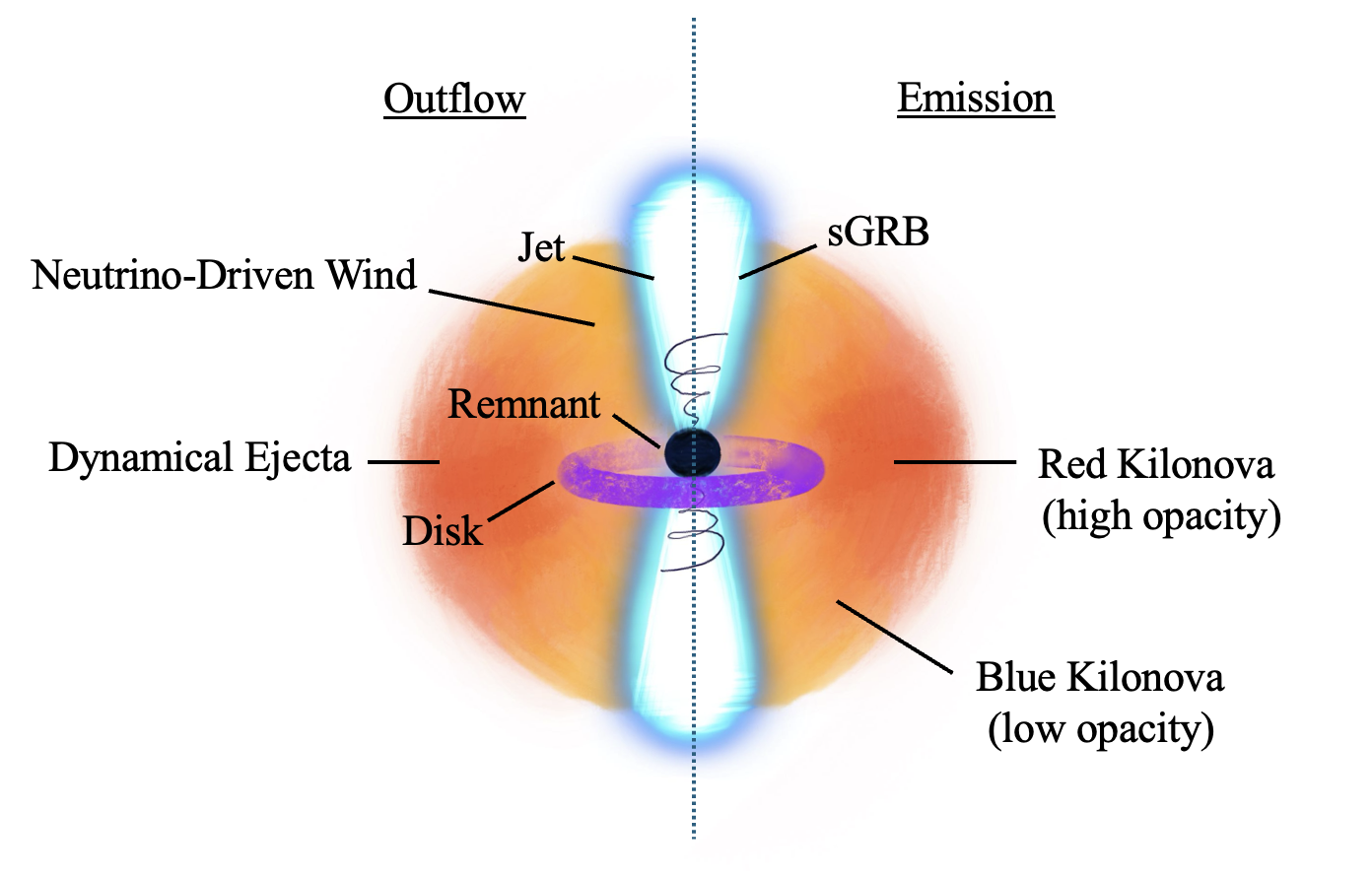}
    \caption{Sketch of the merger remnant, showing outflows (\textit{left}) and their associated (non-)thermal emission components (\textit{right}), including red and blue kilonova emission.}
    \label{fig:ejecta}
\end{figure}

The various distinct outflows of matter associated with BNS mergers (shown in Figure \ref{fig:ejecta}) are sensitive to factors such as the EOS, the masses of the progenitor neutron stars, electromagnetic fields, and the evolutionary channel of the binary system (\citealt{Gottlieb_2025}). As the stars approach their final orbits of the inspiral, they reach velocities nearing $0.5c$, and emit copious amount of energy in the form of gravitational waves. 
During merger, some of the orbital angular momentum is converted into the spin of the remnant, while some NS material is expelled as dynamical ejecta on millisecond timescales. This ejecta typically has a mass of $10^{-3}-10^{-2}M_{\odot}$, characteristic velocities of up to $0.4c$, and an electron fraction ($Y_e$) between $0.1-0.4$ (\citealt{Hotokezaka_2013, Wanajo2014,Sekiguchi_2015,Radice_2016,Rosswog_2018, Shibata_2019, Kullmann_2022}). 

After merger, the remnant gradually takes on a more spherical shape and may drive a relativistic jetted outflow along its rotation axis. While the exact mechanism behind jet formation and the processes responsible for the prompt gamma-ray emission remain an area of active research (\citealt{Kiuchi_2015_sGRB, Kiuchi2024,Mösta_2020,  Combi_2023jets, Pais_2024}), it is well established that the UV/optical afterglow results from the interaction of the jet with the interstellar medium \citep{Nagakura_2014, Murguia_Berthier_2014,Duffell_2015,Kumar_2015, Combi_2023jets}.

Part of the material ejected during the merger remains gravitationally bound, falls back towards the remnant, and forms an accretion torus that can persist even if the remnant neutron star collapses to a black hole (BH). Matter from the accretion disk can also be partially ejected in the form of winds, with velocities ranging between $0.05c-0.1c$. Disk masses can reach up to $0.2M_{\odot}$, and simulations suggest that $10 - 40\%$ of the remnant disk can become unbound over a timescale of a few seconds (see \cite{Radice_2020} and references therein.)

The extreme densities and neutron-rich conditions in the ejecta give rise to the $r$-process (\citealt{ Freiburghaus_1999,Pian2017,Rosswog2019}). In this process, seed nuclei rapidly capture neutrons within a timescale shorter than the $\beta$-decay timescale (\citealt{Metzger_2019}). A key factor in the viability of the $r$-process is the $Y_e$ of the ejecta, which quantifies its neutron-richness:

\begin{equation}
    Y_e=\frac{n_p}{n_n+n_p},
\end{equation}
where $n_p$ and $n_n$ are the proton and neutron densities, respectively. The $r$-process occurs most strongly in neutron-rich ejecta with $Y_e\lesssim 0.24$ for the entropy values typical in outflows from BNS mergers \citep{Lippuner_2015}. 

The merger also produces large amounts of neutrinos, which influence both jet dynamics and matter composition. Neutrino-matter interactions alter the neutron-to-proton ratio in the ejected material through charged current interactions (\citealt{Wanajo2014, Radice_2018, Foucart2020, Foucart_2024,  Curtis_2022}). The neutrino flux increases with latitude, as higher equatorial densities allow more effective neutrino irradiation near the polar axis (\citealt{Rosswog_2003, Foucart_2016}). Consequently, $Y_e$ varies with latitude, with higher latitude ejecta having larger $Y_e$ (\citealt{Cusinato2022}). The choice of neutrino transport scheme is thus of importance in accurately estimating the $Y_e$ and hence the ejecta composition. 

The medium surrounding the merger remnant is primarily heated by the radioactive decay of $r$-process elements. This heating plays a key role in shaping the ejecta morphology and, consequently, the emergent light curves, as demonstrated by \cite{Rosswog2014}, \cite{Grossman2014} and \cite{Foucart_2021}. Accurately modeling the heating rate in simulations is therefore crucial, yet non-trivial, as it depends on multiple factors such as the nuclear abundances, decay properties, and interactions of the synthesized nuclides. Uncertainties in nuclear physics, particularly in nuclear mass models, reaction rates, and fission fragment distributions, can lead to substantial variations in the predicted heating rates—by up to an order of magnitude at low $Y_e$ (\citealt{Rosswog_2017,Zhu_2021,Wu_2022}). 
Nevertheless, some broad trends are robust across a range of ejecta conditions. Typically, the heating rate can be approximated by a power-law time dependence that exhibits an early plateau lasting from seconds to minutes, followed by a gradual decline as neutron captures subside and $\beta$-decay becomes dominant (\citealt{Korobkin_2012,Lippuner_2015}).

In neutron-rich, i.e., low-$Y_e$ ejecta, the heaviest elements are synthesized, including lanthanides ($140< A < 176$;   \citealt{Lippuner_2015, Klion2022}). These elements play a crucial role in shaping the kilonova light curve due to their high opacities, as they cause strong optical line blanketing by absorbing and re-emitting light at longer wavelengths, thus shifting the spectral peak from the optical/UV range to the near-infrared (NIR) band \citep{Kasen_2013}. This produces the so-called "red" kilonova component, which typically peaks around a week after merger (\citealt{Barnes_2013,Tanaka_2018,Even_2020}). This component is thought to arise from tidal dynamical ejecta or disk outflows with low $Y_e\lesssim0.2$, a mass of $M_{\textrm{red}}\approx 0.05 M_{\odot}$, and velocity $v_{\textrm{red}}\approx0.1c$ (\citealt{Villar_2017}). "Blue" kilonova emission is typically associated with higher electron fraction material ($Y_e\gtrsim 0.2$) that is lanthanide-poor and therefore has relatively low opacity. It peaks earlier (within $\sim1$ day) in the UV/optical bands (\citealt{Metzger2014,Perego2014}). The exact origin of this blue component is not yet well understood but recent work points to either outflows from the merger remnant (\citealt{Mösta_2020, Combi_2023jets, Curtis_2024}) or spiral disk outflows (\citealt{Nedora_2019, Radice_2024, Jacobi_2025}).

As the dynamical ejecta is expelled at high velocities during the merger, it creates an outer layer of rapidly expanding, neutron-rich material, creating a “lanthanide curtain” that can obscure emission from inner layers (\citealt{Barnes_2013, Kasen2015, Wollaeger_2018}). Particularly for equatorial observers, it can obscure the blue emission component. Hence, an observer's viewing angle relative to the merger is of importance in interpreting the kilonova signal.

Several prior studies have gauged the influence of ejecta morphology and viewing-angle effects on kilonova emission using asymmetric 2D or 3D geometries constructed to mimic typical merger ejecta components and their properties. For example, \cite{Kawaguchi_2018} performed a 2D radiative transfer simulation that accounts for the interplay of multiple, non-spherical ejecta components to reproduce kilonova light curves and photospheric velocities. \cite{Wollaeger_2018} employed multidimensional radiative transfer with realistic opacities to analyze the role of varying morphologies and compositions on the resulting kilonova. \cite{Darbha_2020} investigated emission from idealized asymmetric configurations—such as ellipsoids and tori—and showed that viewing-angle dependence is primarily determined by the projected surface area of the ejecta along the line of sight. \cite{Korobkin_2021} demonstrated that the geometric distribution of the ejecta can significantly alter both the spectra and light curves, with particularly strong angular dependence in certain two-component models due to lanthanide curtaining. 

However, modeling kilonova evolution, in particular within a self-consistent merger simulation framework, remains challenging due to the combined effects of ejecta mass, velocity, geometry, electron fraction, and opacity. Additional complications arise from neutrino interactions, magnetic field effects and amplification, and remnant lifetimes, all of which influence the post-merger environment and its observable signatures. Capturing the full dynamics of BNS mergers requires sophisticated modeling. General-relativistic magnetohydrodynamics (GRMHD) forms the foundation of this approach, as it couples Einstein’s theory of gravity to the dynamics of magnetized, relativistic fluids. Numerous GRMHD simulations have been carried out over the past two decades to study these systems (e.g., \citealt{Duez_2006,Kiuchi_2015, Kiuchi_2018, Ruiz_2016, Mösta_2020, Most_2023, deHaas2024, Curtis_2024}).
While these simulations are adept at capturing the merger phase itself, extending them to kilonova timescales is computationally infeasible even on today's most powerful supercomputers. 

Therefore, to study the longer-term evolution of the ejecta and self-consistently generate kilonova light curves, it is necessary to map data from early post-merger GRMHD simulations into follow-up simulations that are computationally-affordable while also including the relevant physics for these later phases. Since the detection of GW170817, significant progress has been made in such modeling efforts, with many simulations now numerically evolving the ejecta and incorporating radiative transfer with realistic nuclear heating rates (e.g., \citealt{Kasen2015,Wu_2019,Wu_2022,Zhu_2021,Kawaguchi_2021, Kawaguchi2022,Combi_2023, Collins2023}). Such end-to-end kilonova modeling pipelines frequently rely on reduced dimensionality or spherical symmetry to maintain computational feasibility (\citealt{Curtis_2022,Curtis_2024, deHaas2024}). While efficient, these simplifications neglect important effects like asymmetric mass distributions, anisotropic ejecta velocities, and latitudinal $Y_e$ variations. Full 3D simulations, such as the 3D end-to-end BH-NS kilonova model recently presented by \cite{Kawaguchi_2024} 
, capture these effects but remain computationally expensive. A key objective of current research is thus to develop efficient yet physically consistent modeling pipelines that can bridge the gap between high-fidelity simulations and observational signatures---e.g., \cite{Just_2023} mapped 3D smoothed-particle GR mergers onto 2D post-merger models. 

This work contributes to that effort by constructing an end-to-end modeling pipeline that couples state-of-the-art 3D GRMHD simulations to 2D hydrodynamics and radiative-transfer codes (\citealt{Liekethesis}). We map high-resolution GRMHD data into the hydrodynamics code \texttt{FLASH}, which evolves the ejecta in 2D axisymmetry, and subsequently pass our data to \texttt{Sedona}, to generate multi-band kilonova light curves and spectra  (\citealt{Kasen_2006, Kasen2015}). Our goal is to use a consistent, 2D end-to-end kilonova modeling pipeline that tracks ejecta evolution and viewing-angle dependent behavior of the resulting transient, while remaining more computationally efficient than a 3D pipeline. This setup also allows us to explore the parameter space of relevant input physics uncertainties, such as various $r$-process heating rates and remnant collapse times. 

The paper is organized as follows: Section \ref{sec:methods} describes our input models, mapping procedure and numerical codes. Our results are presented in Section \ref{sec:results}. We discuss broader implications and future directions in Section \ref{sec:summary}.

\section{Methods}
\label{sec:methods}

\begin{figure*}
    \centering
    \includegraphics[width=1.0\textwidth]{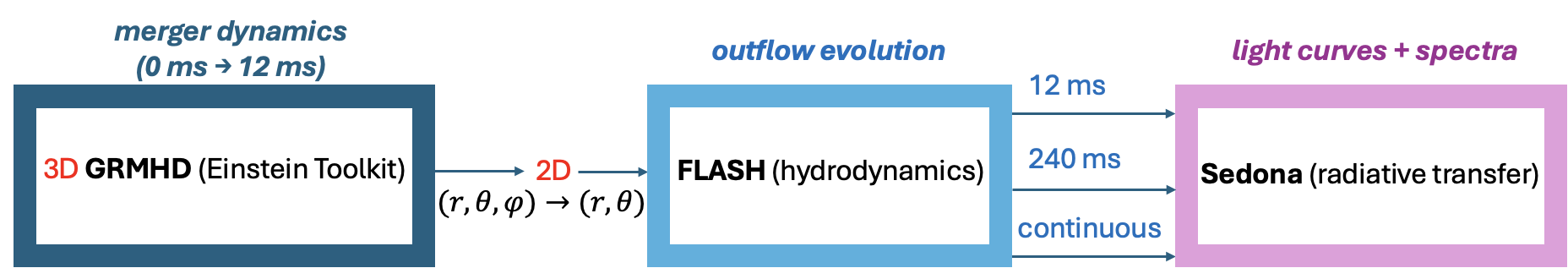}
    \caption{Schematic of the simulation pipeline. A 3D GRMHD simulation models the binary neutron star merger from the hypermassive remnant phase up to black hole formation at 12ms post-mapping. The resulting outflow is angle-averaged to 2D and used as input for the \texttt{FLASH} hydrodynamics code under three ejecta injection scenarios: 12ms cutoff, 240ms injection, and continuous inflow. Outputs are post-processed with the radiative transfer code \texttt{Sedona} to compute light curves and spectra.}
    \label{fig:pipeline}
\end{figure*}

Figure \ref{fig:pipeline} shows a schematic of our simulation pipeline. We start with a 3D GRMHD simulation of a BNS merger, evolved up to black hole formation at 12ms post-mapping. The resulting outflow is azimuthally-averaged to 2D and used as boundary input for the hydrodynamics code \texttt{FLASH}, which we use to simulate three ejecta injection scenarios: 12ms cutoff, 240ms injection, and continuous injection. \texttt{FLASH} evolves the ejecta until homology is attained, at which time the simulation is mapped to the radiative transfer code \texttt{Sedona} to generate light curves and spectra.

\subsection{Initial Data}

Our setup follows the short-lived neutron star configuration simulated in \cite{Curtis_2024}, modeling a remnant formed through the merger of two equal-mass neutron stars ($1.35M_{\odot}$ each, at infinity). The outflows produced in the simulation serve as our initial data. The details of the main setup and the results from this 3D GRMHD simulation are presented in \cite{Curtis_2024}. We summarize the key points below.

The early inspiral and merger phase was simulated by \cite{Radice_2018} using the \texttt{WhiskyTHC} code in a GRHD framework (model LS135135M0), which resulted in the formation of a hypermassive neutron star remnant. To model the subsequent evolution of the remnant, following \cite{Mösta_2020}, we introduced a poloidal magnetic field of strength $B_0=10^{15} \textrm{ G}$, resembling the fields expected from MRI and potential dynamo effects \citep{Kiuchi2024}. Due to the strong post-merger oscillations of the newly formed remnant, the field was added at $t-t_{\textrm{merger}}=17 \textrm{ ms}$, by which time the remnant had settled down enough to map the simulation into the GRMHD setup. The remnant was then simulated in 3D GRMHD using the Einstein Toolkit’s  (\texttt{ETK}) (\url{http://einsteintoolkit.org}) \texttt{GRHydro} module (\citealt{Mösta_2013, Mösta_2015}) under ideal GRMHD with the LS220 (Lattimer-Swesty) EOS (\citealt{LATTIMER1991331}). For neutrino transport, we adopted a two-moment M1 scheme (\citealt{ Radice_2022, Curtis_2024}), which improves upon neutrino leakage schemes by taking into account neutrino heating, pressure, and reabsorption, providing a more realistic treatment of the $Y_e$ evolution.

The simulation presented in \cite{Curtis_2024} spans 12ms, during which the evolution of the hypermassive neutron star remnant and the launching of magnetized winds from the remnant are modeled. At the end of this period, the remnant collapses to a black hole. In principle, a surrounding disk can remain after this collapse, continuing to inject material. However, this simulation does not extend beyond the HMNS phase, largely because modeling the BH spacetime—with its singularity—and subsequent disk evolution poses significant computational challenges. It also does not track any tidal ejecta stripped off prior to the HMNS phase. Consequently, our focus in this work is restricted to modeling the kilonova emission due to outflows that occur during the 12ms of the merger remnant’s evolution, with the magnetized winds produced during this phase representing an important component of the overall merger ejecta, as discussed in \cite{Mösta_2020}, \cite{Curtis_2022} and \cite{Curtis_2024}.

\subsection{\texttt{FLASH} Hydrodynamics}
To bridge the gap between the immediate post-merger phase and the on-set of homologous expansion, we map outflow data from the GRMHD simulation to the \texttt{FLASH} code. \texttt{FLASH} is a publicly available, modular, high-performance hydrodynamics code developed by the \texttt{FLASH} Center for Computational Science (\url{https:
//FLASH.rochester.edu}). The GRMHD outflow data serve as the input boundary for \texttt{FLASH}, which we use to conduct 2D axisymmetric hydrodynamics simulations.

The properties of the outflow produced during the course of the GRMHD simulation were recorded using the \href{https://einsteintoolkit.org/thornguide/EinsteinAnalysis/Outflow/documentation.html}{\texttt{ETK Outflow thorn}}. The thorn measures the flux of relevant hydrodynamical quantities such as density, temperature, and velocity through a spherical surface, here chosen to be at radius $r=$ 120$M_{\odot}$ $=177$ km. The pressure is not recorded by the thorn but obtained by evaluating the LS220 EOS using the corresponding temperature, density and $Y_e$.  We reduce the dimensionality of the GRMHD simulation outflow data from 3D spherical coordinates $(r, \theta, \phi)$ to 2D spherical coordinates $(r, \theta)$, by performing a mass-weighted azimuthal averaging. This approach is justified given the approximate azimuthal symmetry of the outflow properties. 

We impose a time-dependent inner boundary condition in the \texttt{FLASH} simulation, placing our inner boundary at a radius of 177 km. The outer radial boundary condition is set to outflow, while the $\theta$ boundaries are reflecting. To determine the properties of material at the boundary at a given timestep in \texttt{FLASH}, outflow data are linearly interpolated in $\theta$ using the closest timestep available from the \texttt{ETK} simulation. As the time-resolution of the \texttt{ETK} simulation is extremely fine, this is a reasonable approach and performing e.g. a linear interpolation with time to determine input properties for \texttt{FLASH} was not necessary. We do not directly manipulate cells within \texttt{FLASH}'s computational domain, opting instead to fill the ghost cells at the boundary with interpolated data, which propagates onto the grid. Through this mapping procedure, we set the density, pressure, and velocity of the material at the boundary, as well as its $Y_e$. The $Y_e$, however, is not evolved in the \texttt{FLASH} simulation, it is simply advected with the flow. We use a gamma-law EOS with $\gamma = 1.4$. We also use a point source of gravity to represent the central neutron star remnant that was excluded from our domain, with a mass of $2.5 M_{\odot}$.

\texttt{FLASH} utilizes an internal block-structured grid, where each coordinate direction is divided into a set number of blocks. We perform our simulations on a 2D spherical grid  with a total of 24 blocks, with a grid resolution of 6 blocks in \textit{r} and 4 in $\theta$. Each block is further divided into $16 \times 16$ computational cells. The centers of each block are spaced logarithmically in $r$ to achieve finer resolution in the inner regions where we inject material, while coarser resolution suffices in the outer parts of the grid. While this resolution is not sufficient to fully resolve sharp gradients, it provides an adequate level of detail for our current analysis. All simulations are run on a grid with an outer radius of $4.5\cdot10^{10}$ cm up to a time of 2.5 s, at which point the data are transferred to \texttt{Sedona}. 

Our implementation currently does not include magnetic fields, as we prioritize a foundational setup with lower computational costs before adding further complexities. Additionally, neutrino transport is not incorporated. This is not a significant limitation for the purpose of modeling long-term evolution of ejecta once they leave the merging system. Electron neutrino and antineutrino luminosities decay on a timescale of 100ms after the merger \citep{Cusinato2022}, whereas our simulations focus on timescales of the order of seconds when neutrino effects are not as significant.

We conduct three simulations to explore different remnant lifetime scenarios. In the first case, we simulate ejecta from the short-lived remnant that collapses after 12ms, matching the GRMHD simulation. The second scenario extends the remnant lifetime to 240 milliseconds, consistent with the collapse time established in \cite{Curtis_2024}. Finally, we consider a third boundary case, in which the remnant continues to supply ejecta indefinitely beyond 12ms. Since the end time of our simulations is $\sim$2.5s, this effectively corresponds to a remnant that survives for roughly 2.5 seconds. Typical calculations require $\sim$70\text{--}700 core hours, with longer injection durations demanding more computational time.

Extending the lifetime of the remnant beyond the 12ms of simulated time, as in the 240ms and continuous injection scenarios, is motivated by the fact that hypermassive neutron star lifetimes are highly uncertain (\citealt{Bernuzzi2020, Sarin2021}). Our goal here is to assess the impact of longer-lived remnants on the kilonova signal, exploring a significant source of uncertainty in kilonova modeling. For the 240ms and continuous cases, we maintain a steady inflow by continuously injecting the final 12ms boundary conditions from the GRMHD simulation, which are representative of a quasi-steady-state phase of mass ejection from the remnant.

As previously mentioned, the $Y_e$ is not evolved in \texttt{FLASH} and is assumed to remain constant for material that passes the outflow extraction surface. In particular, for the longer-lived remnants, we continuously inject material with the same $Y_e$ as at $t=12$ ms, so that the average $Y_e$ at this time ($Y_e\sim0.43$) dominates the abundances in the longer-lived remnant scenarios. This introduces uncertainties in the composition, as in reality the $Y_e$ is expected to evolve over time. Recent work suggests that later-time outflows from HMNS remnants can even become proton-rich \citep{Bernuzzi2025}, highlighting the need for caution when interpreting the resulting nucleosynthesis and kilonova signal for the extended-lifetime models. Setting the extraction surface at large radii and simulating the evolution of longer-lived remnants in 3D GRMHD will enable us to follow the $Y_e$ evolution of ejecta more closely in future work.

To account for radioactive heating of the ejecta due to nuclear decays, we implement a time-dependent $r$-process heating rate based on the \texttt{H3} and \texttt{H4} heating rate models from \cite{Darbha_2021}. The heating rate $Q(t)$ is defined as:

\begin{equation}\label{eqn:heating}
   Q(t)=Q_0
       \begin{cases}
      1, & \ t<t_b \\
      \left(\frac{t}{t_b}\right)^{\alpha}, & \text{otherwise},
    \end{cases}
\end{equation}
where $Q_0$ denotes the specific heating scale, $t_b=10^{-5}$ days represents the break time, and $\alpha$ the exponent. This form exhibits an early plateau followed by a power-law decay. For \texttt{H3}, $Q_{0,\texttt{H3}}=1\cdot10^{18} \textrm{erg}\textrm{ s}^{-1}\textrm{g}^{-1}$, and $\alpha_{\texttt{H3}}=-2.3$. For \texttt{H4}, $Q_{0,\texttt{H4}}=1\cdot10^{19} \textrm{erg}\textrm{ s}^{-1}\textrm{g}^{-1}$, and $\alpha_{\texttt{H4}}=-3.3$. Although \texttt{H3} provides more realistic heating for our case, given large fractions of our ejecta have $Y_e>0.22$, we also investigate \texttt{H4} as it represents the maximum heating scenario with $Y_e \sim 0.1$, allowing us to assess the effects of stronger $r$-process heating on the ejecta. For a detailed motivation behind these heating rates, refer to \cite{Darbha_2021}.  

To quantitatively determine when the ejecta reach homologous expansion—a requirement for post-processing in \texttt{Sedona}—we calculate the averaged deviation of the radial velocity $\left \langle v_{r}\right\rangle$ as:

\begin{equation}
    \langle  \Delta v_{r} \rangle=\frac{\int D\left(v_{r}-r / t\right) d^{3} x}{\int D d^{3} x}~,
    \label{eqn:homology}
\end{equation}
where $D$ is the mass density and the deviation is computed as a weighted volume-integral over the ejecta. This definition is adapted from (\citealt{Kawaguchi_2021,Neuweiler_2023}). While they adopted the absolute deviation, $\langle |\Delta v_r| \rangle$, we omit the absolute value here. In our simulations, the velocity deviation exhibits a natural cross-over from negative to positive values, and enforcing an absolute value introduces misleading 'kinks' in the time evolution. For homologous expansion, $\langle \Delta v_{r} \rangle \rightarrow 0$.

\subsection{\texttt{Sedona} Radiative Transfer}

When the outflows approach homologous expansion, the output from \texttt{FLASH} is passed to the radiative transfer code \texttt{Sedona} to compute synthetic kilonova light curves and spectra. In time-dependent calculations, \texttt{Sedona} assumes homologous expansion of the ejecta and operates on a velocity-based grid. 
It uses a Monte Carlo scheme, in which radiation is modeled as packets of photon energy. These packets continuously undergo absorption, emission, and scattering events until they escape the ejecta. After each interaction, the code updates each packet’s wavelength and polarization state. The cummulative behavior of all photon packets determines the local radiation fields, which ultimately determine the emergent kilonova spectra. Photons that reach the simulation’s outer boundary—defined by the maximum velocity—escape the ejecta and are subsequently binned based on time, frequency, and viewing angle. This binning process generates the model’s spectral time series. For a detailed description of \texttt{Sedona} and its underlying methods, the reader is referred to \cite{Kasen_2006} and \cite{Roth_2015}. 

We use a snapshot of the 2D \texttt{FLASH} simulations at 2.5 s as initial data for 2D \texttt{Sedona} simulations.  We employ a cylindrical grid consisting of $80\times160$ cells in the $r$ and $z$ directions. The number of MC particles emitted through radioactivity per timestep is set to 5 $\times 10^{5}$, with typical calculations requiring 1500\text{--}2000 core hours.

The input density, temperature, velocities, and $Y_e$ are interpolated from the available quadrant of the \texttt{FLASH} simulation. These data are then mirrored along the $z$-axis to reconstruct the full outflow structure. During the mapping, a relatively small amount of material with negative radial velocities, typically present close to the inner radial boundary of the \texttt{FLASH} simulation, is removed from the grid since it does not satisfy the homologous expansion condition and is expected to fall back through the boundary. We also ensure that the homologous expansion approximation does not introduce superluminal velocities for any zone in the simulation domain. 

An additional parameter introduced at this stage is the ejecta composition. We approximate the composition as a mass-fraction distribution consisting of 33 representative nuclei that sample different regions of the $r$-process pattern. As we do not have the capability to track on-the-fly $r$-process composition evolution in \texttt{FLASH}, the composition of a zone is assigned based on its (advected) $Y_e$. We extract relative abundances of the representative nuclei from prior nucleosynthesis calculations performed in \cite{Curtis_2024}, where the original 3D GRMHD post-merger simulation was post-processed using tracer particles. The composition of a zone is set using the final abundances obtained for the tracer particle with the closest $Y_e$ value at a temperature of 5 GK (when material leaves nuclear statistical equilibrium and the full $r$-process calculation begins), and normalized so that the mass-fractions in each zone sum to 1. 

To bridge the gap in time between the end of the \texttt{FLASH} simulation (2.5 s) and the start time for \texttt{Sedona} calculations (typically 0.01 days here), the input data are extrapolated within the code under the assumption of homologous expansion. Photon-matter interactions in the \texttt{Sedona} simulation are modeled under the assumption of local thermodynamic equilibrium (LTE), which becomes less accurate as the ejecta become optically thin. We use theoretical atomic data from \cite{Tanaka2020}. Our implementation includes bound-bound opacities treated in the Sobolev approximation (\citealt{Sobolev_1960}) within an expansion opacity formalism. We also include free-free and electron-scattering opacities. \texttt{Sedona} calculates these opacities for all isotopes included in the simulation, capturing the wavelength-dependent photon absorption associated with each electronic transition. We use the heating rate and local thermalization efficiency prescription described in \cite{Brethauer_2024} to determine the effective heating by radioactive decay products. Light curves and spectra are computed for different observer viewing angles at a distance of 40 Mpc, the reported distance of the AT2017gfo kilonova.

While it is outside the scope of this work to study the impact of employing different atomic datasets on kilonova features, we caution the reader that the choice of atomic data can have a strong influence on predicted spectra. The atomic data employed here have not been calibrated to match experimental atomic transition wavelengths and only focus on general statistical properties, as stated in \cite{Tanaka2020}, and are therefore not suitable for comparing detailed spectral features with observed spectra. \cite{Shingles2023} find large differences in the spectra obtained when using calibrated data for Sr, Y, and Zr from \cite{Kurucz2018ASPC} instead of uncalibrated data from \cite{Tanaka2020}. Also see \cite{Brethauer_2024} for a comprehensive exploration of the impact of opacity data on kilonova light curves simulated with \texttt{Sedona}.

\section{Results}
\label{sec:results}

\subsection{\texttt{FLASH} Results}\label{sec:results_FLASH}
We present the ejecta mass, velocity structure, and composition for the three distinct central-engine scenarios (12ms, 240ms, continuous). In addition, for each scenario, we quantify the total energy contribution from $r$-process heating by comparing simulations with and without $r$-process heating. We also investigate the onset of homologous expansion across the entire set of simulations.

\subsubsection{Ejecta Evolution and Morphology}

\begin{figure*}
    \centering \includegraphics[width=0.9\textwidth]{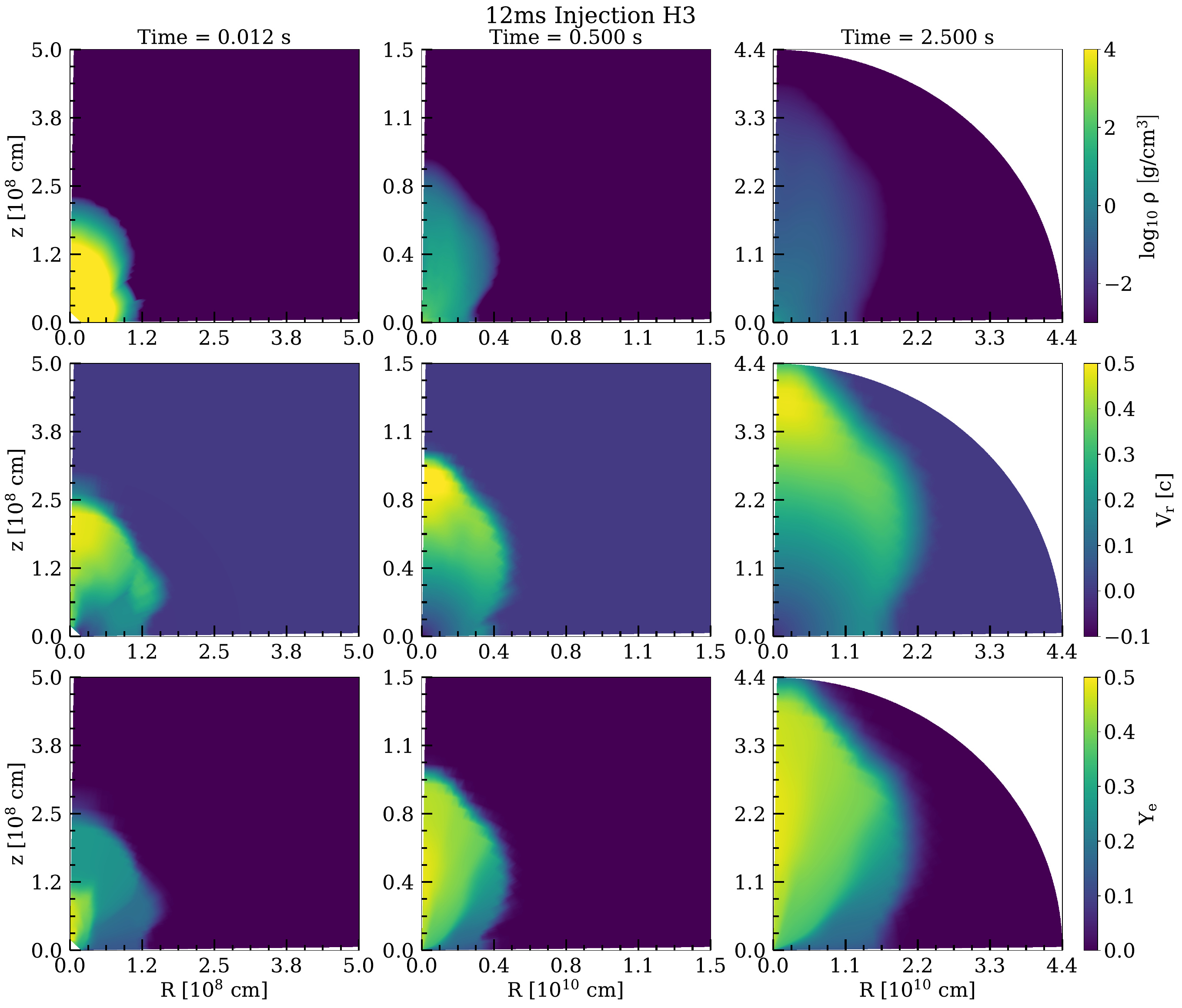}
    \caption{Evolution of the ejecta in the 12ms injection scenario with heating mode \texttt{H3}, shown at three different times: 12ms (left column), 0.5 s (middle), and 2.5 s (right). Each row corresponds to a different quantity: mass density (top), radial velocity (middle), and electron fraction $Y_e$ (bottom). The $v_r$ colorbar ranges from $-0.1c$ to $0.5c$, with negative values indicating fallback, though these are not apparent here because they occur close to the remnant and are obscured due to the large spatial domain used to highlight the ejecta evolution. The spatial domain increases across columns to ensure that key features remain visible, with the final column (2.5 s) covering the full \texttt{FLASH} domain.
}
    \label{fig:12ms_H3_3x3}
\end{figure*}

\begin{figure*}
    \centering\includegraphics[width=0.9\textwidth]{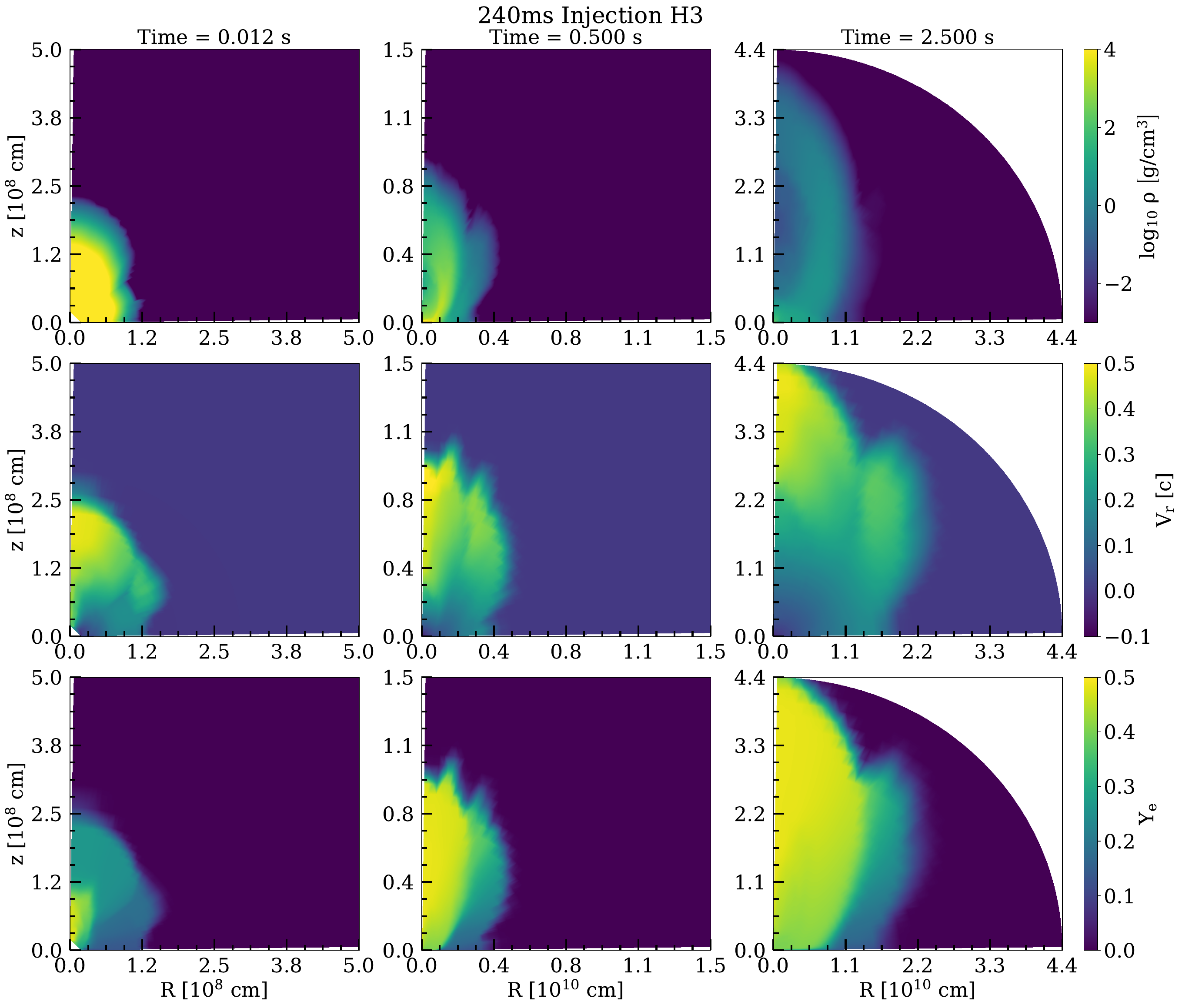}
    \caption{Evolution of the ejecta in the 240ms injection scenario with heating mode \texttt{H3}, shown at three different times: 12ms (left column), 0.5 s (middle), and 2.5 s (right). Each row corresponds to a different quantity: mass density (top), radial velocity (middle), and electron fraction $Y_e$ (bottom). The $v_r$ colorbar ranges from $-0.1c$ to $0.5c$, with negative values indicating fallback, though these are not apparent here because they occur close to the remnant and are obscured due to the large spatial domain used to highlight the ejecta evolution. The spatial domain increases across columns to ensure that key features remain visible, with the final column (2.5 s) covering the full \texttt{FLASH} domain.
}
    \label{fig:240ms_H3_3x3}
\end{figure*}

\begin{figure*}
    \centering\includegraphics[width=0.9\textwidth]{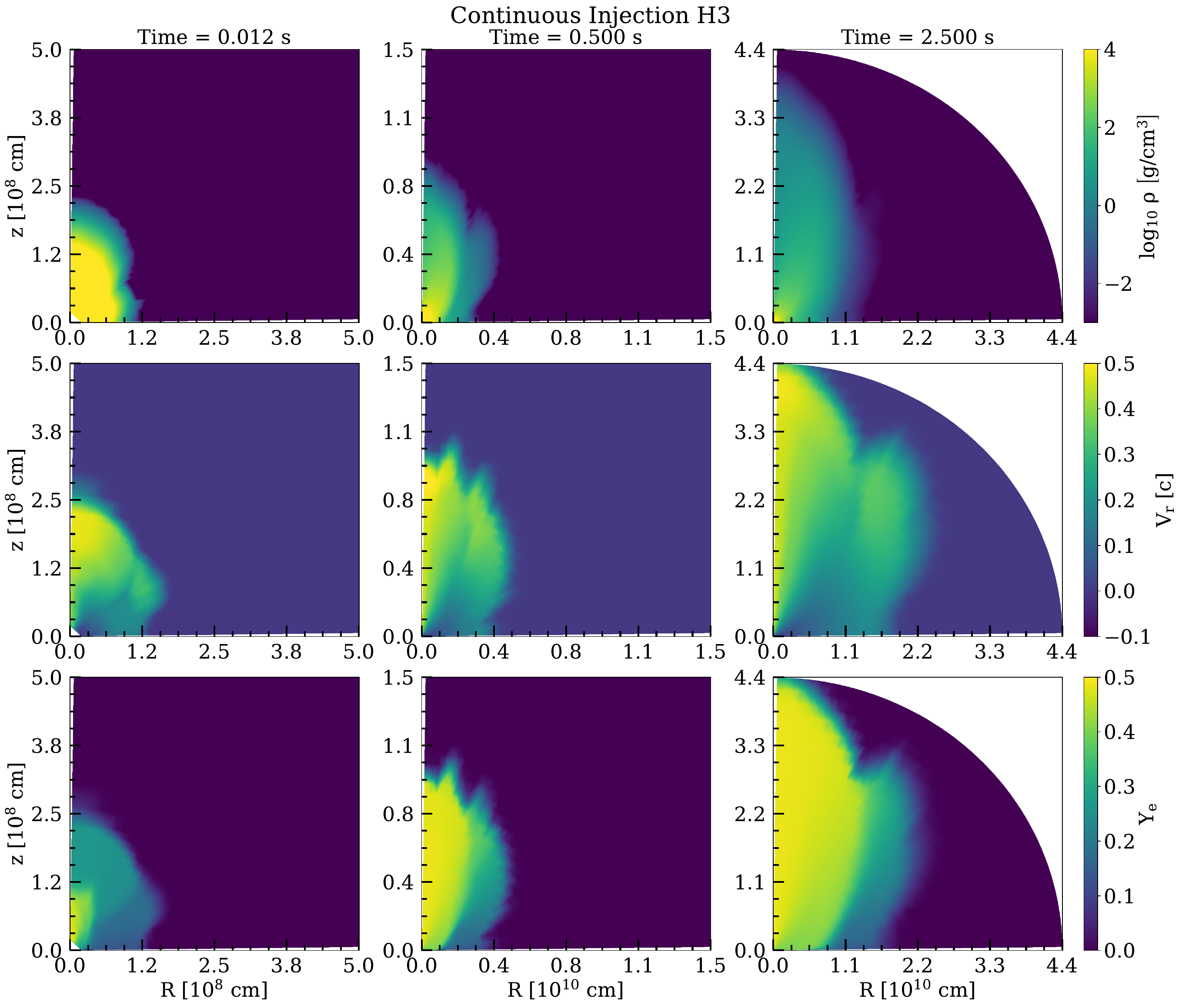}
    \caption{Evolution of the ejecta in the continuous injection scenario with heating mode \texttt{H3}, shown at three different times: 12ms (left column), 0.5 s (middle), and 2.5 s (right). Each row corresponds to a different quantity: mass density (top), radial velocity (middle), and electron fraction $Y_e$ (bottom). The $v_r$ colorbar ranges from $-0.1c$ to $0.5c$, with negative values indicating fallback, though these are not apparent here because they occur close to the remnant and are obscured due to the large spatial domain used to highlight the ejecta evolution. The spatial domain increases across columns to ensure that key features remain visible, with the final column (2.5 s) covering the full \texttt{FLASH} domain.
}
    \label{fig:cont_H3_3x3}
\end{figure*}

Figures \crefrange{fig:12ms_H3_3x3}{fig:cont_H3_3x3} show the evolution of the density $\rho$, radial velocity $v_r$ and electron fraction $Y_e$ for all three injection scenarios.
The 12ms, 240ms and continuous simulation correspond to roughly $10^{-3}M_{\odot}$, $10^{-2}M_{\odot}$ , and $10^{-1}M_{\odot}$ on the grid, respectively. For conciseness, we restrict our primary analysis to the \texttt{H3} model---as \texttt{H4} is more suited for lower-$Y_e$ ejecta with $Y_e\sim0.1$--- and focus on the dependence on remnant lifetime. We note that the morphology for \texttt{H0} is almost indistinguishable from \texttt{H3} in these figures. While \texttt{H4} has a more significant impact on the ejecta morphology, we present these figures in the Appendix (Figures \crefrange{fig:12ms_H4_3x3}{fig:cont_H4_3x3}).

Figure \ref{fig:12ms_H3_3x3} shows the evolution of the 12ms injection scenario with \texttt{H3} heating. The highest densities are concentrated near the remnant. The outflow exhibits an asymmetry as material ejected in the polar regions propagates further outward due to its higher radial velocities. While material in the polar region attains velocities of $\sim0.5c$, equatorial velocities remain lower, around $0.1-0.2c$. As the engine ceases injection past 12ms, no new material is expelled from the remnant, and most of the already ejected material continues propagating outward. The highest electron fraction values ($Y_e\approx0.5$) are concentrated near the pole and gradually decrease toward the equator.

In the 240ms injection case, shown in Figure \ref{fig:240ms_H3_3x3}, the 2.5-second density snapshot reveals a thin outer shell of higher-density material. This shell corresponds to ejecta launched during the 240ms injection period and outlines the leading edge of the outflow. A secondary feature—referred to here as the “second plume”—emerges in this simulation, most prominently in the $v_r$ and $Y_e$ panels. This plume likely originates from an early mass ejection episode within the first few  milliseconds, during which a quasi-spherical shell of material is expelled as the remnant undergoes initial oscillations and begins to settle. Although this component remains visible at later times, it becomes increasingly diffuse and is kinematically distinct from the main polar outflow due to its lower velocity. As a result, it appears spatially separated from the more energetic, polar ejecta. Given its low mass and minimal dynamical impact in our simulation, the second plume is unlikely to play a significant role in shaping observable features, such as kilonova emission.

Figure \ref{fig:cont_H3_3x3} presents the ejecta evolution for the continuous-injection scenario. As time goes on, we consistently see high densities and velocities present close to the remnant due to the sustained injection of material. The overall morphology remains similar to the 240ms scenario, along with the appearance of the low-density secondary plume. However, the total mass on the grid is roughly an order of magnitude higher compared to the 240ms scenario.

\begin{figure}
    \centering
   \begin{tabular}{c}
    \includegraphics[width=0.5\columnwidth]{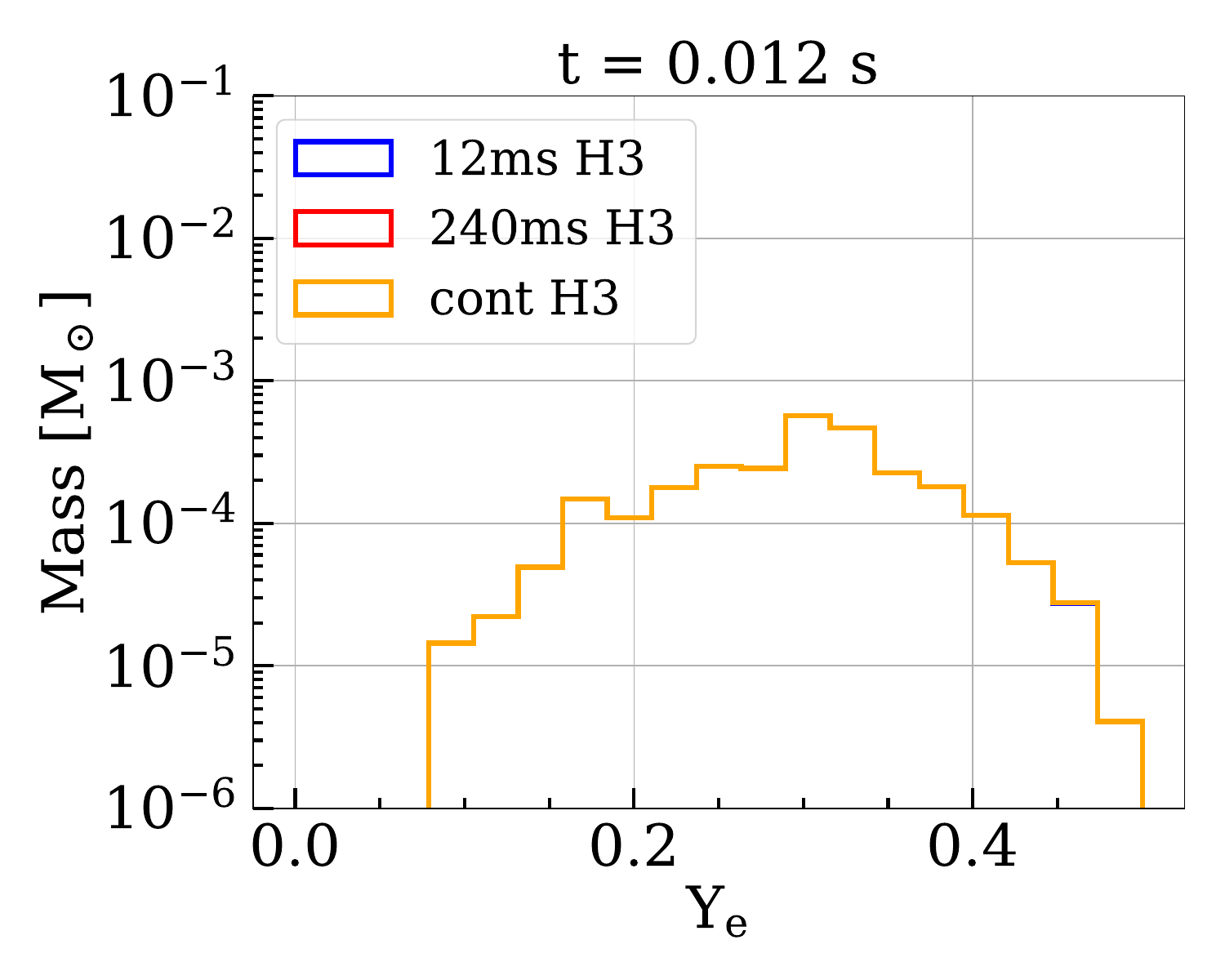}
    \includegraphics[width=0.5\columnwidth]{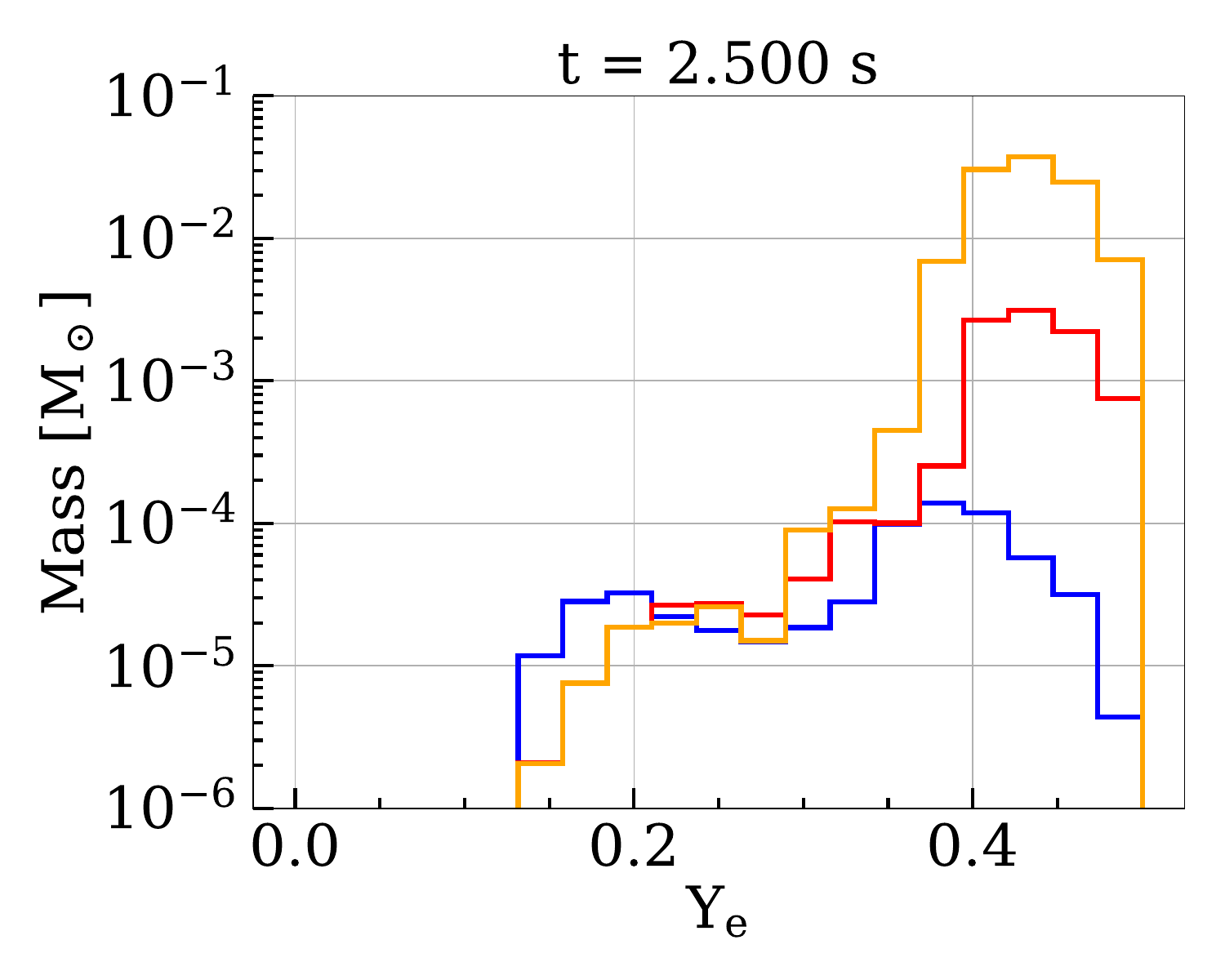}\\

    \includegraphics[width=0.5\columnwidth]{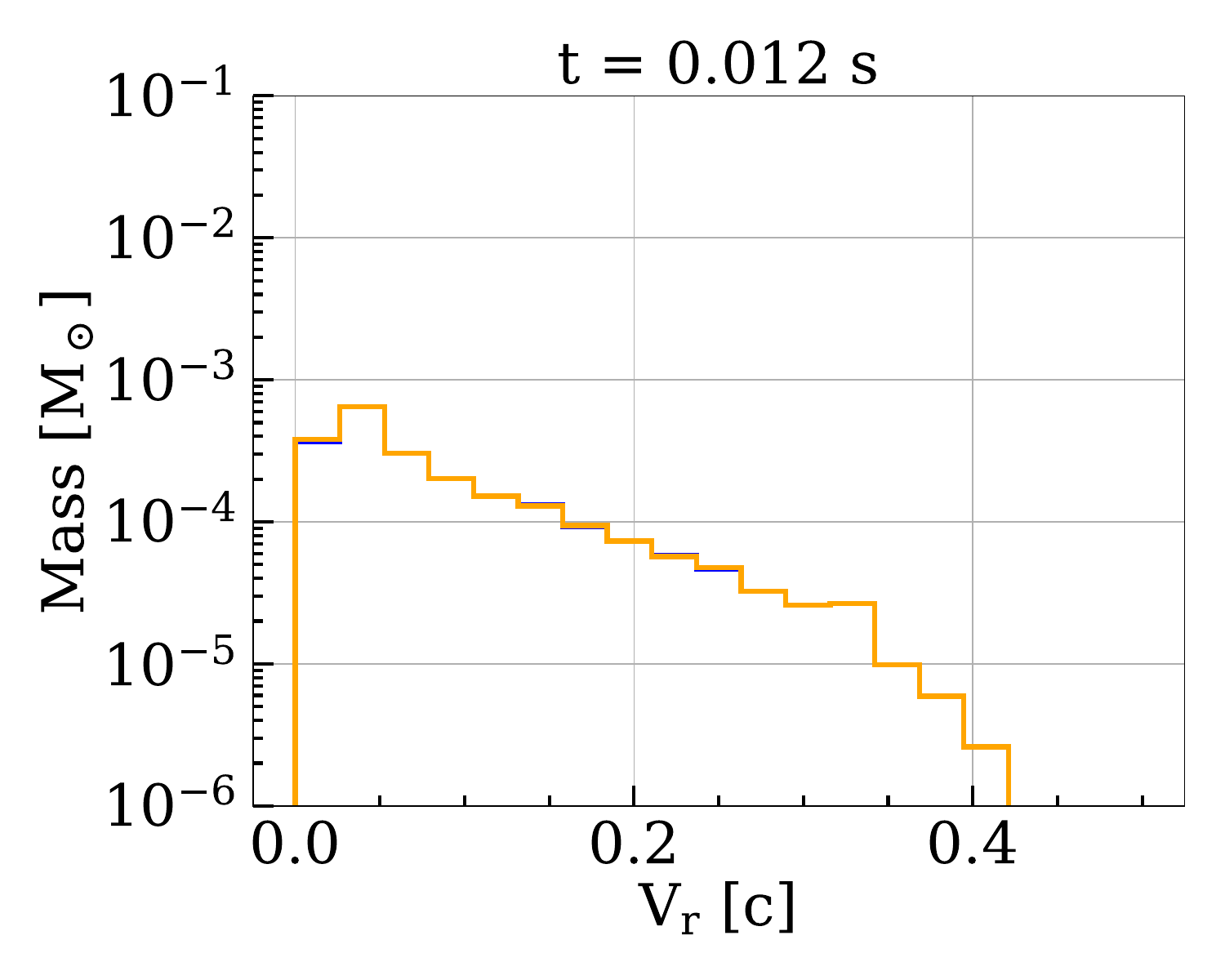}
    \includegraphics[width=0.5\columnwidth]{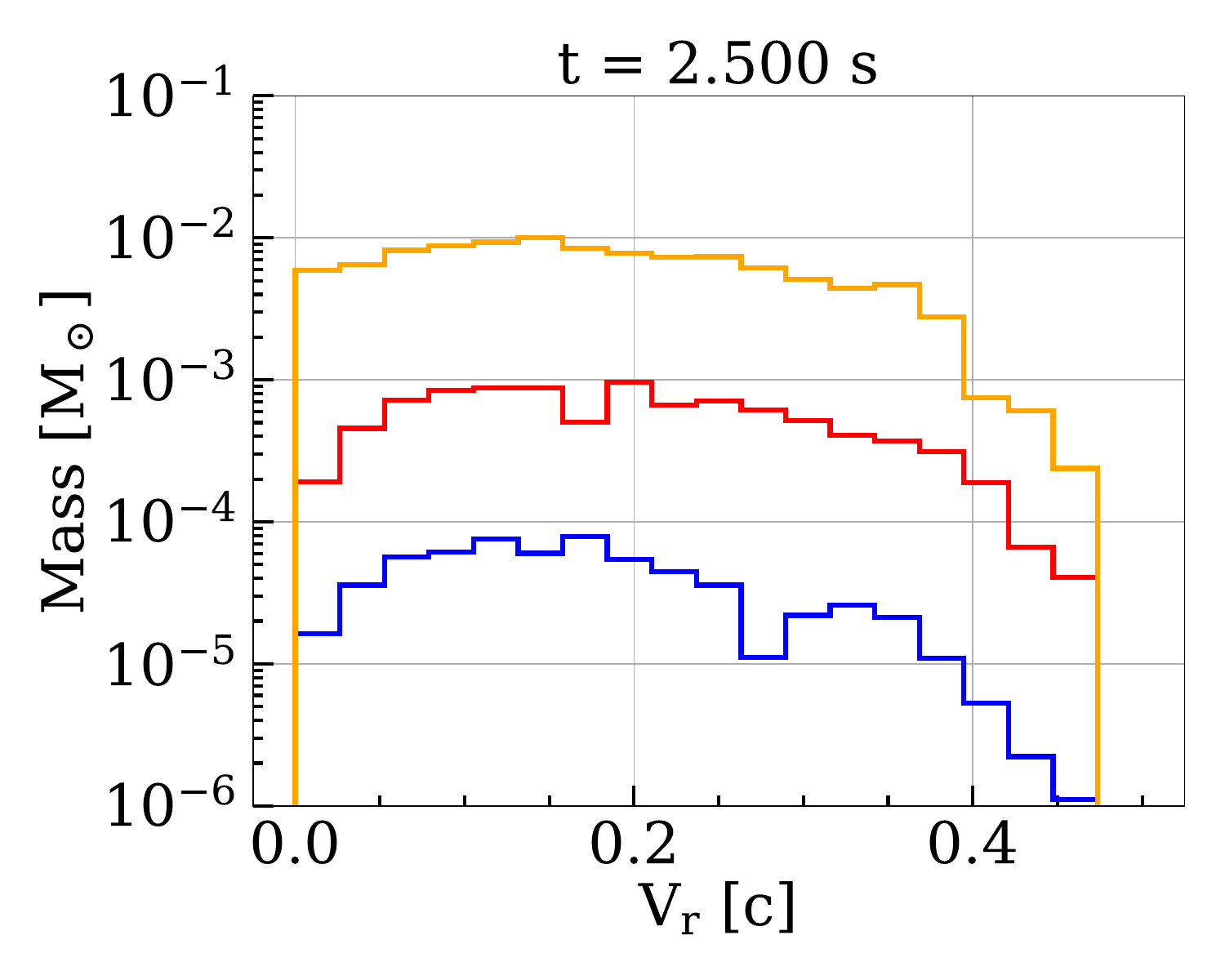}
    \end{tabular}        
    \caption{Histograms of ejecta properties for the three injection scenarios with heating \texttt{H3}. The distributions shown are for electron fraction $Y_e$ (top panels) and radial velocity $v_r$ (bottom panels), at 12ms after merger (left) and at the mapping time of 2.5 s (right).}
    \label{fig:injec_histograms}
\end{figure}

\begin{figure}
    \centering
    \subfigure[Mass inflow rate for the 12ms \texttt{H3} injection scenario. 
    Positive values indicate material entering the domain due to the imposed inflow. 
    After the injection stops at $t=0.012$ s (red dashed line), the rate becomes negative, 
    corresponding to material leaving the simulation domain as it falls back toward the remnant.]{
        \includegraphics[width=1.0\columnwidth]{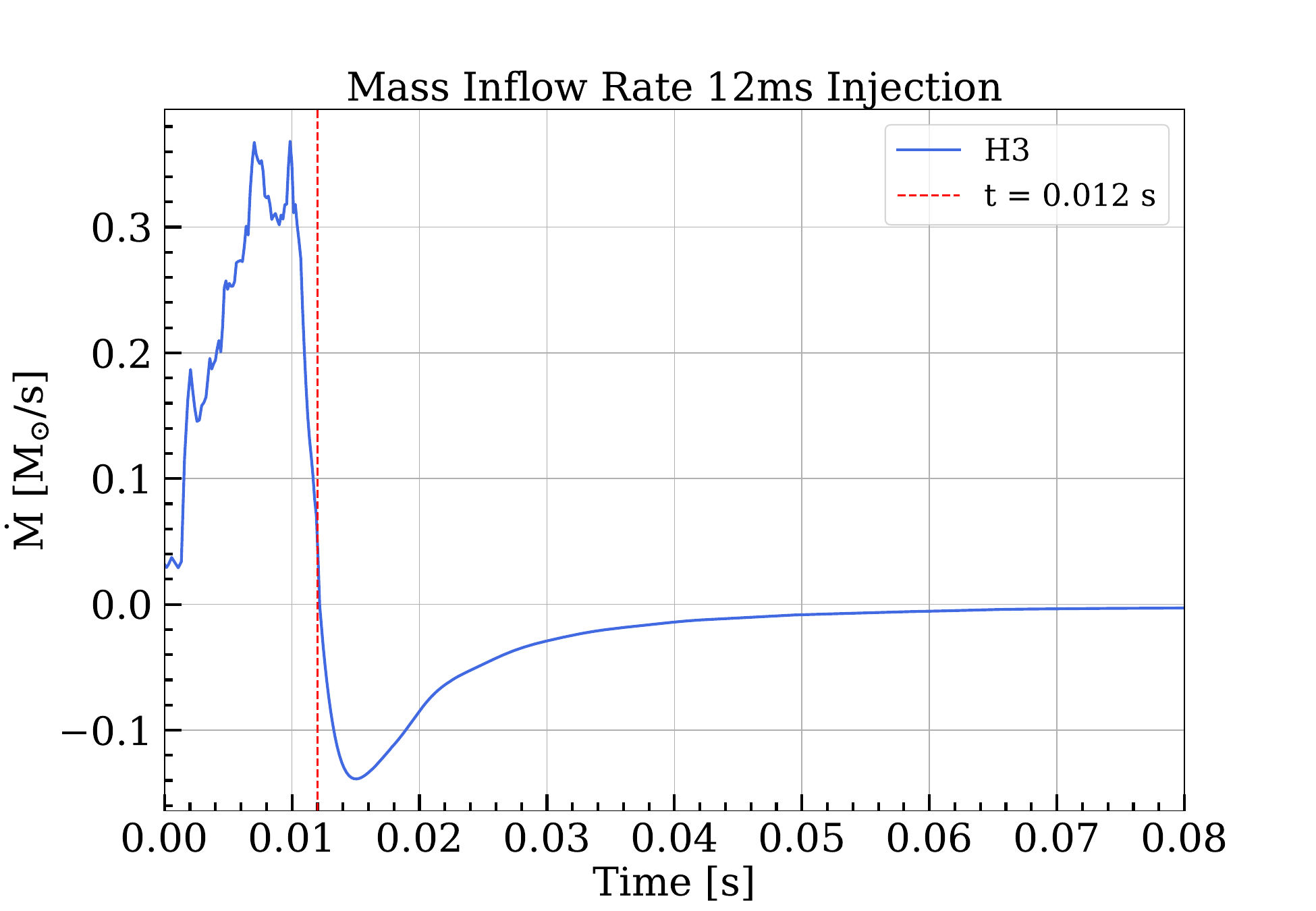}
        \label{fig:mass_inflow_rate}
    }
    
    \vspace{1em}
    
    \subfigure[Regions with negative radial velocity at $t = 0.015$ s (time of strongest fallback). 
    The $v_r$ color scale is restricted to $-0.1c$ to $0.1c$ to emphasize fallback material, 
    while higher positive velocities (up to $0.5c$) appear saturated.]{
        \includegraphics[width=0.9\columnwidth]{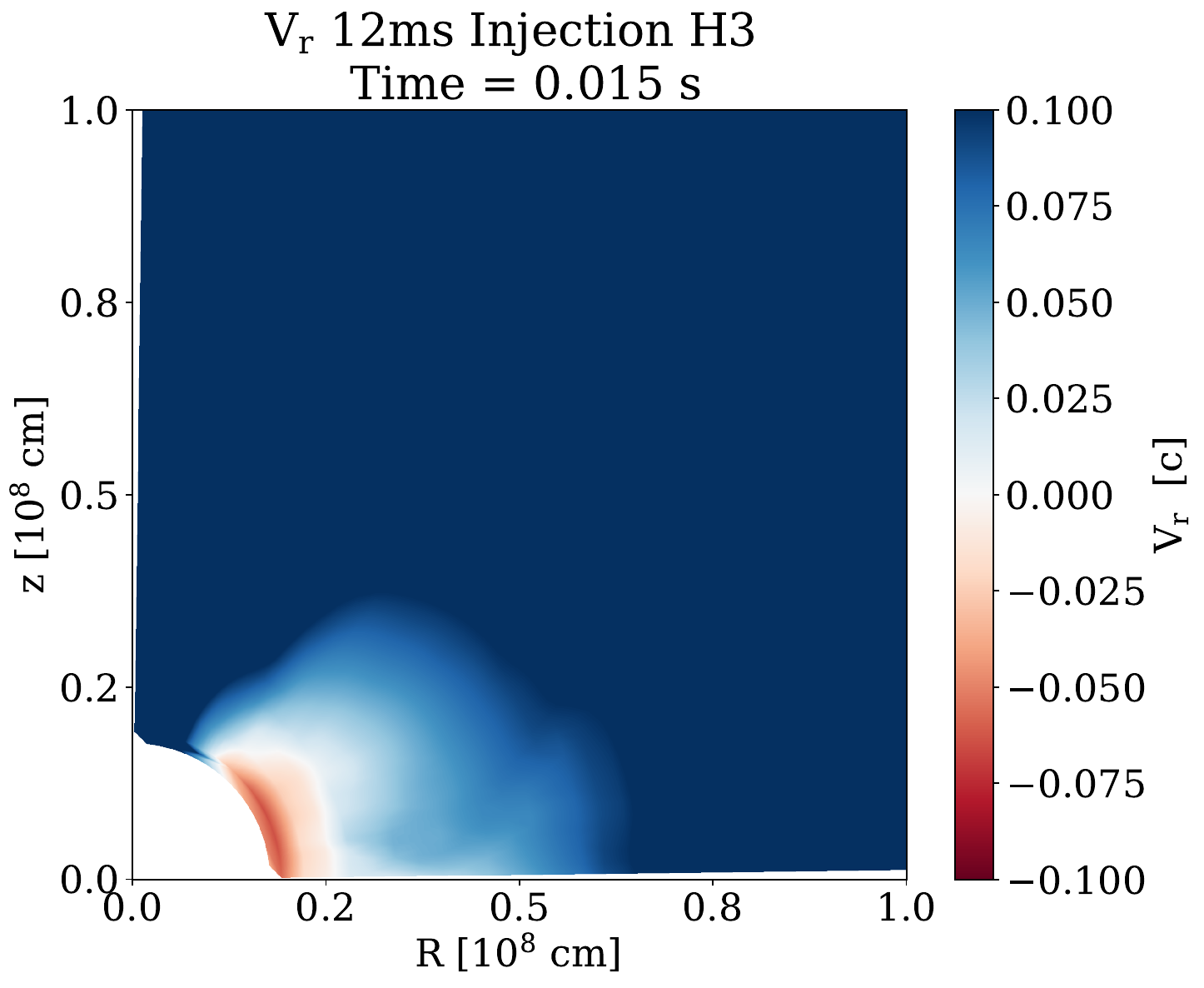}
        \label{fig:vr_12ms_H3}
    }
    
    \caption{Mass inflow and fallback in the 12ms \texttt{H3} injection scenario.}
    \label{fig:mass_inflow_combined}
\end{figure}

Figure \ref{fig:injec_histograms} compares the mass distributions of $v_r$ and $Y_e$ for all three injection cases. At 12ms, the distributions are identical, as expected given the identical outflow data injected up to this point, with most of the material having  $V_r<0.2c$ and $Y_e\sim0.3$. At later times, the total ejected mass increases according to the duration of the injection: the continuous injection case reaches $10^{-2} M_{\odot}$ per velocity bin, the 240ms reaches $10^{-3}M_{\odot}$, and the 12ms case $10^{-4}M_{\odot}$. The shape of the radial velocity distribution itself remains broadly similar across the three scenarios, with changes primarily reflecting the different total masses injected. In contrast, the $Y_e$ distribution evolves more noticeably. In the 12ms case, it develops a double-peaked structure with peaks near $Y_e\sim0.2$ and $Y_e\sim0.4$, whereas for the longer-lived 240ms and continuous injection scenarios the distribution becomes single-peaked, centered at higher values $Y_e\sim0.4-0.5$.
For the 12ms case, no additional material is injected between 12ms and 2.5 s; instead, a significant fraction of the ejecta—approximately half—falls back toward the remnant, as indicated by the mass inflow rate in Figure \ref{fig:mass_inflow_rate}. The radial velocity profile in Figure~\ref{fig:vr_12ms_H3} demonstrates that negative velocities develop mainly in the equatorial region, leading to the selective removal of low-$Y_e$ material and reshaping the distribution into a more bimodal structure. For the other two cases, the shift of the $Y_e$ distributions to higher values largely reflects the additional high-$Y_e$ mass injection.

\subsubsection{Heating Effects}\label{sec:heating_effects}

To ensure a concise presentation of results, we limit our analysis of heating-rate effects to the 240ms injection scenario. This scenario ejects a moderate amount of $r$-process material, of the order of $\sim$ 10$^{-2}M_\odot$, and is consistent with the lifetime estimated to produce a blue kilonova similar to AT2017gfo in \cite{Curtis_2024}.  

\begin{figure}
    \centering
   \begin{tabular}{c}
    \includegraphics[width=0.5\columnwidth]{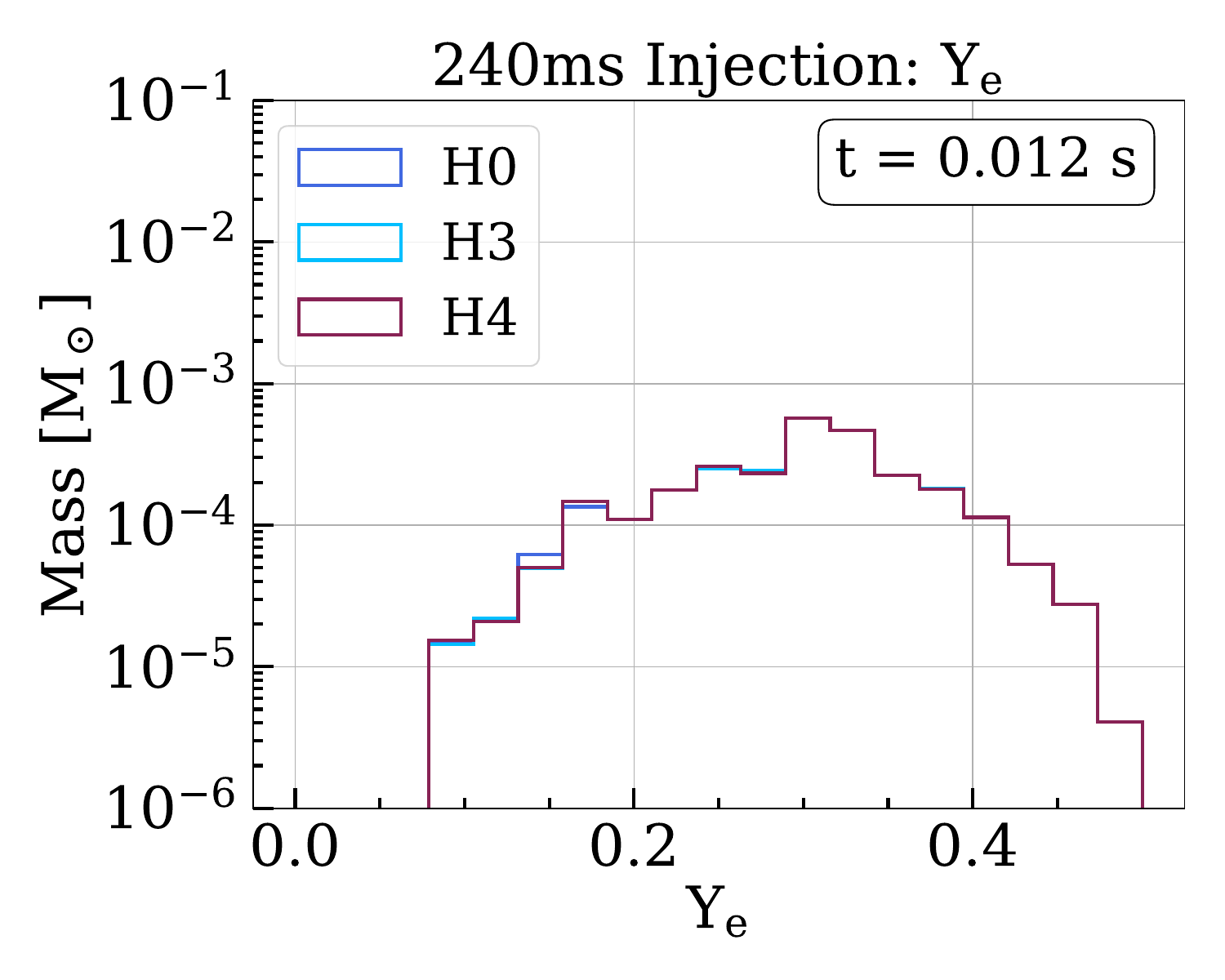}
    \includegraphics[width=0.5\columnwidth]{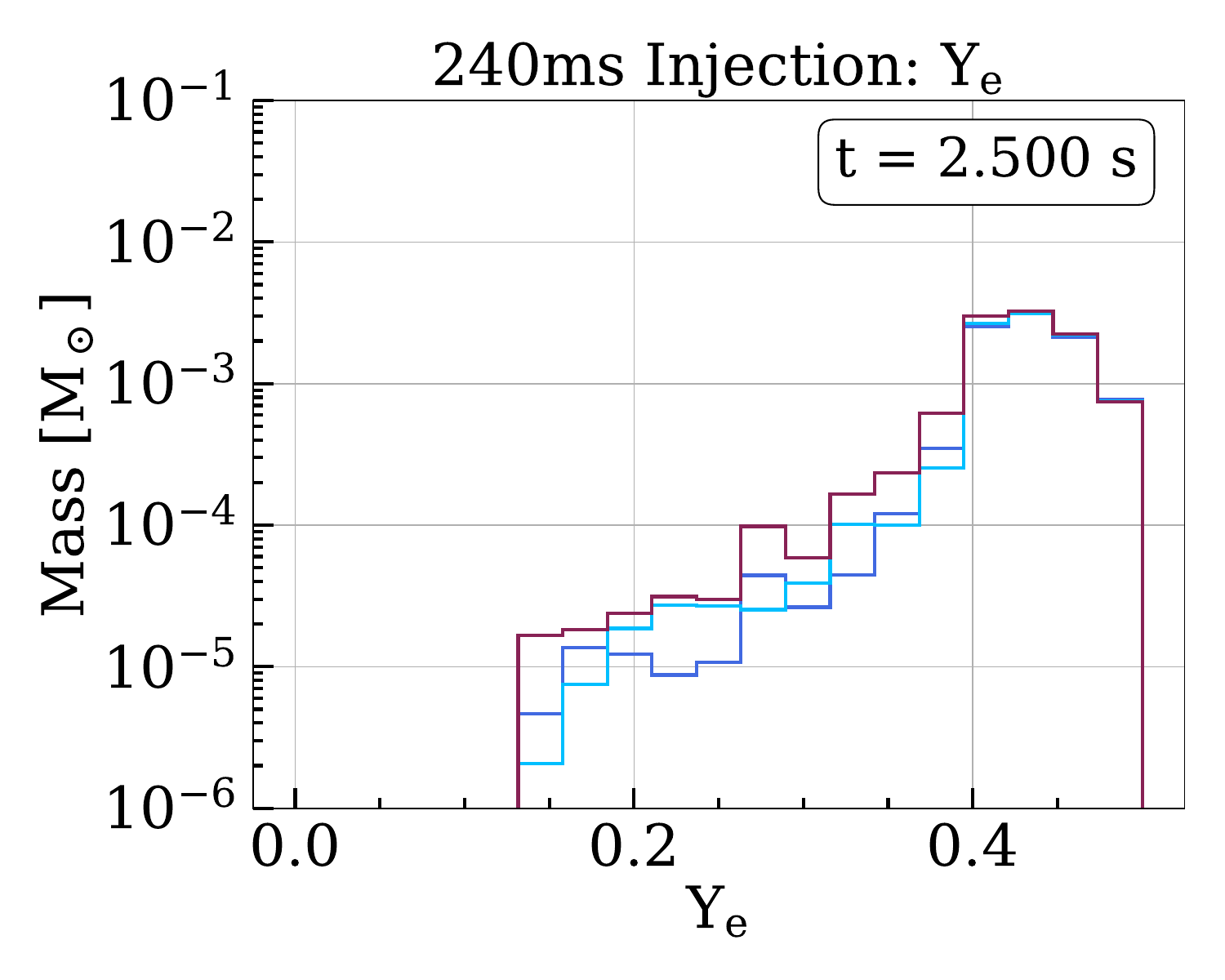}\\

    \includegraphics[width=0.5\columnwidth]{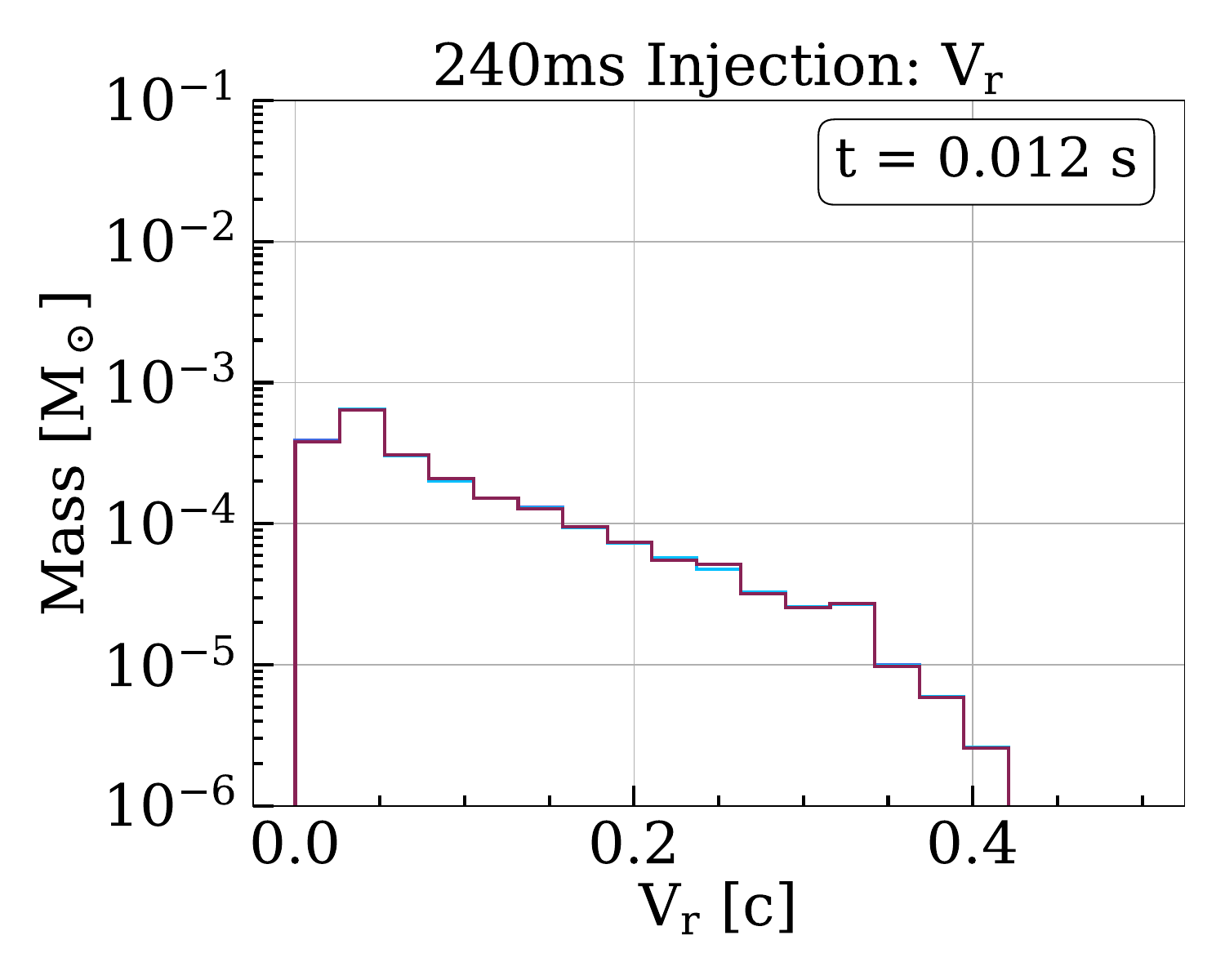}
    \includegraphics[width=0.5\columnwidth]{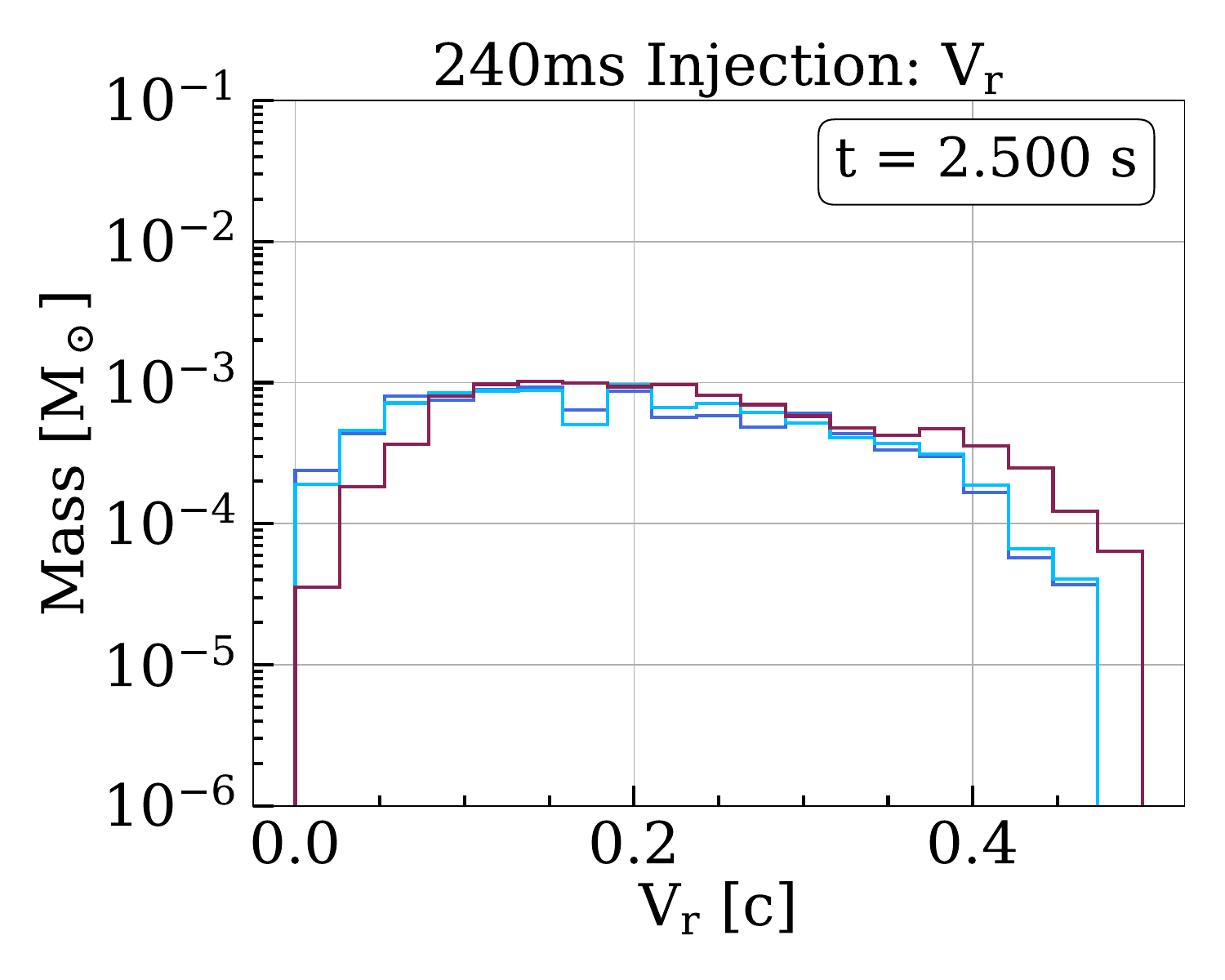}
    \end{tabular}        
    \caption{Histograms of ejecta properties from the 240ms injection simulation for the heating (\texttt{H3}, \texttt{H4}) and no heating (\texttt{H0}) cases. The distributions shown are for electron fraction $Y_e$ (top panels) and radial velocity $v_r$ (bottom panels), at 12ms after merger (left) and at the mapping time of 2.5 s (right).}
    \label{fig:histograms}
\end{figure}

Figure \ref{fig:histograms} shows the evolution of the $Y_e$ and radial velocity $v_r$ distributions for the no-heating case \texttt{H0} and the heating-inclusive scenarios \texttt{H3} and \texttt{H4}. At 12ms, most of the ejecta exhibits relatively low radial velocities and a moderately neutron-rich composition with $Y_e\sim0.3$. At this early stage, heating has not yet significantly altered the outflow, resulting in nearly identical distributions across all three heating cases for both $v_r$ and $Y_e$. By 2.5 s, the radial velocity distribution has broadened and shifted toward higher values, in part owing to the injection of additional high velocity material until 240ms, and in part due to acceleration of the material as it evolves. The effect of heating on the velocity distribution becomes apparent at this time. While the difference between \texttt{H0} and \texttt{H3} remains negligible, the stronger heating in \texttt{H4} produces a small but noticeable shift of the $v_r$ distribution toward higher velocities. At 2.5s, the $Y_e$ distribution has also shifted toward higher values, with the bulk of the material reaching $Y_e\sim0.4-0.5$. This is expected since the bulk of the material injected post-12ms carries the relatively higher $Y_e$ values typical of neutrino-processed ejecta. We also observe a drop in the total amount of low-$Y_e$ material present on the grid, largely due to the fallback of slow-moving, low-$Y_e$ material through the inner boundary. Heating can change the amount of fallback and in turn affect the final $Y_e$ distribution, though the effect here is fairly small.

\begin{figure*}
    \centering
    \subfigure{\includegraphics[width=0.33\textwidth]{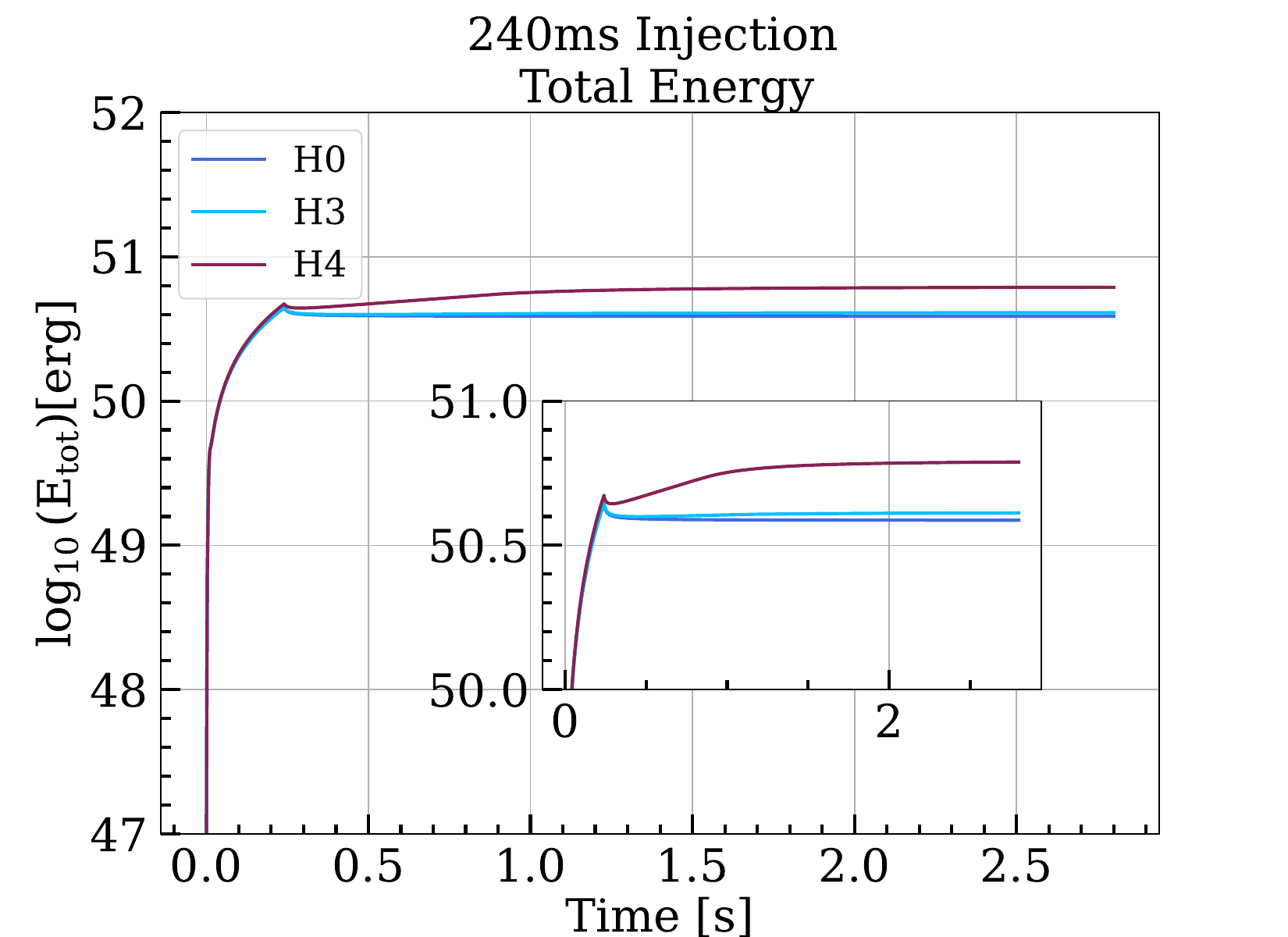}}
    \subfigure{\includegraphics[width=0.33\textwidth]{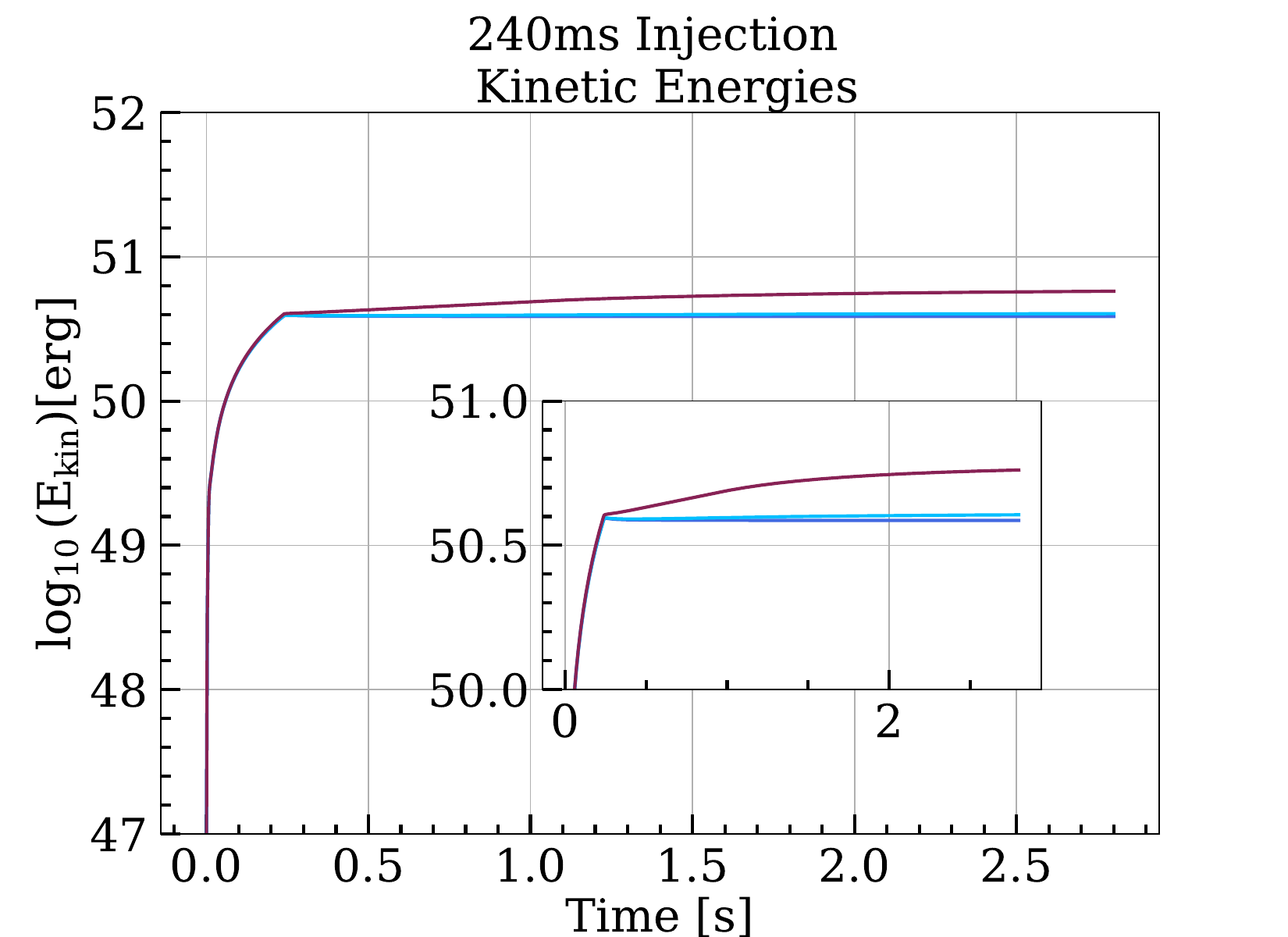}}
    \subfigure{\includegraphics[width=0.33\textwidth]{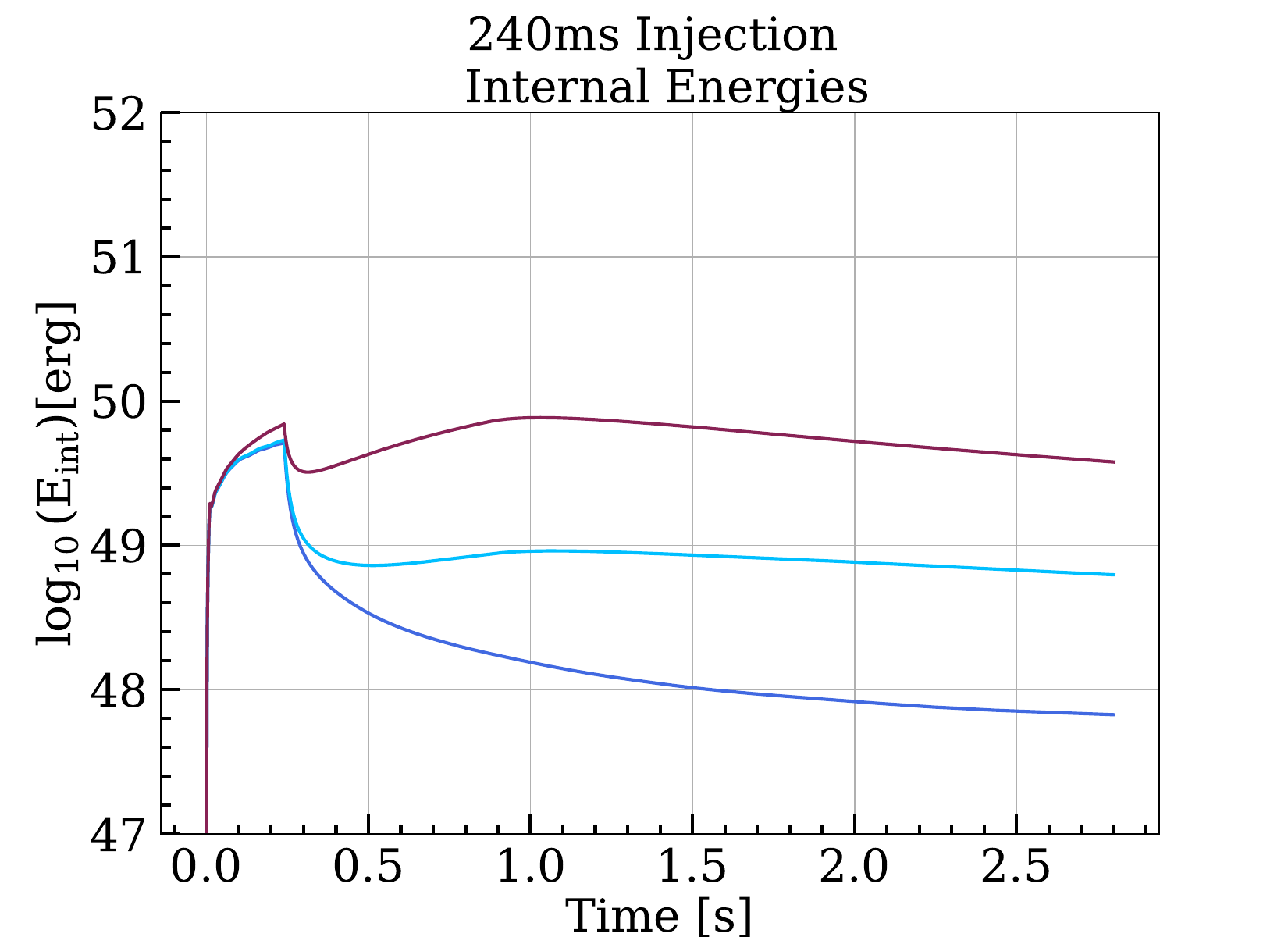}}
    \caption{Total, kinetic, and internal energy evolution for runs without heating (H0) and with heating prescriptions \texttt{H3} and \texttt{H4} in the 240ms injection scenario. Insets in the kinetic and total energy panels provide a zoomed view to highlight the energy differences between runs \texttt{H0}, \texttt{H3}, and \texttt{H4}.}
    \label{fig:energies}
\end{figure*}

The impact of $r$-process heating on the total, kinetic, and internal energy evolution in the 240ms injection scenario is shown in Figure \ref{fig:energies}. The total energy in the simulations with $r$-process heating (\texttt{H3}, \texttt{H4}) remains consistently higher than
in the run without heating (\texttt{H0}). For \texttt{H3} the increase is modest, whereas \texttt{H4} reaches total energies about 60\% higher than \texttt{H0}, with a peak around $10^{50.8}$ erg compared to $\sim10^{50.6}$ erg for \texttt{H0}. A similar trend is observed in the kinetic energies. The most significant effect appears in the internal energies. The energy decrease at 240ms corresponds to the injection shutoff. After this shutoff, the internal energy in the \texttt{H3} case is roughly an order of magnitude higher compared to the \texttt{H0} case, with \texttt{H3} peaking around $\sim10^{49}$ erg compared to $\sim10^{48}$ erg for \texttt{H0}. \texttt{H4}, in turn, reaches almost an order of magnitude higher internal energy at its peak than \texttt{H3}, approaching $10^{50}$ erg before gradually decreasing. 

As the ejecta expand, the internal energy will gradually be converted into kinetic energy. In our simulations, only the strongest heating case (\texttt{H4}) results in a noticeable increase in the total energy budget. While the resulting increase in radial velocity is unlikely to substantially affect the bulk kilonova emission, it could have implications for radio emission when high-velocity ($v_r>0.4c$) ejecta components interact with the surrounding circumstellar medium.

As discussed in the previous section, \texttt{H0} and \texttt{H3} exhibit similar overall morphologies, whereas \texttt{H4} significantly changes the ejecta structure (see Appendix \ref{app:A}: Figures \crefrange{fig:12ms_H4_3x3}{fig:cont_H4_3x3}).

\subsubsection{Homology}\label{homology}

\begin{figure}
    \centering
    \includegraphics[width=0.9\columnwidth]{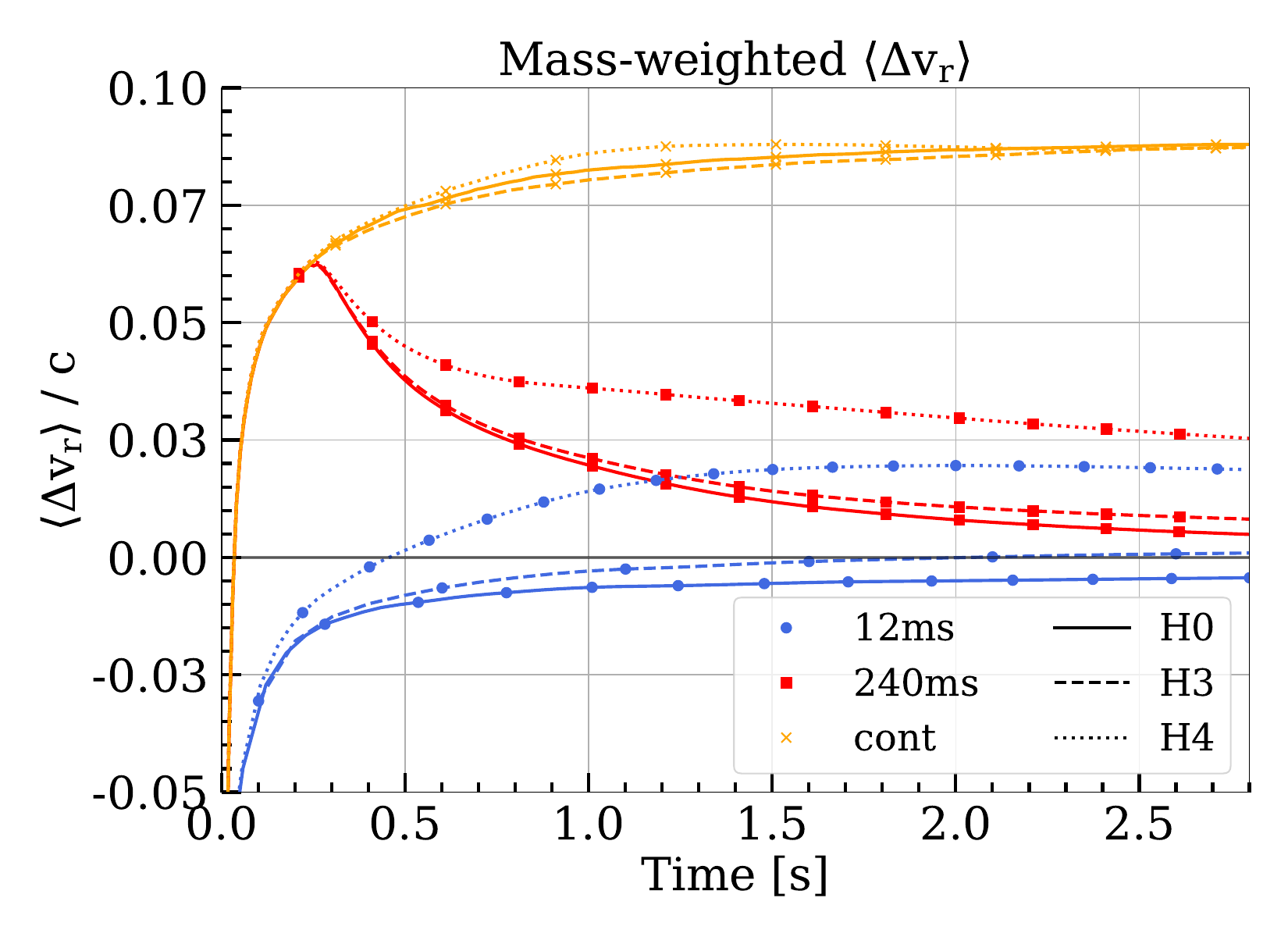}
    \caption{Figure showing the homology calculations as per Equation \ref{eqn:homology} for each of the three injection scenarios (12ms, 240ms, continuous) for no heating (\texttt{H0}) and heating (\texttt{H3}, \texttt{H4}).}
    \label{fig:homology}
\end{figure}

A qualitative assessment of homologous expansion can be made by examining the evolution of the $v_r$ field. Figures~\ref{fig:12ms_H3_3x3}--\ref{fig:cont_H3_3x3} show that high-velocity regions expand outward while largely preserving their spatial structure, indicating a gradual progression toward homologous expansion. This trend is quantified in Figure \ref{fig:homology}.

At early times, all three injection cases exhibit negative $\langle \Delta v_r \rangle$, indicating that the outflow expands more slowly than $r/t$. 

In the 12ms injection case, the ejected material initially exhibits a relatively high $v_r$ dispersion and gradually approaches a homologous profile as it expands. By contrast, in the 240ms and continuous cases the prolonged energy input initially drives the outflow away from the homology condition. 

In the continuous injection scenario, the sustained injection of high-$\langle v_r\rangle$ material leads to a persistent offset from the homologous condition. As a result, the outflow does not converge to homology and $\langle v_r\rangle$ saturates at a non-zero offset even at later times. The intermediate 240ms case initially behaves similarly to the continuous injection scenario but transitions toward homologous expansion once the energy injection ceases.

Heating effects are also evident, though relatively modest between \texttt{H0} and \texttt{H3}. In the continuous injection scenario, heating has only a secondary influence on the approach to homology, as its effect is dominated by the ongoing injection of non-homologous material. For the 12ms case, particularly at late times, model \texttt{H3} reaches homology slightly sooner than \texttt{H0}. This can be understood by noting that the short injection duration yields a relatively small ejected mass and a lower overall energy budget. The modest additional heating in \texttt{H3} provides just enough energy for the outflow to attain homology more efficiently for this case. In contrast, the stronger heating in \texttt{H4} drives the 12ms simulation toward $\langle \Delta v_r\rangle/c\sim0.02$, with a similar effect observed in the 240ms case. These simulations may require much longer timescales to fully converge to zero, as after 1 s heating does not stop but decays as a power law, maintaining an influence on the outflow. 

As heating rate \texttt{H4} is suitable for material with $Y_e \sim$ 0.1 whereas the bulk of our ejecta have relatively high $Y_e$ values, we focus on results obtained for models with heating rate \texttt{H3} for the discussion of kilonovae. 

\subsection{\texttt{Sedona} Results}\label{res:Sedona}

For each of the three injection scenarios with heating rate \texttt{H3}, we map \texttt{FLASH} data at 2.5 seconds  to \texttt{Sedona} and compute spectra. 

As discussed in subsection \ref{homology}, the ejecta in the 12ms and 240ms \texttt{FLASH} simulations are expanding nearly homologously by the time of mapping. When we map the outflow from \texttt{FLASH} to \texttt{Sedona} and impose homologous expansion for these models, \texttt{Sedona} finds a small discrepancy of $\sim$4-6\% with respect to the total ejecta mass and $\sim$2-5\% with respect to the total kinetic energy as computed in the \texttt{FLASH} simulation. We expect this discrepancy to have negligible impact on our results. However, for the continuous injection case, the \texttt{FLASH} simulation is, and remains by construction, non-homologous. Here, we see a $\sim$20\% drop in total ejecta mass and $\sim$15-17\% in kinetic energy when mapping to \texttt{Sedona}. Increasing the \texttt{Sedona} spatial grid resolution to twice the fiducial resolution employed here does improve this discrepancy, but a $\sim 10\%$ drop in total mass and kinetic energy still persists. 
We present kilonova light curves from the higher resolution simulation as a potential outcome for the case of a seconds-long engine lifetime, and note that they differ noticeably from the results obtained with the fiducial resolution only at $\lesssim$0.5 days. The total amount of mass on the \texttt{Sedona} grid post-mapping is of the order of $\sim 10^{-3},10^{-2}$ and $10^{-1} M_{\odot}$ respectively for the 12ms, 240ms, and continuous injection simulations. 

The start time for the \texttt{Sedona} simulations is chosen as 0.01 days for the 12ms and 240ms injection cases in order to adequately include the effects of radioactive heating and capture the peaks of the resulting light curves. For the continuous injection case, we opt for a start time of 0.1 days instead, since no significant spectral or light curve features are expected to occur before this time, and the delayed start time significantly speeds up the simulations by minimizing the computational cost of modeling the early, optically-thick phases of the ejecta. Within the scope of this paper, we do not attempt to accurately model the first few hours of emission or the early UV component in detail; instead, our focus is on the phase that would be observable by LSST. Our choice of start time for the \texttt{Sedona} calculation does affect the early time ($\lesssim$ 1 day) behavior of the light curves, with earlier start times yielding slightly brighter emission. However, the effect is small, and does not alter the qualitative trends described below. As an additional check, we ran a simulation for the continuous injection scenario using a smaller MC particle count, which allowed us to start the calculation at 0.01 days instead of 0.1 days, albeit with noisier results. We found the resulting magnitudes varied by at most 0.3 mag, mainly at early times and primarily in the shorter wavelength bands.

\begin{figure}
    \centering
   \begin{tabular}{c}
    \includegraphics[width=0.45\textwidth]{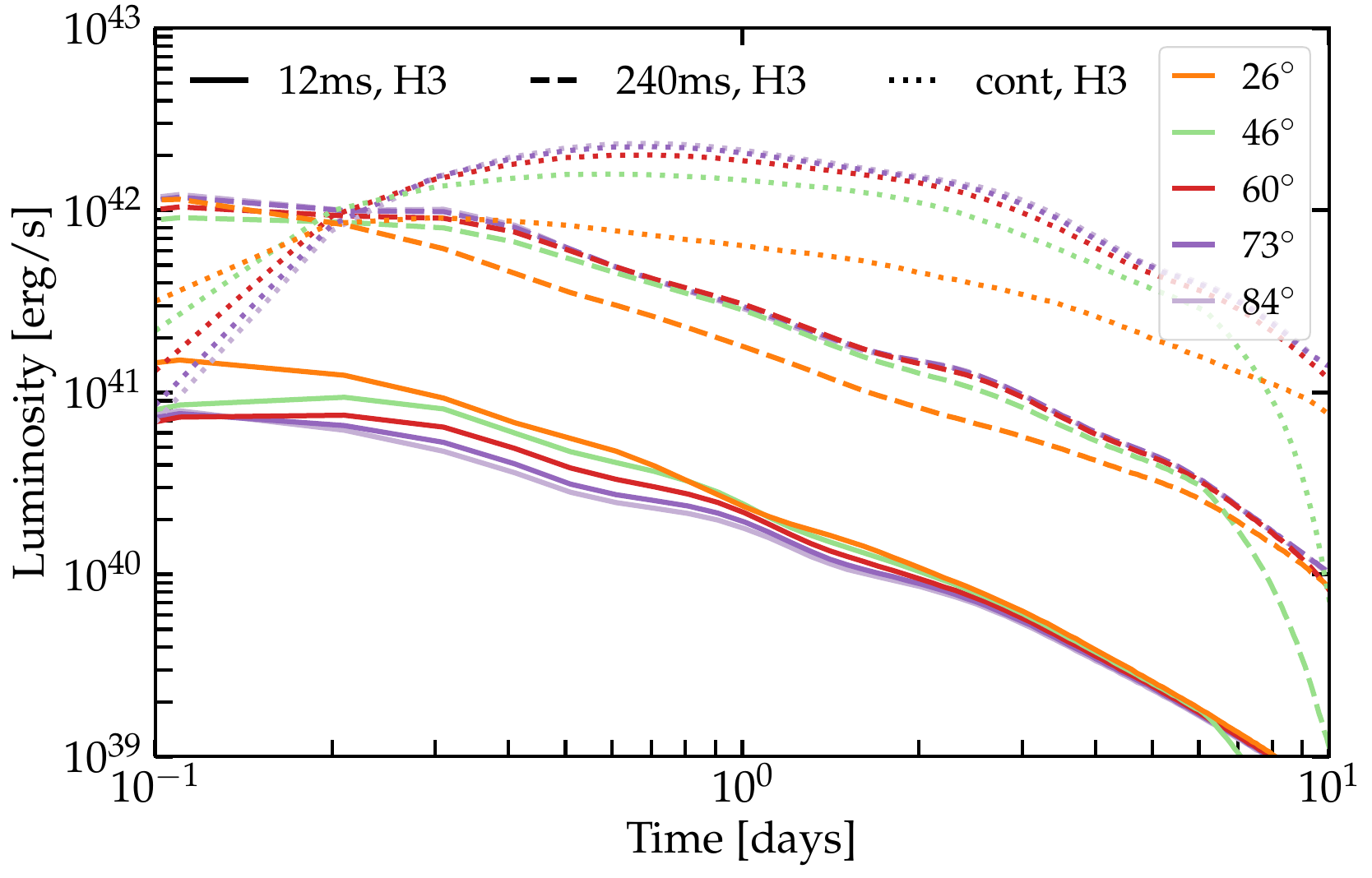}\\

    \end{tabular}        
    \caption{Bolometric luminosity as a function of time for the three injection scenarios, with heating \texttt{H3}. Different colors correspond to different viewing angles, where $0^\circ$ is aligned with the poles and $90^\circ$ is aligned with the equator.   
    }
    \label{fig:bolometric}
\end{figure}

The bolometric luminosities for the 12ms, 240ms, and continuous injection scenarios, employing heating rate \texttt{H3} in the respective \texttt{FLASH} simulations, are shown in Figure \ref{fig:bolometric}. The peak luminosities for the three cases are $\sim 1.5 \times 10^{41}$ erg s$^{-1}$, $1.22 \times 10^{42}$ erg s$^{-1}$, and $2.33 \times 10^{42}$ erg s$^{-1}$ respectively, increasing with longer injection times as expected due to the corresponding increase in total ejecta mass. For each scenario, the time evolution of observed luminosity also depends on the viewing angle, persisting until the ejecta become transparent on a timescale of days to weeks. For the 12ms case, the emission is brightest for polar viewing angles, with the peak luminosity roughly a factor of 2 higher at the poles compared to the equator. The opposite effect is seen for the 240ms and continuous injection cases. Here, except at very early times, the emission is brighter for equatorial viewing angles than for polar viewing angles.  

These trends arise from a combination of factors. For an equatorial viewer, the projected surface area observed is larger compared to the pole, which results in brighter emission \citep{Darbha_2020}. This effect is responsible for the behavior seen in the 240ms and continuous injection scenarios. However, luminosity is also enhanced by Doppler boosting in viewing directions aligned with higher velocities the bulk of the outflow, resulting in increased luminosity for polar observing angles \citep{Darbha_2021}. 

Additionally, in the 12ms scenario, a non-negligible fraction of the total ejected material lies in the equatorial plane and has lower $Y_e$ values. This material has higher lanthanide fractions and correspondingly higher opacities (See Figure \ref{fig:xla} in the Appendix). The observed luminosities in the 12ms case are thus lower when viewed from the equator. If we switch off detailed opacities and simulate the 12ms ejecta using a gray opacity that does not take composition into account, the correlation between luminosity and viewing angle reverts to the same trend seen for longer injection scenarios. For the longer injection cases, higher opacity material represents a relatively small fraction of ejecta and the projected surface area is the dominant factor setting the observed luminosity.

\begin{figure*}
    \centering
    \subfigure[$u$-band]
    {\includegraphics[width=0.33\textwidth]{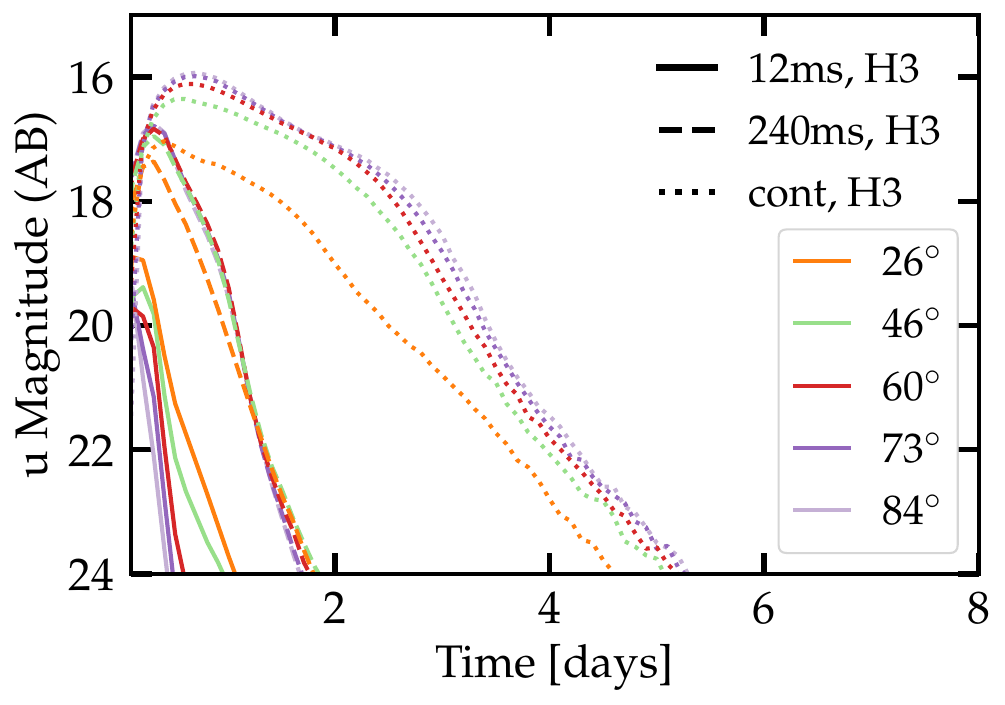}}
    \subfigure[$g$-band]{\includegraphics[width=0.33\textwidth]{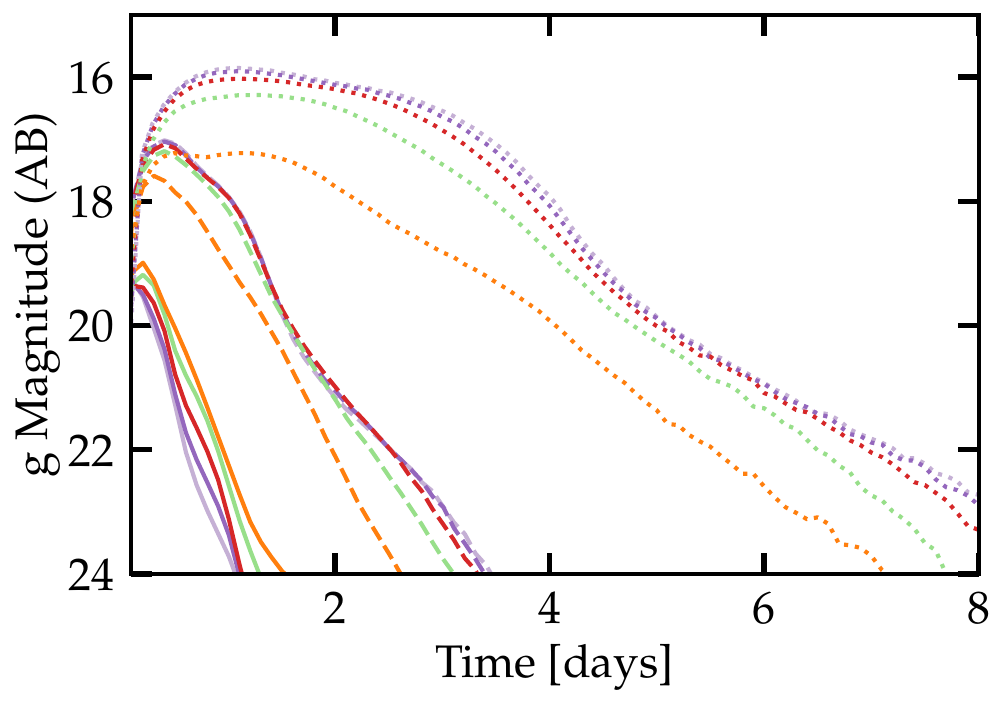}}
    \subfigure[$r$-band]{\includegraphics[width=0.33\textwidth]{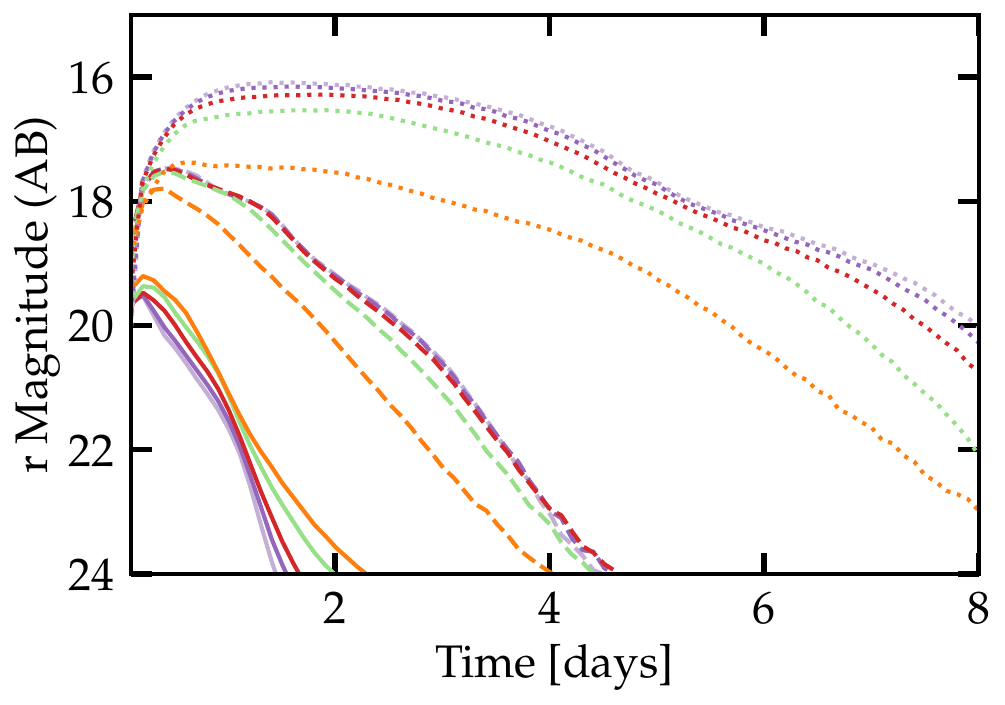}}
    \subfigure[$i$-band]{\includegraphics[width=0.33\textwidth]{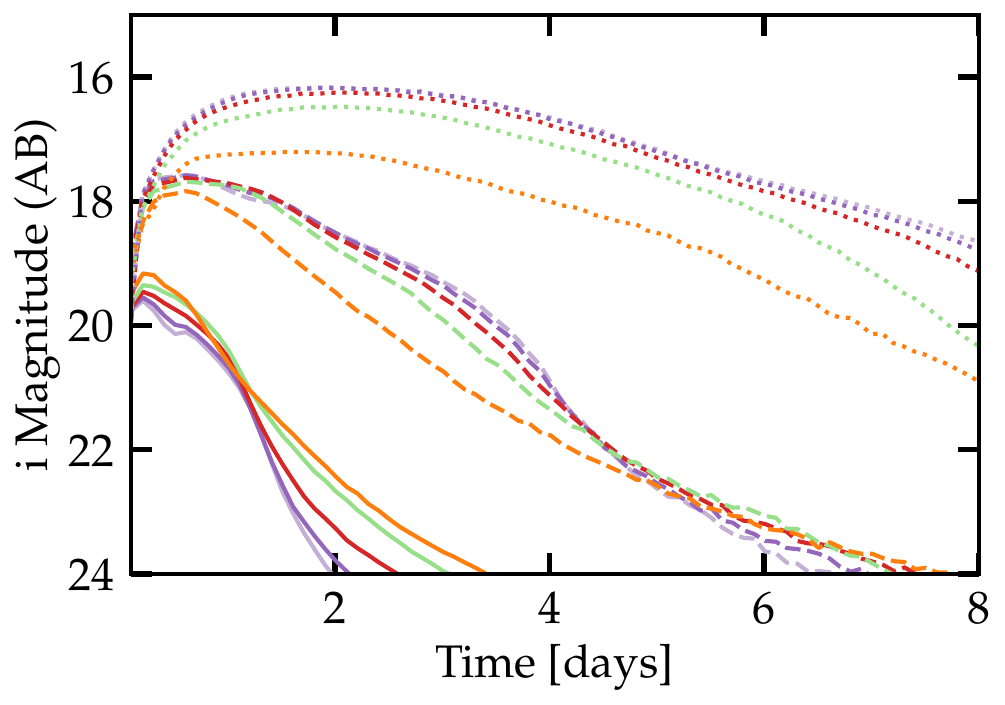}}
    \subfigure[$z$-band]{\includegraphics[width=0.33\textwidth]{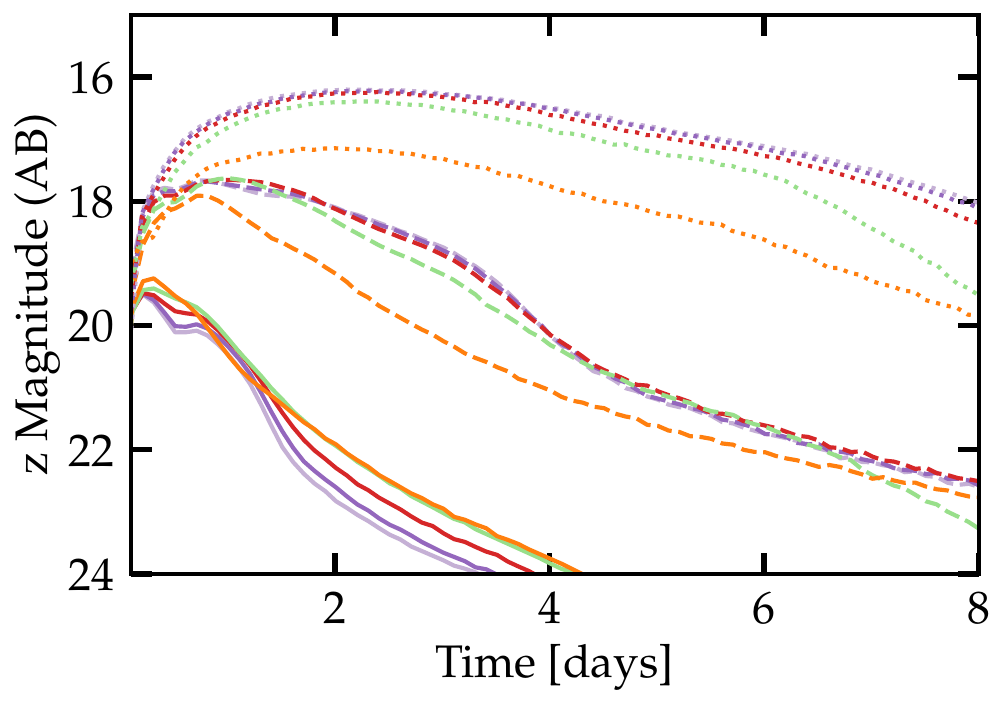}}
    \subfigure[$y$-band]{\includegraphics[width=0.33\textwidth]{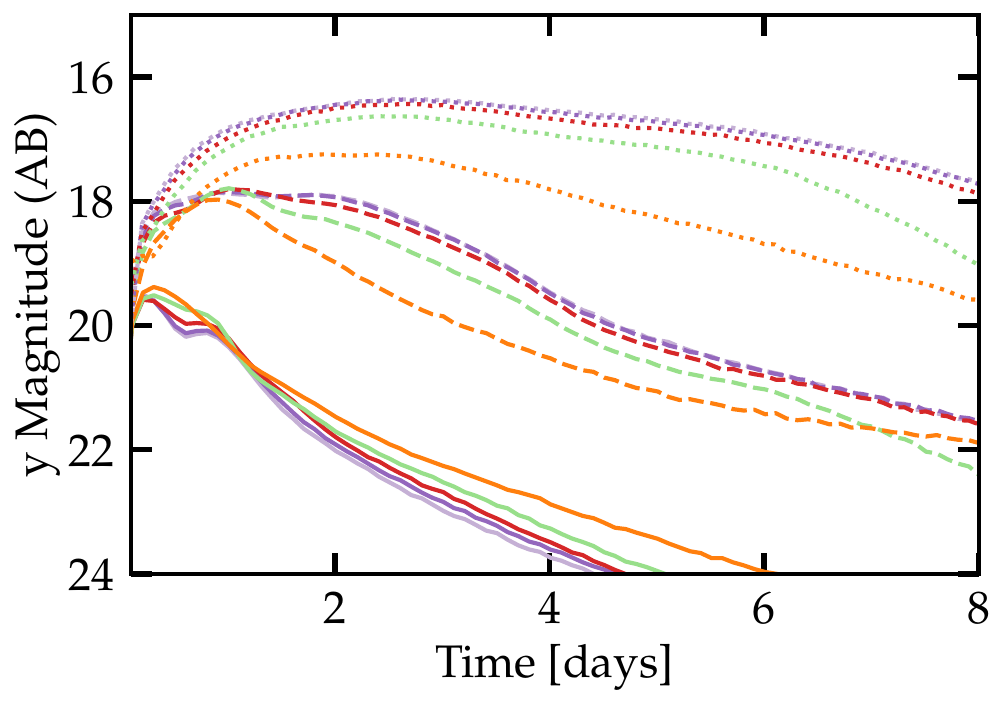}}
    \caption{AB magnitudes in LSST filters for the three injection scenarios, employing heating mode \texttt{H3}. Different colors indicate a range of viewing angles, where  $0^\circ$ is aligned with the poles and $90^\circ$ is aligned with the equator. The distance is set to 40 Mpc, similar to AT2017gfo.}
    \label{fig:band_lightcurves}
\end{figure*}

Figure \ref{fig:band_lightcurves} presents the kilonova light curves in individual LSST bands, covering the ultraviolet ($u$ band), the optical ($g, r$, and $i$ bands), and the near-infrared ($z$ and $y$ bands). Across all filters, the light curves exhibit a rapid rise to peak brightness followed by a steady decline. In general, the $u$-band peaks earliest and fades rapidly, while emission at longer wavelengths persists longer. Similar to the bolometric light curves, the band light curves also depend on viewing angle, reflecting the same overall trend as the bolometric luminosity. Across all bands, the 12ms engine produces brighter emission viewed from the polar direction while the reverse is true for the longer-lived engines. For the 12ms case, depending on the band and varying with the viewing angle, the peak magnitudes lie between $\sim$19--20 mag and the peak times lie between $\sim$0.1-0.3 days. For the 240ms scenario, emission in various bands tends to peak around $\sim$17--18 mag, with peak timescales ranging from $\sim$0.3 days for the $u$-band to roughly 1 day for the $y$-band. The $u$-band peaks earliest for the continuous injection run as well, between $\sim$0.4--0.7 days,  with peak magnitudes of $\sim$16--17 mag. The longer wavelength bands peak in a similar magnitude range on a timescale of $\sim$1--3 days. 

Thus, of the three scenarios presented here, the continuous injection case produces the highest peak magnitudes and longest peak times in all bands, followed by the 240ms case, while the 12ms case produces the dimmest and fastest evolving transient. One interesting effect is that, for the short wavelength bands, the peak magnitudes attained for the continuous injection scenario and for the 240ms scenario can overlap if the viewing angle in the former case is polar and the viewing angle in the latter case is equatorial, illustrating the interplay between projected surface area and observed luminosities. The two scenarios can nonetheless be distinguished based on the post-peak observations, as the more massive outflow in the continuous injection case remains bright for a longer time, i.e. exhibits a slower decline rate, compared to the 240ms case.

\begin{figure*}
    \centering
    \includegraphics[width=0.9\textwidth]{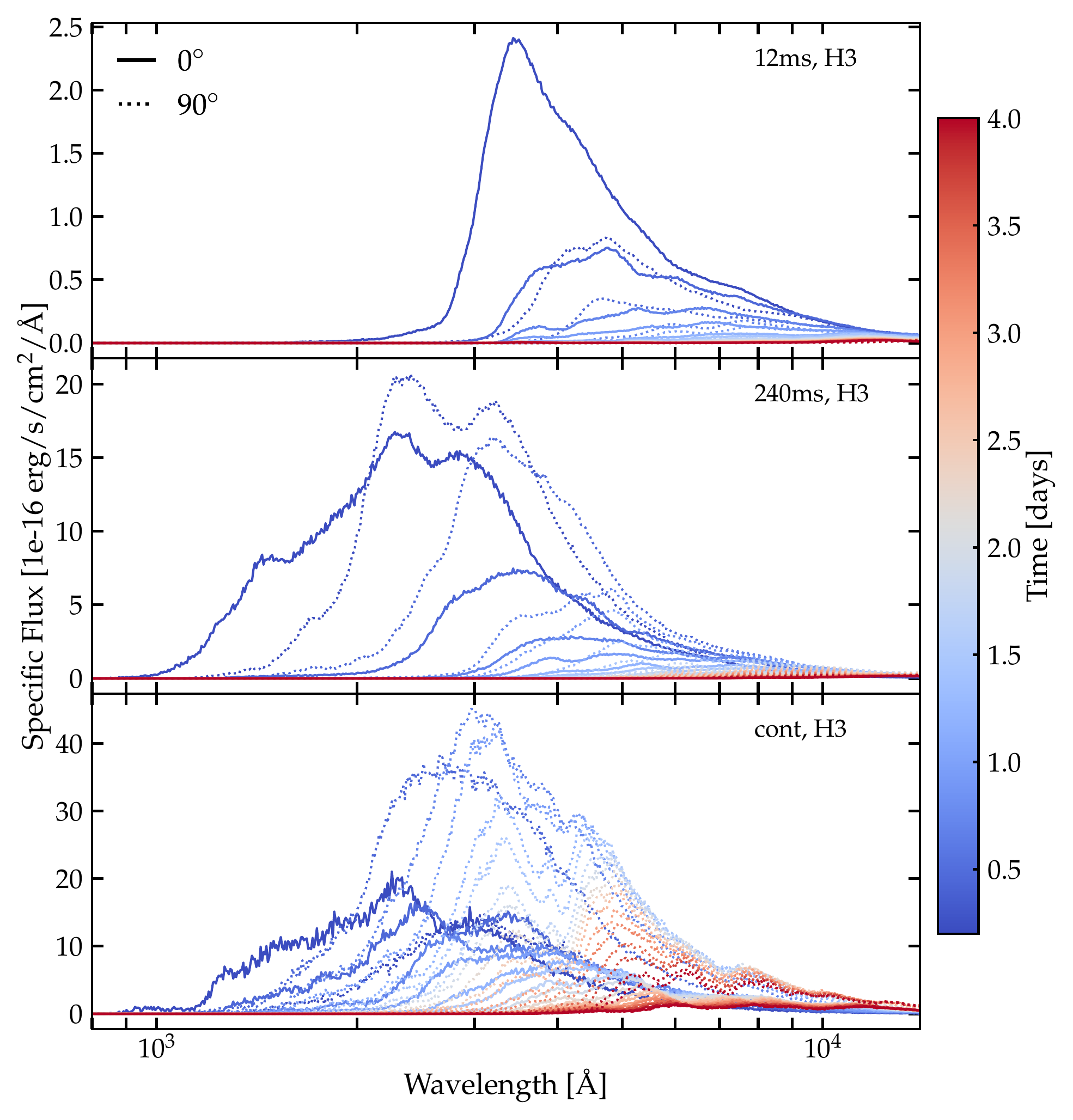}
    \caption{Spectral evolution for three injection scenario, employing heating mode \texttt{H3}, shown over the first four days for two viewing angles, where $0^\circ$ aligned with the poles and $90^\circ$ aligned with the equator. Note the different y-axis ranges of the specific flux in the three scenarios.}
    \label{fig:spectra}
\end{figure*}

In Figure \ref{fig:spectra}, we present the spectral evolution for each injection scenario discussed in Figure \ref{fig:band_lightcurves}, focusing on the period from $t=0.2$ days to $t=4$ days. To illustrate the viewing angle dependence, for each scenario, the spectra are plotted as observed at the poles and at the equator, reflecting the trends evident in the band light curves. Overall, although the composition contains a small amount of lanthanide-bearing material, it is insufficient to induce strong reddening of the emission. The spectra remains relatively blue for days, extending into the UV range (approximately $100\text{--}4000$ \r{A}).

\begin{figure*}
    \centering
    \subfigure{\includegraphics[width=0.33\textwidth]{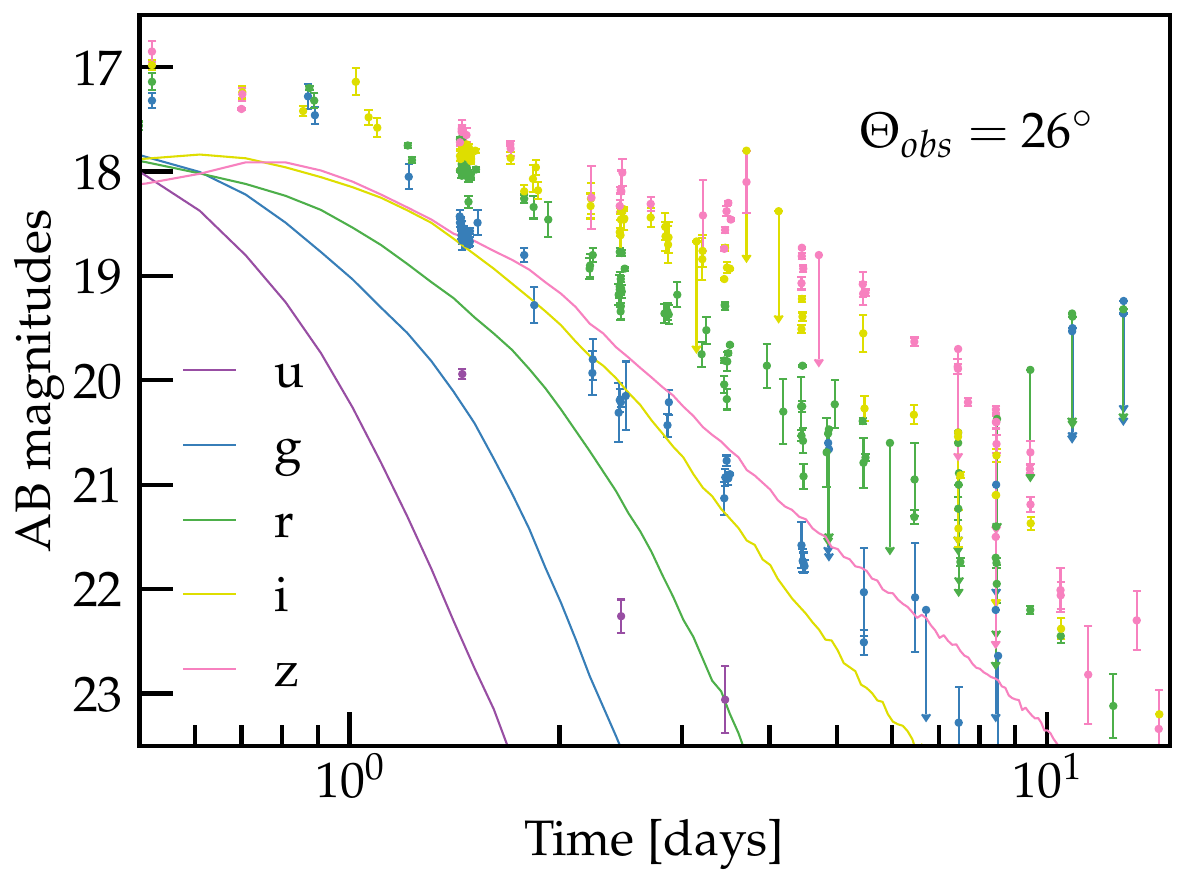}}
    \subfigure{\includegraphics[width=0.33\textwidth]{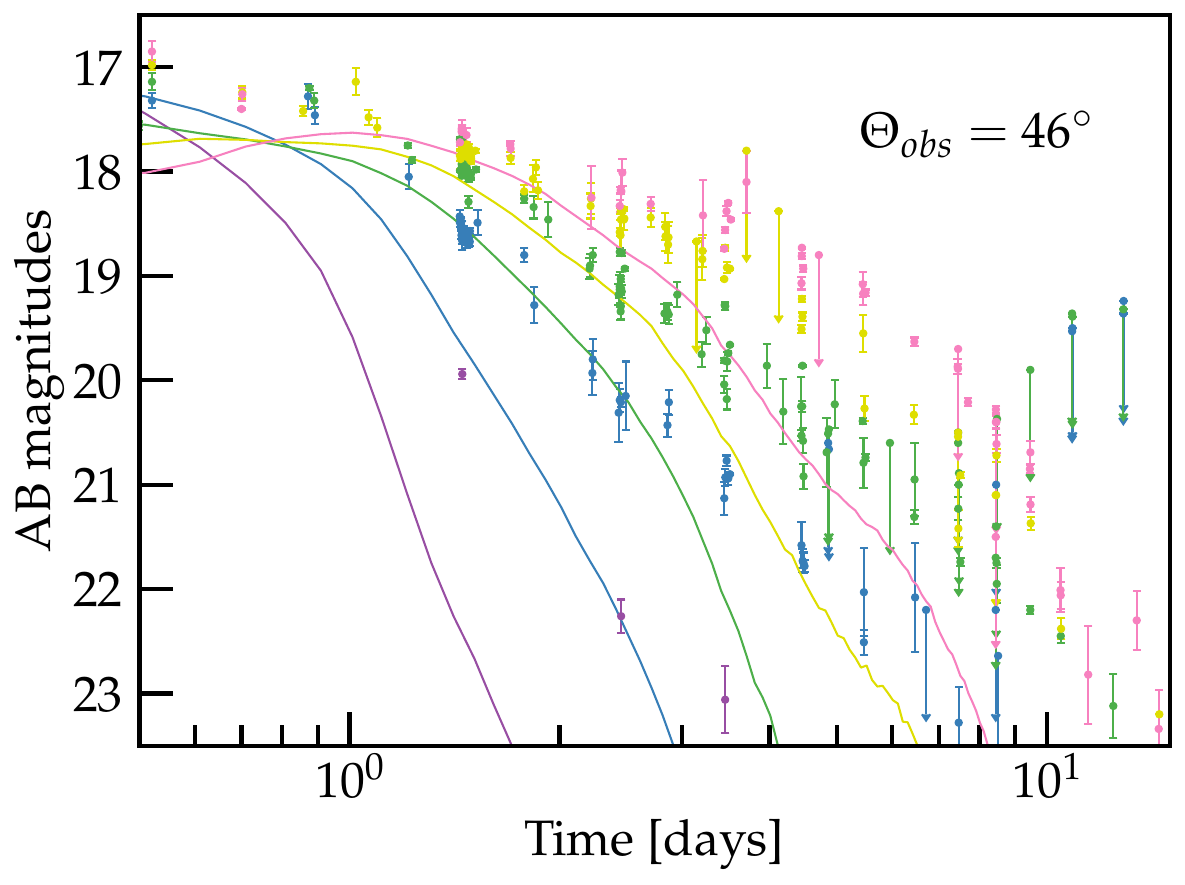}}
    \subfigure{\includegraphics[width=0.33\textwidth]{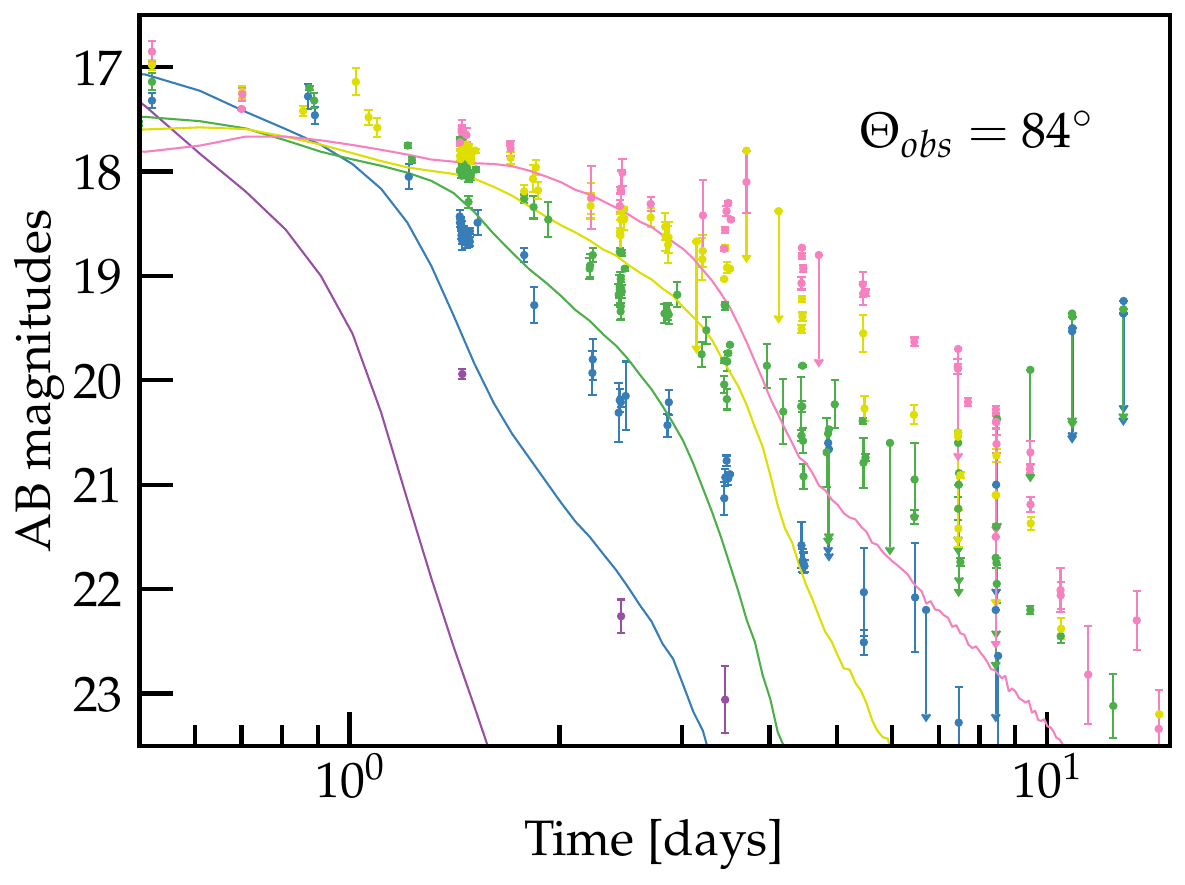}}
    \caption{AB magnitudes in LSST filters for the 240ms injection scenario, employing heating mode \texttt{H3}. The $ugriz$ magnitudes are plotted for three viewing angles $\Theta_{obs}=26^{\circ}, 46^{\circ}$, and $84^{\circ}$. The points represent observed photometric data for AT2017gfo obtained by multiple teams using a variety of instruments \protect\citep{Coulter2017, Andreoni2017, Arcavi2017, Cowperthwaite2017, Diaz2017, Drout2017, Evans2017, Hu2017, Kasliwal2017, Lipunov2017, Pian2017, Pozanenko2018, Smartt2017, Tanvir2017, Troja2017, Utsumi2017, Valenti2017}, compiled and presented in Table 3 of \protect\cite{Villar_2017}.}
    \label{fig:at2017gfo}
\end{figure*}

In Figure \ref{fig:at2017gfo}, we show magnitudes in LSST $ugriz$ bands computed for the 240ms injection scenario, with heating rate \texttt{H3} employed in the \texttt{FLASH} simulation, along with the observed data for AT2017gfo (see Table 3 of \cite{Villar_2017}, and references therein) added for comparison. Although we do not expect to perfectly reproduce the spectral evolution of AT2017gfo for this particular remnant, and the choice of atomic opacities also strongly affects specific spectral features, we can still comment on the question of early blue emission and, in the context of these particular models, on the preferred viewing angle for AT2017gfo. For the 240ms remnant lifetime scenario simulated here, viewing angles close to the poles ($\lesssim$ 26$^{\circ}$) produce a dimmer kilonova than the blue component of AT2017gfo. This places the 240ms model simulated here in tension with the results of \cite{Mooley2022}, who estimate a viewing angle of $19-25^{\circ}$. However, viewed from a 46$^{\circ}$ polar angle, the kilonova does approach the brightness of the blue component of AT2017gfo at early times, though it fades quickly post $\sim$1 day. This suggests better agreement of the kilonova model for this particular remnant with the wider estimated viewing angle range of $\sim20-40^{\circ}$ in \cite{Alexander2017} and \cite{Margutti2017}. 

\section{Summary and Discussion}
\label{sec:summary}

We have developed an end-to-end kilonova modeling pipeline that couples 3D GRMHD simulations conducted with  \texttt{ETK} to 2D hydrodynamic and radiative transfer simulations. This pipeline bridges the gap from the early moments of ejecta launch to the homologous expansion phase. In this work, we present the first application of this pipeline to the case of kilonovae produced by HMNS remnants. Starting from a 3D GRMHD simulation of a HMNS remnant including realistic neutrino transport, we map the remnant outflows into the \texttt{FLASH} hydrodynamics code to follow their evolution in axisymmetry for 2.5 seconds. Once the ejecta reached near-homologous expansion, we post-process the simulations with the radiative transfer code \texttt{Sedona} to generate multi-band kilonova light curves and spectra. We also assess the effects of different engine lifetimes and $r$-process heating rates.

Our results show that the remnant lifetime determines the total ejecta mass, morphology, and viewing-angle dependence of the resulting transient. We find that moderate $r$‑process heating (\texttt{H3}) only modestly affects ejecta energetics and morphology. A high heating rate typical of $Y_e \sim 0.1$ ejecta (\texttt{H4}) drives them toward a more spherical morphology, however, this rate is unlikely to be realistic for our ejecta composition. 

We present kilonova light curves which show a strong dependence of the observed luminosity on the observer's viewing angle, and peak luminosities can differ by up to a factor of 2 or more depending on the viewing angle. The observed luminosity is set mainly by the projected surface area of the emitting material, with polar viewing angles appearing dimmer than equatorial directions given the elongated plume-like morphology of our ejecta. However, the presence of lanthanides can change this picture, dimming the equatorial emission as seen for the 12ms scenario, and detailed ejecta composition must be taken into account when predicting kilonova light curves.  

Our axisymmetric setup combining hydrodynamics and detailed radiative transfer provides a computationally efficient framework for exploring how key physical parameters—such as remnant lifetime, $r$-process heating—affect outflow geometry and kilonova signals. The relatively low-cost simulations presented here allow us to identify interesting regions of the parameter space and determine approximate ejecta dynamics, such as the timescale for homology, both motivating and aiding future 3D explorations. In future work, we will employ a fully 3D pipeline using the same set of codes (\texttt{ETK, FLASH, Sedona}), and explore azimuthal viewing-angle dependence of the kilonova signal as well, with the goal of reproducing AT2017gfo. 

In order to compare directly with the AT2017gfo kilonova, it will be necessary to model all phases of the merger self-consistently and track all components of merger ejecta. In particular, the tidal ejecta launched during the merger and winds driven from the accretion disk can strongly modify the kilonova signal ultimately observed, and also contribute to redder emission, generally referred to as a red kilonova component. Different ejecta components may also interact with each other, resulting in shocks and mixing. While our focus in this work was on characterizing the magnetized HMNS wind component, we will include additional ejecta components and explore these interactions in future work. We may find that emission from the winds we model here is obscured by fast-moving, high-opacity, lanthanide-rich dynamical ejecta produced during the merger, in particular along equatorial viewing angles. However, polar viewing angles are likely to remain unaffected and the kilonova light curves presented here could be compared directly with such observations. 

Overall, our results support the production of blue kilonovae by magnetized outflows driven from temporarily stable HMNS remnants for cases when the remnant survives long enough to produce massive ejecta, in agreement with our previous work \citep{Curtis_2024}, and motivate detailed modeling of such longer-lived remnants ($\sim$ 200 -- 300 ms) as promising progenitors of the blue emission in AT2017gfo.

There are some limitations to our current approach which will be improved in forthcoming work. The assumption of axisymmetry neglects any existing azimuthal motions of the fluid. We believe this to be a reasonable assumption given the large radial velocities of ejecta (relative to polar and azimuthal) as well as overall azimuthal symmetry of various other ejecta properties. Nonetheless, a 3D simulation of the same system is needed to estimate the impact of this approximation. In simulations where we extend the remnant lifetime, we extrapolate the ejecta properties based on data available from the 12ms long 3D GRMHD simulation. While the focus of this work is to establish the 2D pipeline for kilonova modeling, in future 3D GRMHD work, we will simulate additional mergers with longer-lived remnants, and directly map ejecta data to long-term simulations with \texttt{FLASH}. This will help determine whether ejecta properties diverge significantly from our extrapolated values, and whether results from the particular remnant simulated here can be generalized to apply for mergers with different properties. 

Heating rate prescriptions and approximate compositions are often employed in kilonova calculations as it is currently not possible to include on-the-fly $r$-process nucleosynthesis in hydrodynamics calculations. Here, we have opted to explore the effect of different heating prescriptions on our ejecta dynamics and morphology. When comparing with observations for the purpose of inference, however, we will select a heating rate appropriate for our ejecta \citep{Sarin2024}, ideally derived from the detailed nucleosynthesis calculations performed during post-processing. Finally, the composition of our ejecta, while set using abundances of tracer particles from the GRMHD simulation, is approximate. Though it would be desirable to compute detailed abundances and corresponding opacities for the entire ensemble of $r$-process nuclei produced in mergers, this remains a significant challenge for current simulation and modelling work, even without accounting for uncertainties in neutrino transport and atomic opacities. As long as lanthanide fractions are appropriately set, we do not expect broad features of the kilonova light curves generated using our pipeline to be significantly affected by small changes in individual isotopic abundances. 

The work presented here enables us to tie kilonova signals to the progenitor binary and map the role of different merger phases and resulting ejecta components. It complements kilonova modeling performed using simplified geometries and compositions inspired by numerical simulations by directly modeling the relevant phases based on detailed simulation output and nucleosynthesis calculations. Combining simplified kilonova models with such end-to-end models across a variety of merging systems can help populate the entire kilonova zoo, and determine observing strategies for transient surveys using e.g. LSST \citep{Chase2022,Ragosta2024,Andrade2025}.

\section*{Acknowledgements}


The software used in this work was developed in part by the DOE NNSA- and DOE Office of Science--supported \texttt{FLASH} Center for Computational Science at the University of Chicago and the University of Rochester. SC acknowledges support from the National Science Foundation AAPF Award (End-to-End Modeling of the Kilonova Zoo, Award Number: 2303869). PM acknowledges funding through NWO under grant No. OCENW.XL21.XL21.038. DK is supported in part by the U.S. Department of Energy, Office of Science, Office of Nuclear Physics, DE-AC02-05CH11231, DE-SC0004658, and DE-SC0024388, and by a grant from the Simons Foundation (622817DK).
We acknowledge support from the Gordon and Betty Moore Foundation through Grant GBMF5076. DB is partially supported by a NASA Future Investigators in NASA Earth and Space Science and Technology (FINESST) award No. 80NSSC23K1440. We thank Benny Tsang for helpful discussions about the \texttt{FLASH} code and for contributing the logarithmic radial grid setup employed in these simulations. We thank Carla Fr\"ohlich and Jonah Miller for useful discussions on merger remnants, hydrodynamics, and nucleosynthesis.

\textit{Software:} Einstein Toolkit, FLASH, Sedona, Python, NumPy, SciPy, Matplotlib
\section*{Data Availability}
The data underlying this article will be shared on reasonable request to the corresponding author. 
 



\bibliographystyle{mnras}
\bibliography{main} 

@ARTICLE{Andreoni2017,
       author = {{Andreoni}, I. and {Ackley}, K. and {Cooke}, J. and {Acharyya}, A. and {Allison}, J.~R. and {Anderson}, G.~E. and {Ashley}, M.~C.~B. and {Baade}, D. and {Bailes}, M. and {Bannister}, K. and {Beardsley}, A. and {Bessell}, M.~S. and {Bian}, F. and {Bland}, P.~A. and {Boer}, M. and {Booler}, T. and {Brandeker}, A. and {Brown}, I.~S. and {Buckley}, D.~A.~H. and {Chang}, S. -W. and {Coward}, D.~M. and {Crawford}, S. and {Crisp}, H. and {Crosse}, B. and {Cucchiara}, A. and {Cup{\'a}k}, M. and {de Gois}, J.~S. and {Deller}, A. and {Devillepoix}, H.~A.~R. and {Dobie}, D. and {Elmer}, E. and {Emrich}, D. and {Farah}, W. and {Farrell}, T.~J. and {Franzen}, T. and {Gaensler}, B.~M. and {Galloway}, D.~K. and {Gendre}, B. and {Giblin}, T. and {Goobar}, A. and {Green}, J. and {Hancock}, P.~J. and {Hartig}, B.~A.~D. and {Howell}, E.~J. and {Horsley}, L. and {Hotan}, A. and {Howie}, R.~M. and {Hu}, L. and {Hu}, Y. and {James}, C.~W. and {Johnston}, S. and {Johnston-Hollitt}, M. and {Kaplan}, D.~L. and {Kasliwal}, M. and {Keane}, E.~F. and {Kenney}, D. and {Klotz}, A. and {Lau}, R. and {Laugier}, R. and {Lenc}, E. and {Li}, X. and {Liang}, E. and {Lidman}, C. and {Luvaul}, L.~C. and {Lynch}, C. and {Ma}, B. and {Macpherson}, D. and {Mao}, J. and {McClelland}, D.~E. and {McCully}, C. and {M{\"o}ller}, A. and {Morales}, M.~F. and {Morris}, D. and {Murphy}, T. and {Noysena}, K. and {Onken}, C.~A. and {Orange}, N.~B. and {Os{\l}owski}, S. and {Pallot}, D. and {Paxman}, J. and {Potter}, S.~B. and {Pritchard}, T. and {Raja}, W. and {Ridden-Harper}, R. and {Romero-Colmenero}, E. and {Sadler}, E.~M. and {Sansom}, E.~K. and {Scalzo}, R.~A. and {Schmidt}, B.~P. and {Scott}, S.~M. and {Seghouani}, N. and {Shang}, Z. and {Shannon}, R.~M. and {Shao}, L. and {Shara}, M.~M. and {Sharp}, R. and {Sokolowski}, M. and {Sollerman}, J. and {Staff}, J. and {Steele}, K. and {Sun}, T. and {Suntzeff}, N.~B. and {Tao}, C. and {Tingay}, S. and {Towner}, M.~C. and {Thierry}, P. and {Trott}, C. and {Tucker}, B.~E. and {V{\"a}is{\"a}nen}, P. and {Krishnan}, V. Venkatraman and {Walker}, M. and {Wang}, L. and {Wang}, X. and {Wayth}, R. and {Whiting}, M. and {Williams}, A. and {Williams}, T. and {Wolf}, C. and {Wu}, C. and {Wu}, X. and {Yang}, J. and {Yuan}, X. and {Zhang}, H. and {Zhou}, J. and {Zovaro}, H.},
        title = "{Follow Up of GW170817 and Its Electromagnetic Counterpart by Australian-Led Observing Programmes}",
      journal = {\pasa},
         year = 2017,
        month = dec,
       volume = {34},
          eid = {e069},
        pages = {e069},
          doi = {10.1017/pasa.2017.65},
archivePrefix = {arXiv},
       eprint = {1710.05846},
 primaryClass = {astro-ph.HE},
       adsurl = {https://ui.adsabs.harvard.edu/abs/2017PASA...34...69A}
}

@ARTICLE{Coulter2017,
       author = {{Coulter}, D.~A. and {Foley}, R.~J. and {Kilpatrick}, C.~D. and {Drout}, M.~R. and {Piro}, A.~L. and {Shappee}, B.~J. and {Siebert}, M.~R. and {Simon}, J.~D. and {Ulloa}, N. and {Kasen}, D. and {Madore}, B.~F. and {Murguia-Berthier}, A. and {Pan}, Y. -C. and {Prochaska}, J.~X. and {Ramirez-Ruiz}, E. and {Rest}, A. and {Rojas-Bravo}, C.},
        title = "{Swope Supernova Survey 2017a (SSS17a), the optical counterpart to a gravitational wave source}",
      journal = {Science},
         year = 2017,
        month = dec,
       volume = {358},
       number = {6370},
        pages = {1556-1558},
          doi = {10.1126/science.aap9811},
archivePrefix = {arXiv},
       eprint = {1710.05452},
 primaryClass = {astro-ph.HE},
       adsurl = {https://ui.adsabs.harvard.edu/abs/2017Sci...358.1556C}
}

@ARTICLE{Arcavi2017,
       author = {{Arcavi}, Iair and {Hosseinzadeh}, Griffin and {Howell}, D. Andrew and {McCully}, Curtis and {Poznanski}, Dovi and {Kasen}, Daniel and {Barnes}, Jennifer and {Zaltzman}, Michael and {Vasylyev}, Sergiy and {Maoz}, Dan and {Valenti}, Stefano},
        title = "{Optical emission from a kilonova following a gravitational-wave-detected neutron-star merger}",
      journal = {\nat},
         year = 2017,
        month = nov,
       volume = {551},
       number = {7678},
        pages = {64-66},
          doi = {10.1038/nature24291},
archivePrefix = {arXiv},
       eprint = {1710.05843},
 primaryClass = {astro-ph.HE},
       adsurl = {https://ui.adsabs.harvard.edu/abs/2017Natur.551...64A}
}

@ARTICLE{Cowperthwaite2017,
       author = {{Cowperthwaite}, P.~S. and {Berger}, E. and {Villar}, V.~A. and {Metzger}, B.~D. and {Nicholl}, M. and {Chornock}, R. and {Blanchard}, P.~K. and {Fong}, W. and {Margutti}, R. and {Soares-Santos}, M. and {Alexander}, K.~D. and {Allam}, S. and {Annis}, J. and {Brout}, D. and {Brown}, D.~A. and {Butler}, R.~E. and {Chen}, H. -Y. and {Diehl}, H.~T. and {Doctor}, Z. and {Drout}, M.~R. and {Eftekhari}, T. and {Farr}, B. and {Finley}, D.~A. and {Foley}, R.~J. and {Frieman}, J.~A. and {Fryer}, C.~L. and {Garc{\'\i}a-Bellido}, J. and {Gill}, M.~S.~S. and {Guillochon}, J. and {Herner}, K. and {Holz}, D.~E. and {Kasen}, D. and {Kessler}, R. and {Marriner}, J. and {Matheson}, T. and {Neilsen}, Jr., E.~H. and {Quataert}, E. and {Palmese}, A. and {Rest}, A. and {Sako}, M. and {Scolnic}, D.~M. and {Smith}, N. and {Tucker}, D.~L. and {Williams}, P.~K.~G. and {Balbinot}, E. and {Carlin}, J.~L. and {Cook}, E.~R. and {Durret}, F. and {Li}, T.~S. and {Lopes}, P.~A.~A. and {Louren{\c{c}}o}, A.~C.~C. and {Marshall}, J.~L. and {Medina}, G.~E. and {Muir}, J. and {Mu{\~n}oz}, R.~R. and {Sauseda}, M. and {Schlegel}, D.~J. and {Secco}, L.~F. and {Vivas}, A.~K. and {Wester}, W. and {Zenteno}, A. and {Zhang}, Y. and {Abbott}, T.~M.~C. and {Banerji}, M. and {Bechtol}, K. and {Benoit-L{\'e}vy}, A. and {Bertin}, E. and {Buckley-Geer}, E. and {Burke}, D.~L. and {Capozzi}, D. and {Carnero Rosell}, A. and {Carrasco Kind}, M. and {Castander}, F.~J. and {Crocce}, M. and {Cunha}, C.~E. and {D'Andrea}, C.~B. and {da Costa}, L.~N. and {Davis}, C. and {DePoy}, D.~L. and {Desai}, S. and {Dietrich}, J.~P. and {Drlica-Wagner}, A. and {Eifler}, T.~F. and {Evrard}, A.~E. and {Fernandez}, E. and {Flaugher}, B. and {Fosalba}, P. and {Gaztanaga}, E. and {Gerdes}, D.~W. and {Giannantonio}, T. and {Goldstein}, D.~A. and {Gruen}, D. and {Gruendl}, R.~A. and {Gutierrez}, G. and {Honscheid}, K. and {Jain}, B. and {James}, D.~J. and {Jeltema}, T. and {Johnson}, M.~W.~G. and {Johnson}, M.~D. and {Kent}, S. and {Krause}, E. and {Kron}, R. and {Kuehn}, K. and {Nuropatkin}, N. and {Lahav}, O. and {Lima}, M. and {Lin}, H. and {Maia}, M.~A.~G. and {March}, M. and {Martini}, P. and {McMahon}, R.~G. and {Menanteau}, F. and {Miller}, C.~J. and {Miquel}, R. and {Mohr}, J.~J. and {Neilsen}, E. and {Nichol}, R.~C. and {Ogando}, R.~L.~C. and {Plazas}, A.~A. and {Roe}, N. and {Romer}, A.~K. and {Roodman}, A. and {Rykoff}, E.~S. and {Sanchez}, E. and {Scarpine}, V. and {Schindler}, R. and {Schubnell}, M. and {Sevilla-Noarbe}, I. and {Smith}, M. and {Smith}, R.~C. and {Sobreira}, F. and {Suchyta}, E. and {Swanson}, M.~E.~C. and {Tarle}, G. and {Thomas}, D. and {Thomas}, R.~C. and {Troxel}, M.~A. and {Vikram}, V. and {Walker}, A.~R. and {Wechsler}, R.~H. and {Weller}, J. and {Yanny}, B. and {Zuntz}, J.},
        title = "{The Electromagnetic Counterpart of the Binary Neutron Star Merger LIGO/Virgo GW170817. II. UV, Optical, and Near-infrared Light Curves and Comparison to Kilonova Models}",
      journal = {\apjl},
         year = 2017,
        month = oct,
       volume = {848},
       number = {2},
          eid = {L17},
        pages = {L17},
          doi = {10.3847/2041-8213/aa8fc7},
archivePrefix = {arXiv},
       eprint = {1710.05840},
 primaryClass = {astro-ph.HE},
       adsurl = {https://ui.adsabs.harvard.edu/abs/2017ApJ...848L..17C}
}

@ARTICLE{Diaz2017,
       author = {{D{\'\i}az}, M.~C. and {Macri}, L.~M. and {Garcia Lambas}, D. and {Mendes de Oliveira}, C. and {Nilo Castell{\'o}n}, J.~L. and {Ribeiro}, T. and {S{\'a}nchez}, B. and {Schoenell}, W. and {Abramo}, L.~R. and {Akras}, S. and {Alcaniz}, J.~S. and {Artola}, R. and {Beroiz}, M. and {Bonoli}, S. and {Cabral}, J. and {Camuccio}, R. and {Castillo}, M. and {Chavushyan}, V. and {Coelho}, P. and {Colazo}, C. and {Costa-Duarte}, M.~V. and {Cuevas Larenas}, H. and {DePoy}, D.~L. and {Dom{\'\i}nguez Romero}, M. and {Dultzin}, D. and {Fern{\'a}ndez}, D. and {Garc{\'\i}a}, J. and {Girardini}, C. and {Gon{\c{c}}alves}, D.~R. and {Gon{\c{c}}alves}, T.~S. and {Gurovich}, S. and {Jim{\'e}nez-Teja}, Y. and {Kanaan}, A. and {Lares}, M. and {Lopes de Oliveira}, R. and {L{\'o}pez-Cruz}, O. and {Marshall}, J.~L. and {Melia}, R. and {Molino}, A. and {Padilla}, N. and {Pe{\~n}uela}, T. and {Placco}, V.~M. and {Qui{\~n}ones}, C. and {Ram{\'\i}rez Rivera}, A. and {Renzi}, V. and {Riguccini}, L. and {R{\'\i}os-L{\'o}pez}, E. and {Rodriguez}, H. and {Sampedro}, L. and {Schneiter}, M. and {Sodr{\'e}}, L. and {Starck}, M. and {Torres-Flores}, S. and {Tornatore}, M. and {Zadro{\.z}ny}, A.},
        title = "{Observations of the First Electromagnetic Counterpart to a Gravitational-wave Source by the TOROS Collaboration}",
      journal = {\apjl},
         year = 2017,
        month = oct,
       volume = {848},
       number = {2},
          eid = {L29},
        pages = {L29},
          doi = {10.3847/2041-8213/aa9060},
archivePrefix = {arXiv},
       eprint = {1710.05844},
 primaryClass = {astro-ph.HE},
       adsurl = {https://ui.adsabs.harvard.edu/abs/2017ApJ...848L..29D}
}

@ARTICLE{Drout2017,
       author = {{Drout}, M.~R. and {Piro}, A.~L. and {Shappee}, B.~J. and {Kilpatrick}, C.~D. and {Simon}, J.~D. and {Contreras}, C. and {Coulter}, D.~A. and {Foley}, R.~J. and {Siebert}, M.~R. and {Morrell}, N. and {Boutsia}, K. and {Di Mille}, F. and {Holoien}, T.~W. -S. and {Kasen}, D. and {Kollmeier}, J.~A. and {Madore}, B.~F. and {Monson}, A.~J. and {Murguia-Berthier}, A. and {Pan}, Y. -C. and {Prochaska}, J.~X. and {Ramirez-Ruiz}, E. and {Rest}, A. and {Adams}, C. and {Alatalo}, K. and {Ba{\~n}ados}, E. and {Baughman}, J. and {Beers}, T.~C. and {Bernstein}, R.~A. and {Bitsakis}, T. and {Campillay}, A. and {Hansen}, T.~T. and {Higgs}, C.~R. and {Ji}, A.~P. and {Maravelias}, G. and {Marshall}, J.~L. and {Moni Bidin}, C. and {Prieto}, J.~L. and {Rasmussen}, K.~C. and {Rojas-Bravo}, C. and {Strom}, A.~L. and {Ulloa}, N. and {Vargas-Gonz{\'a}lez}, J. and {Wan}, Z. and {Whitten}, D.~D.},
        title = "{Light curves of the neutron star merger GW170817/SSS17a: Implications for r-process nucleosynthesis}",
      journal = {Science},
         year = 2017,
        month = dec,
       volume = {358},
       number = {6370},
        pages = {1570-1574},
          doi = {10.1126/science.aaq0049},
archivePrefix = {arXiv},
       eprint = {1710.05443},
 primaryClass = {astro-ph.HE},
       adsurl = {https://ui.adsabs.harvard.edu/abs/2017Sci...358.1570D}
}

@ARTICLE{Evans2017,
       author = {{Evans}, P.~A. and {Cenko}, S.~B. and {Kennea}, J.~A. and {Emery}, S.~W.~K. and {Kuin}, N.~P.~M. and {Korobkin}, O. and {Wollaeger}, R.~T. and {Fryer}, C.~L. and {Madsen}, K.~K. and {Harrison}, F.~A. and {Xu}, Y. and {Nakar}, E. and {Hotokezaka}, K. and {Lien}, A. and {Campana}, S. and {Oates}, S.~R. and {Troja}, E. and {Breeveld}, A.~A. and {Marshall}, F.~E. and {Barthelmy}, S.~D. and {Beardmore}, A.~P. and {Burrows}, D.~N. and {Cusumano}, G. and {D'A{\`\i}}, A. and {D'Avanzo}, P. and {D'Elia}, V. and {de Pasquale}, M. and {Even}, W.~P. and {Fontes}, C.~J. and {Forster}, K. and {Garcia}, J. and {Giommi}, P. and {Grefenstette}, B. and {Gronwall}, C. and {Hartmann}, D.~H. and {Heida}, M. and {Hungerford}, A.~L. and {Kasliwal}, M.~M. and {Krimm}, H.~A. and {Levan}, A.~J. and {Malesani}, D. and {Melandri}, A. and {Miyasaka}, H. and {Nousek}, J.~A. and {O'Brien}, P.~T. and {Osborne}, J.~P. and {Pagani}, C. and {Page}, K.~L. and {Palmer}, D.~M. and {Perri}, M. and {Pike}, S. and {Racusin}, J.~L. and {Rosswog}, S. and {Siegel}, M.~H. and {Sakamoto}, T. and {Sbarufatti}, B. and {Tagliaferri}, G. and {Tanvir}, N.~R. and {Tohuvavohu}, A.},
        title = "{Swift and NuSTAR observations of GW170817: Detection of a blue kilonova}",
      journal = {Science},
         year = 2017,
        month = dec,
       volume = {358},
       number = {6370},
        pages = {1565-1570},
          doi = {10.1126/science.aap9580},
archivePrefix = {arXiv},
       eprint = {1710.05437},
 primaryClass = {astro-ph.HE},
       adsurl = {https://ui.adsabs.harvard.edu/abs/2017Sci...358.1565E}
}

@ARTICLE{Hu2017,
       author = {{Hu}, Lei and {Wu}, Xuefeng and {Andreoni}, Igor and {Ashley}, Michael C.~B. and {Cooke}, Jeff and {Cui}, Xiangqun and {Du}, Fujia and {Dai}, Zigao and {Gu}, Bozhong and {Hu}, Yi and {Lu}, Haiping and {Li}, Xiaoyan and {Li}, Zhengyang and {Liang}, Ensi and {Liu}, Liangduan and {Ma}, Bin and {Shang}, Zhaohui and {Sun}, Tianrui and {Suntzeff}, N.~B. and {Tao}, Charling and {Udden}, Syed A. and {Wang}, Lifan and {Wang}, Xiaofeng and {Wen}, Haikun and {Xiao}, Di and {Su}, Jin and {Yang}, Ji and {Yang}, Shihai and {Yuan}, Xiangyan and {Zhou}, Hongyan and {Zhang}, Hui and {Zhou}, Jilin and {Zhu}, Zonghong},
        title = "{Optical observations of LIGO source GW 170817 by the Antarctic Survey Telescopes at Dome A, Antarctica}",
      journal = {Science Bulletin},
         year = 2017,
        month = oct,
       volume = {62},
        pages = {1433-1438},
          doi = {10.1016/j.scib.2017.10.006},
archivePrefix = {arXiv},
       eprint = {1710.05462},
 primaryClass = {astro-ph.HE},
       adsurl = {https://ui.adsabs.harvard.edu/abs/2017SciBu..62.1433H}
}

@ARTICLE{Kasliwal2017,
       author = {{Kasliwal}, M.~M. and {Nakar}, E. and {Singer}, L.~P. and {Kaplan}, D.~L. and {Cook}, D.~O. and {Van Sistine}, A. and {Lau}, R.~M. and {Fremling}, C. and {Gottlieb}, O. and {Jencson}, J.~E. and {Adams}, S.~M. and {Feindt}, U. and {Hotokezaka}, K. and {Ghosh}, S. and {Perley}, D.~A. and {Yu}, P. -C. and {Piran}, T. and {Allison}, J.~R. and {Anupama}, G.~C. and {Balasubramanian}, A. and {Bannister}, K.~W. and {Bally}, J. and {Barnes}, J. and {Barway}, S. and {Bellm}, E. and {Bhalerao}, V. and {Bhattacharya}, D. and {Blagorodnova}, N. and {Bloom}, J.~S. and {Brady}, P.~R. and {Cannella}, C. and {Chatterjee}, D. and {Cenko}, S.~B. and {Cobb}, B.~E. and {Copperwheat}, C. and {Corsi}, A. and {De}, K. and {Dobie}, D. and {Emery}, S.~W.~K. and {Evans}, P.~A. and {Fox}, O.~D. and {Frail}, D.~A. and {Frohmaier}, C. and {Goobar}, A. and {Hallinan}, G. and {Harrison}, F. and {Helou}, G. and {Hinderer}, T. and {Ho}, A.~Y.~Q. and {Horesh}, A. and {Ip}, W. -H. and {Itoh}, R. and {Kasen}, D. and {Kim}, H. and {Kuin}, N.~P.~M. and {Kupfer}, T. and {Lynch}, C. and {Madsen}, K. and {Mazzali}, P.~A. and {Miller}, A.~A. and {Mooley}, K. and {Murphy}, T. and {Ngeow}, C. -C. and {Nichols}, D. and {Nissanke}, S. and {Nugent}, P. and {Ofek}, E.~O. and {Qi}, H. and {Quimby}, R.~M. and {Rosswog}, S. and {Rusu}, F. and {Sadler}, E.~M. and {Schmidt}, P. and {Sollerman}, J. and {Steele}, I. and {Williamson}, A.~R. and {Xu}, Y. and {Yan}, L. and {Yatsu}, Y. and {Zhang}, C. and {Zhao}, W.},
        title = "{Illuminating gravitational waves: A concordant picture of photons from a neutron star merger}",
      journal = {Science},
         year = 2017,
        month = dec,
       volume = {358},
       number = {6370},
        pages = {1559-1565},
          doi = {10.1126/science.aap9455},
archivePrefix = {arXiv},
       eprint = {1710.05436},
 primaryClass = {astro-ph.HE},
       adsurl = {https://ui.adsabs.harvard.edu/abs/2017Sci...358.1559K}
}

@ARTICLE{Lipunov2017,
       author = {{Lipunov}, V.~M. and {Gorbovskoy}, E. and {Kornilov}, V.~G. and {. Tyurina}, N. and {Balanutsa}, P. and {Kuznetsov}, A. and {Vlasenko}, D. and {Kuvshinov}, D. and {Gorbunov}, I. and {Buckley}, D.~A.~H. and {Krylov}, A.~V. and {Podesta}, R. and {Lopez}, C. and {Podesta}, F. and {Levato}, H. and {Saffe}, C. and {Mallamachi}, C. and {Potter}, S. and {Budnev}, N.~M. and {Gress}, O. and {Ishmuhametova}, Yu. and {Vladimirov}, V. and {Zimnukhov}, D. and {Yurkov}, V. and {Sergienko}, Yu. and {Gabovich}, A. and {Rebolo}, R. and {Serra-Ricart}, M. and {Israelyan}, G. and {Chazov}, V. and {Wang}, Xiaofeng and {Tlatov}, A. and {Panchenko}, M.~I.},
        title = "{MASTER Optical Detection of the First LIGO/Virgo Neutron Star Binary Merger GW170817}",
      journal = {\apjl},
         year = 2017,
        month = nov,
       volume = {850},
       number = {1},
          eid = {L1},
        pages = {L1},
          doi = {10.3847/2041-8213/aa92c0},
archivePrefix = {arXiv},
       eprint = {1710.05461},
 primaryClass = {astro-ph.HE},
       adsurl = {https://ui.adsabs.harvard.edu/abs/2017ApJ...850L...1L}
}

@ARTICLE{Pozanenko2018,
       author = {{Pozanenko}, A.~S. and {Barkov}, M.~V. and {Minaev}, P. Yu. and {Volnova}, A.~A. and {Mazaeva}, E.~D. and {Moskvitin}, A.~S. and {Krugov}, M.~A. and {Samodurov}, V.~A. and {Loznikov}, V.~M. and {Lyutikov}, M.},
        title = "{GRB 170817A Associated with GW170817: Multi-frequency Observations and Modeling of Prompt Gamma-Ray Emission}",
      journal = {\apjl},
         year = 2018,
        month = jan,
       volume = {852},
       number = {2},
          eid = {L30},
        pages = {L30},
          doi = {10.3847/2041-8213/aaa2f6},
archivePrefix = {arXiv},
       eprint = {1710.05448},
 primaryClass = {astro-ph.HE},
       adsurl = {https://ui.adsabs.harvard.edu/abs/2018ApJ...852L..30P}
}

@ARTICLE{Smartt2017,
       author = {{Smartt}, S.~J. and {Chen}, T. -W. and {Jerkstrand}, A. and {Coughlin}, M. and {Kankare}, E. and {Sim}, S.~A. and {Fraser}, M. and {Inserra}, C. and {Maguire}, K. and {Chambers}, K.~C. and {Huber}, M.~E. and {Kr{\"u}hler}, T. and {Leloudas}, G. and {Magee}, M. and {Shingles}, L.~J. and {Smith}, K.~W. and {Young}, D.~R. and {Tonry}, J. and {Kotak}, R. and {Gal-Yam}, A. and {Lyman}, J.~D. and {Homan}, D.~S. and {Agliozzo}, C. and {Anderson}, J.~P. and {Angus}, C.~R. and {Ashall}, C. and {Barbarino}, C. and {Bauer}, F.~E. and {Berton}, M. and {Botticella}, M.~T. and {Bulla}, M. and {Bulger}, J. and {Cannizzaro}, G. and {Cano}, Z. and {Cartier}, R. and {Cikota}, A. and {Clark}, P. and {De Cia}, A. and {Della Valle}, M. and {Denneau}, L. and {Dennefeld}, M. and {Dessart}, L. and {Dimitriadis}, G. and {Elias-Rosa}, N. and {Firth}, R.~E. and {Flewelling}, H. and {Fl{\"o}rs}, A. and {Franckowiak}, A. and {Frohmaier}, C. and {Galbany}, L. and {Gonz{\'a}lez-Gait{\'a}n}, S. and {Greiner}, J. and {Gromadzki}, M. and {Guelbenzu}, A. Nicuesa and {Guti{\'e}rrez}, C.~P. and {Hamanowicz}, A. and {Hanlon}, L. and {Harmanen}, J. and {Heintz}, K.~E. and {Heinze}, A. and {Hernandez}, M. -S. and {Hodgkin}, S.~T. and {Hook}, I.~M. and {Izzo}, L. and {James}, P.~A. and {Jonker}, P.~G. and {Kerzendorf}, W.~E. and {Klose}, S. and {Kostrzewa-Rutkowska}, Z. and {Kowalski}, M. and {Kromer}, M. and {Kuncarayakti}, H. and {Lawrence}, A. and {Lowe}, T.~B. and {Magnier}, E.~A. and {Manulis}, I. and {Martin-Carrillo}, A. and {Mattila}, S. and {McBrien}, O. and {M{\"u}ller}, A. and {Nordin}, J. and {O'Neill}, D. and {Onori}, F. and {Palmerio}, J.~T. and {Pastorello}, A. and {Patat}, F. and {Pignata}, G. and {Podsiadlowski}, Ph. and {Pumo}, M.~L. and {Prentice}, S.~J. and {Rau}, A. and {Razza}, A. and {Rest}, A. and {Reynolds}, T. and {Roy}, R. and {Ruiter}, A.~J. and {Rybicki}, K.~A. and {Salmon}, L. and {Schady}, P. and {Schultz}, A.~S.~B. and {Schweyer}, T. and {Seitenzahl}, I.~R. and {Smith}, M. and {Sollerman}, J. and {Stalder}, B. and {Stubbs}, C.~W. and {Sullivan}, M. and {Szegedi}, H. and {Taddia}, F. and {Taubenberger}, S. and {Terreran}, G. and {van Soelen}, B. and {Vos}, J. and {Wainscoat}, R.~J. and {Walton}, N.~A. and {Waters}, C. and {Weiland}, H. and {Willman}, M. and {Wiseman}, P. and {Wright}, D.~E. and {Wyrzykowski}, {\L}. and {Yaron}, O.},
        title = "{A kilonova as the electromagnetic counterpart to a gravitational-wave source}",
      journal = {\nat},
         year = 2017,
        month = nov,
       volume = {551},
       number = {7678},
        pages = {75-79},
          doi = {10.1038/nature24303},
archivePrefix = {arXiv},
       eprint = {1710.05841},
 primaryClass = {astro-ph.HE},
       adsurl = {https://ui.adsabs.harvard.edu/abs/2017Natur.551...75S}
}

@ARTICLE{Tanvir2017,
       author = {{Tanvir}, N.~R. and {Levan}, A.~J. and {Gonz{\'a}lez-Fern{\'a}ndez}, C. and {Korobkin}, O. and {Mandel}, I. and {Rosswog}, S. and {Hjorth}, J. and {D'Avanzo}, P. and {Fruchter}, A.~S. and {Fryer}, C.~L. and {Kangas}, T. and {Milvang-Jensen}, B. and {Rosetti}, S. and {Steeghs}, D. and {Wollaeger}, R.~T. and {Cano}, Z. and {Copperwheat}, C.~M. and {Covino}, S. and {D'Elia}, V. and {de Ugarte Postigo}, A. and {Evans}, P.~A. and {Even}, W.~P. and {Fairhurst}, S. and {Figuera Jaimes}, R. and {Fontes}, C.~J. and {Fujii}, Y.~I. and {Fynbo}, J.~P.~U. and {Gompertz}, B.~P. and {Greiner}, J. and {Hodosan}, G. and {Irwin}, M.~J. and {Jakobsson}, P. and {J{\o}rgensen}, U.~G. and {Kann}, D.~A. and {Lyman}, J.~D. and {Malesani}, D. and {McMahon}, R.~G. and {Melandri}, A. and {O'Brien}, P.~T. and {Osborne}, J.~P. and {Palazzi}, E. and {Perley}, D.~A. and {Pian}, E. and {Piranomonte}, S. and {Rabus}, M. and {Rol}, E. and {Rowlinson}, A. and {Schulze}, S. and {Sutton}, P. and {Th{\"o}ne}, C.~C. and {Ulaczyk}, K. and {Watson}, D. and {Wiersema}, K. and {Wijers}, R.~A.~M.~J.},
        title = "{The Emergence of a Lanthanide-rich Kilonova Following the Merger of Two Neutron Stars}",
      journal = {\apjl},
         year = 2017,
        month = oct,
       volume = {848},
       number = {2},
          eid = {L27},
        pages = {L27},
          doi = {10.3847/2041-8213/aa90b6},
archivePrefix = {arXiv},
       eprint = {1710.05455},
 primaryClass = {astro-ph.HE},
       adsurl = {https://ui.adsabs.harvard.edu/abs/2017ApJ...848L..27T}
}

@ARTICLE{Troja2017,
       author = {{Troja}, E. and {Piro}, L. and {van Eerten}, H. and {Wollaeger}, R.~T. and {Im}, M. and {Fox}, O.~D. and {Butler}, N.~R. and {Cenko}, S.~B. and {Sakamoto}, T. and {Fryer}, C.~L. and {Ricci}, R. and {Lien}, A. and {Ryan}, R.~E. and {Korobkin}, O. and {Lee}, S. -K. and {Burgess}, J.~M. and {Lee}, W.~H. and {Watson}, A.~M. and {Choi}, C. and {Covino}, S. and {D'Avanzo}, P. and {Fontes}, C.~J. and {Gonz{\'a}lez}, J. Becerra and {Khandrika}, H.~G. and {Kim}, J. and {Kim}, S. -L. and {Lee}, C. -U. and {Lee}, H.~M. and {Kutyrev}, A. and {Lim}, G. and {S{\'a}nchez-Ram{\'\i}rez}, R. and {Veilleux}, S. and {Wieringa}, M.~H. and {Yoon}, Y.},
        title = "{The X-ray counterpart to the gravitational-wave event GW170817}",
      journal = {\nat},
         year = 2017,
        month = nov,
       volume = {551},
       number = {7678},
        pages = {71-74},
          doi = {10.1038/nature24290},
archivePrefix = {arXiv},
       eprint = {1710.05433},
 primaryClass = {astro-ph.HE},
       adsurl = {https://ui.adsabs.harvard.edu/abs/2017Natur.551...71T}
}

@ARTICLE{Utsumi2017,
       author = {{Utsumi}, Yousuke and {Tanaka}, Masaomi and {Tominaga}, Nozomu and {Yoshida}, Michitoshi and {Barway}, Sudhanshu and {Nagayama}, Takahiro and {Zenko}, Tetsuya and {Aoki}, Kentaro and {Fujiyoshi}, Takuya and {Furusawa}, Hisanori and {Kawabata}, Koji S. and {Koshida}, Shintaro and {Lee}, Chien-Hsiu and {Morokuma}, Tomoki and {Motohara}, Kentaro and {Nakata}, Fumiaki and {Ohsawa}, Ryou and {Ohta}, Kouji and {Okita}, Hirofumi and {Tajitsu}, Akito and {Tanaka}, Ichi and {Terai}, Tsuyoshi and {Yasuda}, Naoki and {Abe}, Fumio and {Asakura}, Yuichiro and {Bond}, Ian A. and {Miyazaki}, Shota and {Sumi}, Takahiro and {Tristram}, Paul J. and {Honda}, Satoshi and {Itoh}, Ryosuke and {Itoh}, Yoichi and {Kawabata}, Miho and {Morihana}, Kumiko and {Nagashima}, Hiroki and {Nakaoka}, Tatsuya and {Ohshima}, Tomohito and {Takahashi}, Jun and {Takayama}, Masaki and {Aoki}, Wako and {Baar}, Stefan and {Doi}, Mamoru and {Finet}, Francois and {Kanda}, Nobuyuki and {Kawai}, Nobuyuki and {Kim}, Ji Hoon and {Kuroda}, Daisuke and {Liu}, Wei and {Matsubayashi}, Kazuya and {Murata}, Katsuhiro L. and {Nagai}, Hiroshi and {Saito}, Tomoki and {Saito}, Yoshihiko and {Sako}, Shigeyuki and {Sekiguchi}, Yuichiro and {Tamura}, Yoichi and {Tanaka}, Masayuki and {Uemura}, Makoto and {Yamaguchi}, Masaki S.},
        title = "{J-GEM observations of an electromagnetic counterpart to the neutron star merger GW170817}",
      journal = {\pasj},
         year = 2017,
        month = dec,
       volume = {69},
       number = {6},
          eid = {101},
        pages = {101},
          doi = {10.1093/pasj/psx118},
archivePrefix = {arXiv},
       eprint = {1710.05848},
 primaryClass = {astro-ph.HE},
       adsurl = {https://ui.adsabs.harvard.edu/abs/2017PASJ...69..101U}
}

@ARTICLE{Valenti2017,
       author = {{Valenti}, Stefano and {Sand}, David J. and {Yang}, Sheng and {Cappellaro}, Enrico and {Tartaglia}, Leonardo and {Corsi}, Alessandra and {Jha}, Saurabh W. and {Reichart}, Daniel E. and {Haislip}, Joshua and {Kouprianov}, Vladimir},
        title = "{The Discovery of the Electromagnetic Counterpart of GW170817: Kilonova AT 2017gfo/DLT17ck}",
      journal = {\apjl},
         year = 2017,
        month = oct,
       volume = {848},
       number = {2},
          eid = {L24},
        pages = {L24},
          doi = {10.3847/2041-8213/aa8edf},
archivePrefix = {arXiv},
       eprint = {1710.05854},
 primaryClass = {astro-ph.HE},
       adsurl = {https://ui.adsabs.harvard.edu/abs/2017ApJ...848L..24V}
}

@INPROCEEDINGS{Kurucz2018ASPC,
       author = {{Kurucz}, R.~L.},
        title = "{Including All the Lines: Data Releases for Spectra and Opacities through 2017}",
    booktitle = {Workshop on Astrophysical Opacities},
         year = 2018,
       series = {Astronomical Society of the Pacific Conference Series},
       volume = {515},
        month = aug,
        pages = {47},
       adsurl = {https://ui.adsabs.harvard.edu/abs/2018ASPC..515...47K}
}

@ARTICLE{Mooley2022,
       author = {{Mooley}, Kunal P. and {Anderson}, Jay and {Lu}, Wenbin},
        title = "{Optical superluminal motion measurement in the neutron-star merger GW170817}",
      journal = {\nat},
         year = 2022,
        month = oct,
       volume = {610},
       number = {7931},
        pages = {273-276},
          doi = {10.1038/s41586-022-05145-7},
archivePrefix = {arXiv},
       eprint = {2210.06568},
 primaryClass = {astro-ph.HE},
       adsurl = {https://ui.adsabs.harvard.edu/abs/2022Natur.610..273M}
}

@ARTICLE{Shingles2023,
       author = {{Shingles}, Luke J. and {Collins}, Christine E. and {Vijayan}, Vimal and {Fl{\"o}rs}, Andreas and {Just}, Oliver and {Leck}, Gerrit and {Xiong}, Zewei and {Bauswein}, Andreas and {Mart{\'\i}nez-Pinedo}, Gabriel and {Sim}, Stuart A.},
        title = "{Self-consistent 3D Radiative Transfer for Kilonovae: Directional Spectra from Merger Simulations}",
      journal = {\apjl},
         year = 2023,
        month = sep,
       volume = {954},
       number = {2},
          eid = {L41},
        pages = {L41},
          doi = {10.3847/2041-8213/acf29a},
archivePrefix = {arXiv},
       eprint = {2306.17612},
 primaryClass = {astro-ph.HE},
       adsurl = {https://ui.adsabs.harvard.edu/abs/2023ApJ...954L..41S}
}

@ARTICLE{Tanaka2020,
       author = {{Tanaka}, Masaomi and {Kato}, Daiji and {Gaigalas}, Gediminas and {Kawaguchi}, Kyohei},
        title = "{Systematic opacity calculations for kilonovae}",
      journal = {\mnras},
         year = 2020,
        month = aug,
       volume = {496},
       number = {2},
        pages = {1369-1392},
          doi = {10.1093/mnras/staa1576},
archivePrefix = {arXiv},
       eprint = {1906.08914},
 primaryClass = {astro-ph.HE},
       adsurl = {https://ui.adsabs.harvard.edu/abs/2020MNRAS.496.1369T}
}

@ARTICLE{Ragosta2024,
       author = {{Ragosta}, Fabio and {Ahumada}, Tom{\'a}s and {Piranomonte}, Silvia and {Andreoni}, Igor and {Melandri}, Andrea and {Colombo}, Alberto and {Coughlin}, Michael W.},
        title = "{Kilonova Parameter Estimation with LSST at Vera C. Rubin Observatory}",
      journal = {\apj},
         year = 2024,
        month = may,
       volume = {966},
       number = {2},
          eid = {214},
        pages = {214},
          doi = {10.3847/1538-4357/ad35c1},
archivePrefix = {arXiv},
       eprint = {2403.14016},
 primaryClass = {astro-ph.HE},
       adsurl = {https://ui.adsabs.harvard.edu/abs/2024ApJ...966..214R}
}

@ARTICLE{Chase2022,
       author = {{Chase}, Eve A. and {O'Connor}, Brendan and {Fryer}, Christopher L. and {Troja}, Eleonora and {Korobkin}, Oleg and {Wollaeger}, Ryan T. and {Ristic}, Marko and {Fontes}, Christopher J. and {Hungerford}, Aimee L. and {Herring}, Angela M.},
        title = "{Kilonova Detectability with Wide-field Instruments}",
      journal = {\apj},
         year = 2022,
        month = mar,
       volume = {927},
       number = {2},
          eid = {163},
        pages = {163},
          doi = {10.3847/1538-4357/ac3d25},
archivePrefix = {arXiv},
       eprint = {2105.12268},
 primaryClass = {astro-ph.HE},
       adsurl = {https://ui.adsabs.harvard.edu/abs/2022ApJ...927..163C}
}

@ARTICLE{Bernuzzi2025,
       author = {{Bernuzzi}, Sebastiano and {Magistrelli}, Fabio and {Jacobi}, Maximilian and {Logoteta}, Domenico and {Perego}, Albino and {Radice}, David},
        title = "{Long-lived neutron-star remnants from asymmetric binary neutron star mergers: element formation, kilonova signals and gravitational waves}",
      journal = {\mnras},
         year = 2025,
        month = jul,
          doi = {10.1093/mnras/staf1147},
archivePrefix = {arXiv},
       eprint = {2409.18185},
 primaryClass = {astro-ph.HE},
       adsurl = {https://ui.adsabs.harvard.edu/abs/2025MNRAS.tmp.1175B}
}

@ARTICLE{Margutti2017,
       author = {{Margutti}, R. and {Berger}, E. and {Fong}, W. and {Guidorzi}, C. and {Alexander}, K.~D. and {Metzger}, B.~D. and {Blanchard}, P.~K. and {Cowperthwaite}, P.~S. and {Chornock}, R. and {Eftekhari}, T. and {Nicholl}, M. and {Villar}, V.~A. and {Williams}, P.~K.~G. and {Annis}, J. and {Brown}, D.~A. and {Chen}, H. and {Doctor}, Z. and {Frieman}, J.~A. and {Holz}, D.~E. and {Sako}, M. and {Soares-Santos}, M.},
        title = "{The Electromagnetic Counterpart of the Binary Neutron Star Merger LIGO/Virgo GW170817. V. Rising X-Ray Emission from an Off-axis Jet}",
      journal = {\apjl},
         year = 2017,
        month = oct,
       volume = {848},
       number = {2},
          eid = {L20},
        pages = {L20},
          doi = {10.3847/2041-8213/aa9057},
archivePrefix = {arXiv},
       eprint = {1710.05431},
 primaryClass = {astro-ph.HE},
       adsurl = {https://ui.adsabs.harvard.edu/abs/2017ApJ...848L..20M}
}

@ARTICLE{Alexander2017,
       author = {{Alexander}, K.~D. and {Berger}, E. and {Fong}, W. and {Williams}, P.~K.~G. and {Guidorzi}, C. and {Margutti}, R. and {Metzger}, B.~D. and {Annis}, J. and {Blanchard}, P.~K. and {Brout}, D. and {Brown}, D.~A. and {Chen}, H. -Y. and {Chornock}, R. and {Cowperthwaite}, P.~S. and {Drout}, M. and {Eftekhari}, T. and {Frieman}, J. and {Holz}, D.~E. and {Nicholl}, M. and {Rest}, A. and {Sako}, M. and {Soares-Santos}, M. and {Villar}, V.~A.},
        title = "{The Electromagnetic Counterpart of the Binary Neutron Star Merger LIGO/Virgo GW170817. VI. Radio Constraints on a Relativistic Jet and Predictions for Late-time Emission from the Kilonova Ejecta}",
      journal = {\apjl},
         year = 2017,
        month = oct,
       volume = {848},
       number = {2},
          eid = {L21},
        pages = {L21},
          doi = {10.3847/2041-8213/aa905d},
archivePrefix = {arXiv},
       eprint = {1710.05457},
 primaryClass = {astro-ph.HE},
       adsurl = {https://ui.adsabs.harvard.edu/abs/2017ApJ...848L..21A}
}

@ARTICLE{Sarin2024,
       author = {{Sarin}, Nikhil and {Rosswog}, Stephan},
        title = "{Cautionary Tales on Heating-rate Prescriptions in Kilonovae}",
      journal = {\apjl},
         year = 2024,
        month = sep,
       volume = {973},
       number = {1},
          eid = {L24},
        pages = {L24},
          doi = {10.3847/2041-8213/ad739d},
archivePrefix = {arXiv},
       eprint = {2404.07271},
 primaryClass = {astro-ph.HE},
       adsurl = {https://ui.adsabs.harvard.edu/abs/2024ApJ...973L..24S}
}

@ARTICLE{Andrade2025,
       author = {{Andrade}, Cristina and {Alserkal}, Raiyah and {Salazar Manzano}, Luis and {Martin}, Emma and {Andreoni}, Igor and {Coughlin}, Michael W. and {Guessoum}, Nidhal and {Rivera Sandoval}, Liliana},
        title = "{The Effect of Vera C. Rubin Observatory Cadence Selections on Kilonova Detectability}",
      journal = {\pasp},
         year = 2025,
        month = mar,
       volume = {137},
       number = {3},
          eid = {034102},
        pages = {034102},
          doi = {10.1088/1538-3873/adbfbc},
archivePrefix = {arXiv},
       eprint = {2502.14124},
 primaryClass = {astro-ph.HE},
       adsurl = {https://ui.adsabs.harvard.edu/abs/2025PASP..137c4102A}
}

@ARTICLE{Collins2023,
       author = {{Collins}, Christine E. and {Bauswein}, Andreas and {Sim}, Stuart A. and {Vijayan}, Vimal and {Mart{\'\i}nez-Pinedo}, Gabriel and {Just}, Oliver and {Shingles}, Luke J. and {Kromer}, Markus},
        title = "{3D radiative transfer kilonova modelling for binary neutron star merger simulations}",
      journal = {\mnras},
         year = 2023,
        month = may,
       volume = {521},
       number = {2},
        pages = {1858-1870},
          doi = {10.1093/mnras/stad606},
archivePrefix = {arXiv},
       eprint = {2209.05246},
 primaryClass = {astro-ph.HE},
       adsurl = {https://ui.adsabs.harvard.edu/abs/2023MNRAS.521.1858C}
}

@ARTICLE{Kawaguchi2022,
       author = {{Kawaguchi}, Kyohei and {Fujibayashi}, Sho and {Hotokezaka}, Kenta and {Shibata}, Masaru and {Wanajo}, Shinya},
        title = "{Electromagnetic Counterparts of Binary-neutron-star Mergers Leading to a Strongly Magnetized Long-lived Remnant Neutron Star}",
      journal = {\apj},
         year = 2022,
        month = jul,
       volume = {933},
       number = {1},
          eid = {22},
        pages = {22},
          doi = {10.3847/1538-4357/ac6ef7},
archivePrefix = {arXiv},
       eprint = {2202.13149},
 primaryClass = {astro-ph.HE},
       adsurl = {https://ui.adsabs.harvard.edu/abs/2022ApJ...933...22K}
}

@ARTICLE{Foucart2020,
       author = {{Foucart}, Francois and {Duez}, Matthew D. and {Hebert}, Francois and {Kidder}, Lawrence E. and {Pfeiffer}, Harald P. and {Scheel}, Mark A.},
        title = "{Monte-Carlo Neutrino Transport in Neutron Star Merger Simulations}",
      journal = {\apjl},
         year = 2020,
        month = oct,
       volume = {902},
       number = {1},
          eid = {L27},
        pages = {L27},
          doi = {10.3847/2041-8213/abbb87},
archivePrefix = {arXiv},
       eprint = {2008.08089},
 primaryClass = {astro-ph.HE},
       adsurl = {https://ui.adsabs.harvard.edu/abs/2020ApJ...902L..27F}
}

@ARTICLE{Foucart_2024,
       author = {{Foucart}, Francois and {Cheong}, Patrick Chi-Kit and {Duez}, Matthew D. and {Kidder}, Lawrence E. and {Pfeiffer}, Harald P. and {Scheel}, Mark A.},
        title = "{Robustness of neutron star merger simulations to changes in neutrino transport and neutrino-matter interactions}",
      journal = {\prd},
         year = 2024,
        month = oct,
       volume = {110},
       number = {8},
          eid = {083028},
        pages = {083028},
          doi = {10.1103/PhysRevD.110.083028},
archivePrefix = {arXiv},
       eprint = {2407.15989},
 primaryClass = {astro-ph.HE},
       adsurl = {https://ui.adsabs.harvard.edu/abs/2024PhRvD.110h3028F}
}

@article{Kasen2015,
  author = {Daniel Kasen and Rodrigo Fernández and Brian D. Metzger},
  title = {Kilonova light curves from the disc wind outflows of compact object mergers},
  journal = {Monthly Notices of the Royal Astronomical Society},
  volume = {450},
  number = {2},
  pages = {1777--1786},
  year = {2015},
  month = {June},
  doi = {10.1093/mnras/stv721}
}

@article{Nedora_2019,
doi = {10.3847/2041-8213/ab5794},
url = {https://dx.doi.org/10.3847/2041-8213/ab5794},
year = {2019},
month = {nov},
publisher = {The American Astronomical Society},
volume = {886},
number = {2},
pages = {L30},
author = {Nedora, Vsevolod and Bernuzzi, Sebastiano and Radice, David and Perego, Albino and Endrizzi, Andrea and Ortiz, Néstor},
title = {Spiral-wave Wind for the Blue Kilonova},
journal = {The Astrophysical Journal Letters}
}

@article{Wu_2019,
  title = {Fingerprints of Heavy-Element Nucleosynthesis in the Late-Time Lightcurves of Kilonovae},
  author = {Wu, Meng-Ru and Barnes, J. and Mart\'{\i}nez-Pinedo, G. and Metzger, B. D.},
  journal = {Phys. Rev. Lett.},
  volume = {122},
  issue = {6},
  pages = {062701},
  numpages = {6},
  year = {2019},
  month = {Feb},
  publisher = {American Physical Society},
  doi = {10.1103/PhysRevLett.122.062701},
  url = {https://link.aps.org/doi/10.1103/PhysRevLett.122.062701}
}

@article{Kasen_2006,
doi = {10.1086/506190},
url = {https://dx.doi.org/10.1086/506190},
year = {2006},
month = {nov},
publisher = {},
volume = {651},
number = {1},
pages = {366},
author = {Kasen, Daniel and Thomas, R. C. and Nugent, P.},
title = {Time-dependent Monte Carlo Radiative Transfer Calculations for Three-dimensional Supernova Spectra, Light Curves, and Polarization},
journal = {The Astrophysical Journal}
}

@article{Hotokezaka_2013,
  title = {Mass ejection from the merger of binary neutron stars},
  author = {Hotokezaka, Kenta and Kiuchi, Kenta and Kyutoku, Koutarou and Okawa, Hirotada and Sekiguchi, Yu-ichiro and Shibata, Masaru and Taniguchi, Keisuke},
  journal = {Phys. Rev. D},
  volume = {87},
  issue = {2},
  pages = {024001},
  numpages = {27},
  year = {2013},
  month = {Jan},
  publisher = {American Physical Society},
  doi = {10.1103/PhysRevD.87.024001},
  url = {https://link.aps.org/doi/10.1103/PhysRevD.87.024001}
}

@article{Shibata_2019,
   author = "Shibata, Masaru and Hotokezaka, Kenta",
   title = "Merger and Mass Ejection of Neutron Star Binaries", 
   journal= "Annual Review of Nuclear and Particle Science",
   year = "2019",
   volume = "69",
   number = "Volume 69, 2019",
   pages = "41-64",
   doi = "https://doi.org/10.1146/annurev-nucl-101918-023625",
   url = "https://www.annualreviews.org/content/journals/10.1146/annurev-nucl-101918-023625",
   publisher = "Annual Reviews",
   issn = "1545-4134",
   type = "Journal Article"
}

@ARTICLE{Kawaguchi_2024,
       author = {{Kawaguchi}, Kyohei and {Domoto}, Nanae and {Fujibayashi}, Sho and {Hamidani}, Hamid and {Hayashi}, Kota and {Shibata}, Masaru and {Tanaka}, Masaomi and {Wanajo}, Shinya},
        title = "{Three dimensional end-to-end simulation for kilonova emission from a black hole neutron star merger}",
      journal = {\mnras},
         year = 2024,
        month = dec,
       volume = {535},
       number = {4},
        pages = {3711-3731},
          doi = {10.1093/mnras/stae2594},
archivePrefix = {arXiv},
       eprint = {2404.15027},
 primaryClass = {astro-ph.HE},
       adsurl = {https://ui.adsabs.harvard.edu/abs/2024MNRAS.535.3711K}
}

@article{Perego2014,
  author       = {Perego, A. and Rosswog, S. and Cabezón, R. M. and Korobkin, O. and Käppeli, R. and Arcones, A. and Liebendörfer, M.},
  title        = {Neutrino-driven winds from neutron star merger remnants},
  journal      = {Monthly Notices of the Royal Astronomical Society},
  volume       = {443},
  number       = {4},
  pages        = {3134--3156},
  year         = {2014},
  month        = {October},
  doi          = {10.1093/mnras/stu1352},
  url          = {https://doi.org/10.1093/mnras/stu1352}
}

@article{Rosswog2014,
  author    = {S. Rosswog and O. Korobkin and A. Arcones and F.-K. Thielemann and T. Piran},
  title     = {The long-term evolution of neutron star merger remnants – I. The impact of r-process nucleosynthesis},
  journal   = {Monthly Notices of the Royal Astronomical Society},
  volume    = {439},
  number    = {1},
  pages     = {744--756},
  year      = {2014},
  month     = {March},
  doi       = {10.1093/mnras/stt2502},
  url       = {https://doi.org/10.1093/mnras/stt2502}
}

@article{Curtis_2024,
doi = {10.3847/2041-8213/ad0fe1},
url = {https://dx.doi.org/10.3847/2041-8213/ad0fe1},
year = {2024},
month = {jan},
publisher = {The American Astronomical Society},
volume = {961},
number = {1},
pages = {L26},
author = {Curtis, Sanjana and Bosch, Pablo and Mösta, Philipp and Radice, David and Bernuzzi, Sebastiano and Perego, Albino and Haas, Roland and Schnetter, Erik},
title = {Magnetized Outflows from Short-lived Neutron Star Merger Remnants Can Produce a Blue Kilonova},
journal = {The Astrophysical Journal Letters}
}

@article{Klion2022,
  author    = {Hannah Klion and Alexander Tchekhovskoy and Daniel Kasen and Adithan Kathirgamaraju and Eliot Quataert and Rodrigo Fernández},
  title     = {The impact of r-process heating on the dynamics of neutron star merger accretion disc winds and their electromagnetic radiation},
  journal   = {Monthly Notices of the Royal Astronomical Society},
  volume    = {510},
  number    = {2},
  pages     = {2968--2979},
  year      = {2022},
  month     = {February},
  doi       = {10.1093/mnras/stab3583},
  url       = {https://doi.org/10.1093/mnras/stab3583}
}

@article{Tanaka_2018,
doi = {10.3847/1538-4357/aaa0cb},
url = {https://dx.doi.org/10.3847/1538-4357/aaa0cb},
year = {2018},
month = {jan},
publisher = {The American Astronomical Society},
volume = {852},
number = {2},
pages = {109},
author = {Tanaka, Masaomi and Kato, Daiji and Gaigalas, Gediminas and Rynkun, Pavel and Radžiūtė, Laima and Wanajo, Shinya and Sekiguchi, Yuichiro and Nakamura, Nobuyuki and Tanuma, Hajime and Murakami, Izumi and Sakaue, Hiroyuki A.},
title = {Properties of Kilonovae from Dynamical and Post-merger Ejecta of Neutron Star Mergers},
journal = {The Astrophysical Journal}
}

@article{Radice_2016,
   title={Dynamical mass ejection from binary neutron star mergers},
   volume={460},
   ISSN={1365-2966},
   url={http://dx.doi.org/10.1093/mnras/stw1227},
   DOI={10.1093/mnras/stw1227},
   number={3},
   journal={Monthly Notices of the Royal Astronomical Society},
   publisher={Oxford University Press (OUP)},
   author={Radice, David and Galeazzi, Filippo and Lippuner, Jonas and Roberts, Luke F. and Ott, Christian D. and Rezzolla, Luciano},
   year={2016},
   month=may, pages={3255–3271} }

@article{Freiburghaus_1999,
doi = {10.1086/312343},
url = {https://dx.doi.org/10.1086/312343},
year = {1999},
month = {oct},
publisher = {},
volume = {525},
number = {2},
pages = {L121},
author = {Freiburghaus, C. and Rosswog, S. and Thielemann, F.-K.},
title = {r-Process in Neutron Star
Mergers},
journal = {The Astrophysical Journal}
}

@article{Pian2017,
  author       = {Pian, E. and D’Avanzo, P. and Benetti, S. and others},
  title        = {Spectroscopic identification of r-process nucleosynthesis in a double neutron-star merger},
  journal      = {Nature},
  volume       = {551},
  pages        = {67--70},
  year         = {2017},
  doi          = {10.1038/nature24298},
  url          = {https://doi.org/10.1038/nature24298},
  received     = {2017-09-12},
  accepted     = {2017-09-20},
  published    = {2017-10-16},
  note         = {Issue Date: 02 November 2017}
}

@article{Radice_2024,
doi = {10.1088/1742-6596/2742/1/012009},
url = {https://dx.doi.org/10.1088/1742-6596/2742/1/012009},
year = {2024},
month = {apr},
publisher = {IOP Publishing},
volume = {2742},
number = {1},
pages = {012009},
author = {Radice, David and Bernuzzi, Sebastiano},
title = {Secular Outflows from Long-Lived Neutron Star Merger Remnants},
journal = {Journal of Physics: Conference Series}
}

@article{Rosswog2019,
    author = "Rosswog, Stephan",
    editor = "Formicola, Alba and Junker, Matthias and Gialanella, Lucio and Imbriani, Gianluca",
    title = "{Neutron Star Mergers as r-Process Sources}",
    doi = "10.1007/978-3-030-13876-9_17",
    journal = "Springer Proc. Phys.",
    volume = "219",
    pages = "105--110",
    year = "2019"
}

@article{Grossman2014,
  author       = {Doron Grossman and Oleg Korobkin and Stephan Rosswog and Tsvi Piran},
  title        = {The long-term evolution of neutron star merger remnants – II. Radioactively powered transients},
  journal      = {Monthly Notices of the Royal Astronomical Society},
  volume       = {439},
  number       = {1},
  pages        = {757--770},
  year         = {2014},
  month        = {March},
  doi          = {10.1093/mnras/stt2503},
  url          = {https://doi.org/10.1093/mnras/stt2503}
}

@article{Curtis_2022,
   title={r-process nucleosynthesis and kilonovae from hypermassive neutron star post-merger remnants},
   volume={518},
   ISSN={1365-2966},
   url={http://dx.doi.org/10.1093/mnras/stac3128},
   DOI={10.1093/mnras/stac3128},
   number={4},
   journal={Monthly Notices of the Royal Astronomical Society},
   publisher={Oxford University Press (OUP)},
   author={Curtis, Sanjana and Mösta, Philipp and Wu, Zhenyu and Radice, David and Roberts, Luke and Ricigliano, Giacomo and Perego, Albino},
   year={2022},
   month=nov, pages={5313–5322} }

@article{Mösta_2020,
   title={A Magnetar Engine for Short GRBs and Kilonovae},
   volume={901},
   ISSN={2041-8213},
   url={http://dx.doi.org/10.3847/2041-8213/abb6ef},
   DOI={10.3847/2041-8213/abb6ef},
   number={2},
   journal={The Astrophysical Journal Letters},
   publisher={American Astronomical Society},
   author={Mösta, Philipp and Radice, David and Haas, Roland and Schnetter, Erik and Bernuzzi, Sebastiano},
   year={2020},
   month=oct, pages={L37} }

@mastersthesis{Liekethesis,
  author = {Lieke {Sippens Groenewegen}},
  title = {End-to-End Modeling of Kilonovae from
Binary Neutron Star Mergers},
  school = {University of Amsterdam},
  year = {2025},
  type = {Master's thesis},
  address = {Amsterdam, the Netherlands},
}

@article{Kiuchi2024,
  author       = {Kiuchi, Kenta and Reboul-Salze, Aur{\'e}lien and Shibata, Masaru and others},
  title        = {A large-scale magnetic field produced by a solar-like dynamo in binary neutron star mergers},
  journal      = {Nature Astronomy},
  year         = {2024},
  volume       = {8},
  pages        = {298--307},
  doi          = {10.1038/s41550-024-02194-y},
  url          = {https://doi.org/10.1038/s41550-024-02194-y},
  month        = mar,
  note         = {Published online: 15 February 2024}
}

@ARTICLE{Mösta_2015,
       author = {{M{\"o}sta}, Philipp and {Ott}, Christian D. and {Radice}, David and {Roberts}, Luke F. and {Schnetter}, Erik and {Haas}, Roland},
        title = "{A large-scale dynamo and magnetoturbulence in rapidly rotating core-collapse supernovae}",
      journal = {\nat},
         year = 2015,
        month = dec,
       volume = {528},
       number = {7582},
        pages = {376-379},
          doi = {10.1038/nature15755},
archivePrefix = {arXiv},
       eprint = {1512.00838},
 primaryClass = {astro-ph.HE},
       adsurl = {https://ui.adsabs.harvard.edu/abs/2015Natur.528..376M}
}

@article{deHaas2024,
  author  = {de Haas, Sebastiaan and Bosch, Pablo and M{\"o}sta, Philipp and Curtis, Sanjana and Schut, Nathanyel},
  title   = {Magnetic field effects on nucleosynthesis and kilonovae from neutron star merger remnants},
  journal = {Monthly Notices of the Royal Astronomical Society},
  volume  = {527},
  number  = {2},
  pages   = {2240--2250},
  year    = {2024},
  month   = {jan},
  doi     = {10.1093/mnras/stad2931},
  url     = {https://doi.org/10.1093/mnras/stad2931}
}

@misc{FLASH,
    author = {{FLASH:}},
    howpublished ={\url{https://flash.rochester.edu}},
}

@article{Cusinato2022,
  author    = {M. Cusinato and F. M. Guercilena and A. Perego and others},
  title     = {Neutrino emission from binary neutron star mergers: characterising light curves and mean energies},
  journal   = {European Physical Journal A},
  volume    = {58},
  pages     = {99},
  year      = {2022},
  doi       = {10.1140/epja/s10050-022-00743-5},
  url       = {https://doi.org/10.1140/epja/s10050-022-00743-5},
  received  = {2021-11-21},
  accepted  = {2022-05-05},
  published = {2022-05-23}
}

@article{Kasen_2013,
   title={OPACITIES AND SPECTRA OF THEr-PROCESS EJECTA FROM NEUTRON STAR MERGERS},
   volume={774},
   ISSN={1538-4357},
   url={http://dx.doi.org/10.1088/0004-637X/774/1/25},
   DOI={10.1088/0004-637x/774/1/25},
   number={1},
   journal={The Astrophysical Journal},
   publisher={American Astronomical Society},
   author={Kasen, Daniel and Badnell, N. R. and Barnes, Jennifer},
   year={2013},
   month=aug, pages={25} }

@article{Kasen_2017,
   title={Origin of the heavy elements in binary neutron-star mergers from a gravitational-wave event},
   volume={551},
   ISSN={1476-4687},
   url={http://dx.doi.org/10.1038/nature24453},
   DOI={10.1038/nature24453},
   number={7678},
   journal={Nature},
   publisher={Springer Science and Business Media LLC},
   author={Kasen, Daniel and Metzger, Brian and Barnes, Jennifer and Quataert, Eliot and Ramirez-Ruiz, Enrico},
   year={2017},
   month=oct, pages={80–84} }

@article{Radice_2020,
   title={The Dynamics of Binary Neutron Star Mergers and GW170817},
   volume={70},
   ISSN={1545-4134},
   url={http://dx.doi.org/10.1146/annurev-nucl-013120-114541},
   DOI={10.1146/annurev-nucl-013120-114541},
   number={1},
   journal={Annual Review of Nuclear and Particle Science},
   publisher={Annual Reviews},
   author={Radice, David and Bernuzzi, Sebastiano and Perego, Albino},
   year={2020},
   month=oct, pages={95–119} }

@article{Metzger_2019,
   title={Kilonovae},
   volume={23},
   ISSN={1433-8351},
   url={http://dx.doi.org/10.1007/s41114-019-0024-0},
   DOI={10.1007/s41114-019-0024-0},
   number={1},
   journal={Living Reviews in Relativity},
   publisher={Springer Science and Business Media LLC},
   author={Metzger, Brian D.},
   year={2019},
   month=dec }

@article{Metzger2014,
  author       = {Metzger, Brian D. and Fernández, Rodrigo},
  title        = {Red or blue? A potential kilonova imprint of the delay until black hole formation following a neutron star merger},
  journal      = {Monthly Notices of the Royal Astronomical Society},
  volume       = {441},
  number       = {4},
  pages        = {3444--3453},
  year         = {2014},
  month        = {July},
  doi          = {10.1093/mnras/stu802},
  url          = {https://doi.org/10.1093/mnras/stu802}
}

@ARTICLE{Cowan_2021,
       author = {{Cowan}, John J. and {Sneden}, Christopher and {Lawler}, James E. and {Aprahamian}, Ani and {Wiescher}, Michael and {Langanke}, Karlheinz and {Mart{\'\i}nez-Pinedo}, Gabriel and {Thielemann}, Friedrich-Karl},
        title = "{Origin of the heaviest elements: The rapid neutron-capture process}",
      journal = {Reviews of Modern Physics},
         year = 2021,
        month = jan,
       volume = {93},
       number = {1},
          eid = {015002},
        pages = {015002},
          doi = {10.1103/RevModPhys.93.015002},
archivePrefix = {arXiv},
       eprint = {1901.01410},
 primaryClass = {astro-ph.HE},
       adsurl = {https://ui.adsabs.harvard.edu/abs/2021RvMP...93a5002C}
}

@ARTICLE{Lattimer_1977,
       author = {{Lattimer}, J.~M. and {Mackie}, F. and {Ravenhall}, D.~G. and {Schramm}, D.~N.},
        title = "{The decompression of cold neutron star matter.}",
      journal = {\apj},
         year = 1977,
        month = apr,
       volume = {213},
        pages = {225-233},
          doi = {10.1086/155148},
       adsurl = {https://ui.adsabs.harvard.edu/abs/1977ApJ...213..225L}
}

@article{Eichler_1989,
  author = {Eichler, David and Livio, Mario and Piran, Tsvi and Schramm, David N.},
  title = {Nucleosynthesis, neutrino bursts and $\gamma$-rays from coalescing neutron stars},
  journal = {Nature},
  year = {1989},
  volume = {340},
  pages = {126--128},
  doi = {10.1038/340126a0},
  url = {https://doi.org/10.1038/340126a0}
}

@article{Chornock_2017,
doi = {10.3847/2041-8213/aa905c},
url = {https://dx.doi.org/10.3847/2041-8213/aa905c},
year = {2017},
month = {10},
publisher = {The American Astronomical Society},
volume = {848},
number = {2},
pages = {L19},
author = {R. Chornock and E. Berger and D. Kasen and P. S. Cowperthwaite and M. Nicholl and V. A. Villar and K. D. Alexander and P. K. Blanchard and T. Eftekhari and W. Fong and R. Margutti and P. K. G. Williams and J. Annis and D. Brout and D. A. Brown and H.-Y. Chen and M. R. Drout and B. Farr and R. J. Foley and J. A. Frieman and C. L. Fryer and K. Herner and D. E. Holz and R. Kessler and T. Matheson and B. D. Metzger and E. Quataert and A. Rest and M. Sako and D. M. Scolnic and N. Smith and M. Soares-Santos},
title = {The Electromagnetic Counterpart of the Binary Neutron Star Merger LIGO/Virgo GW170817. IV. Detection of Near-infrared Signatures of r-process Nucleosynthesis with Gemini-South},
journal = {The Astrophysical Journal Letters},
}

@article{Nicholl_2017,
doi = {10.3847/2041-8213/aa9029},
url = {https://dx.doi.org/10.3847/2041-8213/aa9029},
year = {2017},
month = {10},
publisher = {The American Astronomical Society},
volume = {848},
number = {2},
pages = {L18},
author = {M. Nicholl and E. Berger and D. Kasen and B. D. Metzger and J. Elias and C. Briceño and K. D. Alexander and P. K. Blanchard and R. Chornock and P. S. Cowperthwaite and T. Eftekhari and W. Fong and R. Margutti and V. A. Villar and P. K. G. Williams and W. Brown and J. Annis and A. Bahramian and D. Brout and D. A. Brown and H.-Y. Chen and J. C. Clemens and E. Dennihy and B. Dunlap and D. E. Holz and E. Marchesini and F. Massaro and N. Moskowitz and I. Pelisoli and A. Rest and F. Ricci and M. Sako and M. Soares-Santos and J. Strader},
title = {The Electromagnetic Counterpart of the Binary Neutron Star Merger LIGO/Virgo GW170817. III. Optical and UV Spectra of a Blue Kilonova from Fast Polar Ejecta},
journal = {The Astrophysical Journal Letters},
}

@article{tanaka2017kilonova,
  author = {Tanaka, Masaomi and Utsumi, Yousuke and Mazzali, Paolo A. and Tominaga, Nozomu and Yoshida, Michitoshi and Sekiguchi, Yuichiro and Morokuma, Tomoki and Motohara, Kentaro and Ohta, Kouji and Kawabata, Koji S. and Abe, Fumio and Aoki, Kentaro and Asakura, Yuichiro and Baar, Stefan and Barway, Sudhanshu and Bond, Ian A. and Doi, Mamoru and Fujiyoshi, Takuya and Furusawa, Hisanori and Honda, Satoshi and Itoh, Yoichi and Kawabata, Miho and Kawai, Nobuyuki and Kim, Ji Hoon and Lee, Chien-Hsiu and Miyazaki, Shota and Morihana, Kumiko and Nagashima, Hiroki and Nagayama, Takahiro and Nakaoka, Tatsuya and Nakata, Fumiaki and Ohsawa, Ryou and Ohshima, Tomohito and Okita, Hirofumi and Saito, Tomoki and Sumi, Takahiro and Tajitsu, Akito and Takahashi, Jun and Takayama, Masaki and Tamura, Yoichi and Tanaka, Ichi and Terai, Tsuyoshi and Tristram, Paul J. and Yasuda, Naoki and Zenko, Tetsuya},
  title = {Kilonova from post-merger ejecta as an optical and near-Infrared counterpart of GW170817},
  journal = {Publications of the Astronomical Society of Japan},
  volume = {69},
  number = {6},
  pages = {102},
  year = {2017},
  doi = {10.1093/pasj/psx121},
  url = {https://doi.org/10.1093/pasj/psx121},
  issn = {0004-6264}
}

@article{Abbott_2017,
   title={GW170817: Observation of Gravitational Waves from a Binary Neutron Star Inspiral},
   volume={119},
   ISSN={1079-7114},
   url={http://dx.doi.org/10.1103/PhysRevLett.119.161101},
   DOI={10.1103/physrevlett.119.161101},
   number={16},
   journal={Physical Review Letters},
   publisher={American Physical Society (APS)},
   author={Abbott, B. P. and Abbott, R. and Abbott, T. D. Birney},
   year={2017},
   month=10 }

@article{Mösta_2013,
doi = {10.1088/0264-9381/31/1/015005},
url = {https://dx.doi.org/10.1088/0264-9381/31/1/015005},
year = {2013},
month = {11},
publisher = {IOP Publishing},
volume = {31},
number = {1},
pages = {015005},
author = {{M{\"o}sta}, Philipp and Bruno C Mundim and Joshua A Faber and Roland Haas and Scott C Noble and Tanja Bode and Frank Löffler and Christian D Ott and Christian Reisswig and Erik Schnetter},
title = {GRHydro: a new open-source general-relativistic magnetohydrodynamics code for the Einstein toolkit},
journal = {Classical and Quantum Gravity},
}

@article{Rosswog_2018,
   title={The first direct double neutron star merger detection: Implications for cosmic nucleosynthesis},
   volume={615},
   ISSN={1432-0746},
   url={http://dx.doi.org/10.1051/0004-6361/201732117},
   DOI={10.1051/0004-6361/201732117},
   journal={Astronomy &amp; Astrophysics},
   publisher={EDP Sciences},
   author={Rosswog, S. and Sollerman, J. and Feindt, U. and Goobar, A. and Korobkin, O. and Wollaeger, R. and Fremling, C. and Kasliwal, M. M.},
   year={2018},
   month=jul, pages={A132} }

@article{Just_2023,
doi = {10.3847/2041-8213/acdad2},
url = {https://dx.doi.org/10.3847/2041-8213/acdad2},
year = {2023},
month = {jul},
publisher = {The American Astronomical Society},
volume = {951},
number = {1},
pages = {L12},
author = {Just, O. and Vijayan, V. and Xiong, Z. and Goriely, S. and Soultanis, T. and Bauswein, A. and Guilet, J. and Janka, H.-Th. and Martínez-Pinedo, G.},
title = {End-to-end Kilonova Models of Neutron Star Mergers with Delayed Black Hole Formation},
journal = {The Astrophysical Journal Letters}
}

@article{Zhu_2021,
doi = {10.3847/1538-4357/abc69e},
url = {https://dx.doi.org/10.3847/1538-4357/abc69e},
year = {2021},
month = {1},
publisher = {The American Astronomical Society},
volume = {906},
number = {2},
pages = {94},
author = {Y. L. Zhu and K. A. Lund and J. Barnes and T. M. Sprouse and N. Vassh and G. C. McLaughlin and M. R. Mumpower and R. Surman},
title = {Modeling Kilonova Light Curves: Dependence on Nuclear Inputs},
journal = {The Astrophysical Journal},
}

@article{Wu_2022,
   title={Radiation hydrodynamics modelling of kilonovae with SNEC},
   volume={512},
   ISSN={1365-2966},
   url={http://dx.doi.org/10.1093/mnras/stac399},
   DOI={10.1093/mnras/stac399},
   number={1},
   journal={Monthly Notices of the Royal Astronomical Society},
   publisher={Oxford University Press (OUP)},
   author={Wu, Zhenyu and Ricigliano, Giacomo and Kashyap, Rahul and Perego, Albino and Radice, David},
   year={2022},
   month=feb, pages={328–347} }

@article{Rosswog_2017,
doi = {10.1088/1361-6382/aa68a9},
url = {https://dx.doi.org/10.1088/1361-6382/aa68a9},
year = {2017},
month = {4},
publisher = {IOP Publishing},
volume = {34},
number = {10},
pages = {104001},
author = {S Rosswog and U Feindt and O Korobkin and M-R Wu and J Sollerman and A Goobar and G Martinez-Pinedo},
title = {Detectability of compact binary merger macronovae},
journal = {Classical and Quantum Gravity},
}

@article{Darbha_2021,
doi = {10.3847/1538-4357/abff5d},
url = {https://dx.doi.org/10.3847/1538-4357/abff5d},
year = {2021},
month = {7},
publisher = {The American Astronomical Society},
volume = {915},
number = {1},
pages = {69},
author = {Darbha, Siva and Kasen, Daniel and Foucart, Francois and Price, Daniel J.},
title = {Electromagnetic Signatures from the Tidal Tail of a Black Hole—Neutron Star Merger},
journal = {The Astrophysical Journal},
}

@article{Neuweiler_2023,
  title = {Long-term simulations of dynamical ejecta: Homologous expansion and kilonova properties},
  author = {Neuweiler, Anna and Dietrich, Tim and Bulla, Mattia and Chaurasia, Swami Vivekanandji and Rosswog, Stephan and Ujevic, Maximiliano},
  journal = {Phys. Rev. D},
  volume = {107},
  issue = {2},
  pages = {023016},
  numpages = {16},
  year = {2023},
  month = {Jan},
  publisher = {American Physical Society},
  doi = {10.1103/PhysRevD.107.023016},
  url = {https://link.aps.org/doi/10.1103/PhysRevD.107.023016}
}

@article{Wollaeger_2018,
  author    = {Ryan T. Wollaeger and Oleg Korobkin and Christopher J. Fontes and Stephan K. Rosswog and Wesley P. Even and Christopher L. Fryer and Jesper Sollerman and Aimee L. Hungerford and Daniel R. van Rossum and Allan B. Wollaber},
  title     = {Impact of ejecta morphology and composition on the electromagnetic signatures of neutron star mergers},
  journal   = {Monthly Notices of the Royal Astronomical Society},
  volume    = {478},
  number    = {3},
  pages     = {3298--3334},
  year      = {2018},
  month     = {August},
  doi       = {10.1093/mnras/sty1018}
}

@ARTICLE{Sarin2021,
       author = {{Sarin}, Nikhil and {Lasky}, Paul D.},
        title = "{The evolution of binary neutron star post-merger remnants: a review}",
      journal = {General Relativity and Gravitation},
         year = 2021,
        month = jun,
       volume = {53},
       number = {6},
          eid = {59},
        pages = {59},
          doi = {10.1007/s10714-021-02831-1},
archivePrefix = {arXiv},
       eprint = {2012.08172},
 primaryClass = {astro-ph.HE},
       adsurl = {https://ui.adsabs.harvard.edu/abs/2021GReGr..53...59S}
}

@ARTICLE{Wanajo2014,
       author = {{Wanajo}, Shinya and {Sekiguchi}, Yuichiro and {Nishimura}, Nobuya and {Kiuchi}, Kenta and {Kyutoku}, Koutarou and {Shibata}, Masaru},
        title = "{Production of All the r-process Nuclides in the Dynamical Ejecta of Neutron Star Mergers}",
      journal = {\apjl},
         year = 2014,
        month = jul,
       volume = {789},
       number = {2},
          eid = {L39},
        pages = {L39},
          doi = {10.1088/2041-8205/789/2/L39},
archivePrefix = {arXiv},
       eprint = {1402.7317},
 primaryClass = {astro-ph.SR},
       adsurl = {https://ui.adsabs.harvard.edu/abs/2014ApJ...789L..39W}
}

@ARTICLE{Bernuzzi2020,
       author = {{Bernuzzi}, Sebastiano},
        title = "{Neutron star merger remnants}",
      journal = {General Relativity and Gravitation},
         year = 2020,
        month = nov,
       volume = {52},
       number = {11},
          eid = {108},
        pages = {108},
          doi = {10.1007/s10714-020-02752-5},
archivePrefix = {arXiv},
       eprint = {2004.06419},
 primaryClass = {astro-ph.HE},
       adsurl = {https://ui.adsabs.harvard.edu/abs/2020GReGr..52..108B}
}

@article{Kawaguchi_2018,
   title={Radiative Transfer Simulation for the Optical and Near-infrared Electromagnetic Counterparts to GW170817},
   volume={865},
   ISSN={2041-8213},
   url={http://dx.doi.org/10.3847/2041-8213/aade02},
   DOI={10.3847/2041-8213/aade02},
   number={2},
   journal={The Astrophysical Journal Letters},
   publisher={American Astronomical Society},
   author={Kawaguchi, Kyohei and Shibata, Masaru and Tanaka, Masaomi},
   year={2018},
   month=sep, pages={L21} }

@ARTICLE{Korobkin_2021,
       author = {{Korobkin}, Oleg and {Wollaeger}, Ryan T. and {Fryer}, Christopher L. and {Hungerford}, Aimee L. and {Rosswog}, Stephan and {Fontes}, Christopher J. and {Mumpower}, Matthew R. and {Chase}, Eve A. and {Even}, Wesley P. and {Miller}, Jonah and {Misch}, G. Wendell and {Lippuner}, Jonas},
        title = "{Axisymmetric Radiative Transfer Models of Kilonovae}",
      journal = {\apj},
         year = 2021,
        month = apr,
       volume = {910},
       number = {2},
          eid = {116},
        pages = {116},
          doi = {10.3847/1538-4357/abe1b5},
archivePrefix = {arXiv},
       eprint = {2004.00102},
 primaryClass = {astro-ph.HE},
       adsurl = {https://ui.adsabs.harvard.edu/abs/2021ApJ...910..116K}
}

@article{Brethauer_2024,
doi = {10.3847/1538-4357/ad7d83},
url = {https://dx.doi.org/10.3847/1538-4357/ad7d83},
year = {2024},
month = {11},
publisher = {The American Astronomical Society},
volume = {975},
number = {2},
pages = {213},
author = {Brethauer, D. and Kasen, D. and Margutti, R. and Chornock, R.},
title = {Impact of Systematic Modeling Uncertainties on Kilonova Property Estimation},
journal = {The Astrophysical Journal},
}

@article{Roth_2015,
   title={Monte-Carlo Radiation-Hydrodynamics With Implicit Methods},
   volume={217},
   ISSN={1538-4365},
   url={http://dx.doi.org/10.1088/0067-0049/217/1/9},
   DOI={10.1088/0067-0049/217/1/9},
   number={1},
   journal={The Astrophysical Journal Supplement Series},
   publisher={American Astronomical Society},
   author={Roth, Nathaniel and Kasen, Daniel},
   year={2015},
   month=3, pages={9} }

@BOOK{Sobolev_1960,
       author = {{Sobolev}, V.~V.},
        title = "{Moving Envelopes of Stars}",
        year = 1960,
        publisher = {Harvard University Press},
        doi = {10.4159/harvard.9780674864658},
       adsurl = {https://ui.adsabs.harvard.edu/abs/1960mes..book.....S}

}

@article{Radice_2018,
  author = {David Radice and Albino Perego and Kenta Hotokezaka and Steven A. Fromm and Sebastiano Bernuzzi and Luke F. Roberts},
  title = {Binary Neutron Star Mergers: Mass Ejection, Electromagnetic Counterparts, and Nucleosynthesis},
  journal = {The Astrophysical Journal},
  year = {2018},
  volume = {869},
  number = {2},
  pages = {130},
  doi = {10.3847/1538-4357/aaf054},
  url = {https://dx.doi.org/10.3847/1538-4357/aaf054},
  month = {12},
  publisher = {The American Astronomical Society}
}

@article{LATTIMER1991331,
title = {A generalized equation of state for hot, dense matter},
journal = {Nuclear Physics A},
volume = {535},
number = {2},
pages = {331-376},
year = {1991},
issn = {0375-9474},
doi = {https://doi.org/10.1016/0375-9474(91)90452-C},
url = {https://www.sciencedirect.com/science/article/pii/037594749190452C},
author = {James M. Lattimer and F. {Douglas Swesty}},
}

@article{Radice_2022,
    author = {Radice, David and Bernuzzi, Sebastiano and Perego, Albino and Haas, Roland},
    title = "{A new moment-based general-relativistic neutrino-radiation transport code: Methods and first applications to neutron star mergers}",
    journal = {Monthly Notices of the Royal Astronomical Society},
    volume = {512},
    number = {1},
    pages = {1499-1521},
    year = {2022},
    month = {03},
    issn = {0035-8711},
    doi = {10.1093/mnras/stac589},
    url = {https://doi.org/10.1093/mnras/stac589},
    eprint = {https://academic.oup.com/mnras/article-pdf/512/1/1499/42978846/stac589.pdf}
}

@ARTICLE{Foucart_2016,
       author = {{Foucart}, Francois and {O'Connor}, Evan and {Roberts}, Luke and {Kidder}, Lawrence E. and {Pfeiffer}, Harald P. and {Scheel}, Mark A.},
        title = "{Impact of an improved neutrino energy estimate on outflows in neutron star merger simulations}",
      journal = {\prd},
         year = 2016,
        month = dec,
       volume = {94},
       number = {12},
          eid = {123016},
        pages = {123016},
          doi = {10.1103/PhysRevD.94.123016},
       adsurl = {https://ui.adsabs.harvard.edu/abs/2016PhRvD..94l3016F}
}

@ARTICLE{Rosswog_2003,
       author = {{Rosswog}, S. and {Liebend{\"o}rfer}, M.},
        title = "{High-resolution calculations of merging neutron stars - II. Neutrino emission}",
      journal = {\mnras},
         year = 2003,
        month = jul,
       volume = {342},
       number = {3},
        pages = {673-689},
          doi = {10.1046/j.1365-8711.2003.06579.x},
archivePrefix = {arXiv},
       eprint = {astro-ph/0302301},
 primaryClass = {astro-ph},
       adsurl = {https://ui.adsabs.harvard.edu/abs/2003MNRAS.342..673R}
}

@article{Kawaguchi_2021,
doi = {10.3847/1538-4357/abf3bc},
url = {https://dx.doi.org/10.3847/1538-4357/abf3bc},
year = {2021},
month = {may},
publisher = {The American Astronomical Society},
volume = {913},
number = {2},
pages = {100},
author = {{Kawaguchi}, Kyohei and Fujibayashi, Sho and Shibata, Masaru and Tanaka, Masaomi and Wanajo, Shinya},
title = {A Low-mass Binary Neutron Star: Long-term Ejecta Evolution and Kilonovae with Weak Blue Emission},
journal = {The Astrophysical Journal},
}

@article{Lippuner_2015,
doi = {10.1088/0004-637X/815/2/82},
url = {https://dx.doi.org/10.1088/0004-637X/815/2/82},
year = {2015},
month = {dec},
publisher = {The American Astronomical Society},
volume = {815},
number = {2},
pages = {82},
author = {Lippuner, Jonas and Roberts, Luke F.},
title = {r-PROCESS LANTHANIDE PRODUCTION AND HEATING RATES IN KILONOVAE},
journal = {The Astrophysical Journal},
}

@article{Even_2020,
doi = {10.3847/1538-4357/ab70b9},
url = {https://dx.doi.org/10.3847/1538-4357/ab70b9},
year = {2020},
month = {aug},
publisher = {The American Astronomical Society},
volume = {899},
number = {1},
pages = {24},
author = {Even, Wesley and Korobkin, Oleg and Fryer, Christopher L. and Fontes, Christopher J. and Wollaeger, R. T. and Hungerford, Aimee and Lippuner, Jonas and Miller, Jonah and Mumpower, Matthew R. and Misch, G. Wendell},
title = {Composition Effects on Kilonova Spectra and Light Curves. I},
journal = {The Astrophysical Journal}
}

@article{Barnes_2013,
doi = {10.1088/0004-637X/775/1/18},
url = {https://dx.doi.org/10.1088/0004-637X/775/1/18},
year = {2013},
month = {aug},
publisher = {The American Astronomical Society},
volume = {775},
number = {1},
pages = {18},
author = {Barnes, Jennifer and Kasen, Daniel},
title = {EFFECT OF A HIGH OPACITY ON THE LIGHT CURVES OF RADIOACTIVELY POWERED TRANSIENTS FROM COMPACT OBJECT MERGERS},
journal = {The Astrophysical Journal}
}

@article{Duez_2006,
  title = {Evolution of magnetized, differentially rotating neutron stars: Simulations in full general relativity},
  author = {Duez, Matthew D. and Liu, Yuk Tung and Shapiro, Stuart L. and Shibata, Masaru and Stephens, Branson C.},
  journal = {Phys. Rev. D},
  volume = {73},
  issue = {10},
  pages = {104015},
  numpages = {25},
  year = {2006},
  month = {May},
  publisher = {American Physical Society},
  doi = {10.1103/PhysRevD.73.104015},
  url = {https://link.aps.org/doi/10.1103/PhysRevD.73.104015}
}

@article{Kiuchi_2015_sGRB,
   title={High resolution magnetohydrodynamic simulation of black hole-neutron star merger: Mass ejection and short gamma ray bursts},
   volume={92},
   ISSN={1550-2368},
   url={http://dx.doi.org/10.1103/PhysRevD.92.064034},
   DOI={10.1103/physrevd.92.064034},
   number={6},
   journal={Physical Review D},
   publisher={American Physical Society (APS)},
   author={Kiuchi, Kenta and Sekiguchi, Yuichiro and Kyutoku, Koutarou and Shibata, Masaru and Taniguchi, Keisuke and Wada, Tomohide},
   year={2015},
   month=9 }

@article{Kiuchi_2018,
  title = {Global simulations of strongly magnetized remnant massive neutron stars formed in binary neutron star mergers},
  author = {Kiuchi, Kenta and Kyutoku, Koutarou and Sekiguchi, Yuichiro and Shibata, Masaru},
  journal = {Phys. Rev. D},
  volume = {97},
  issue = {12},
  pages = {124039},
  numpages = {16},
  year = {2018},
  month = {Jun},
  publisher = {American Physical Society},
  doi = {10.1103/PhysRevD.97.124039},
  url = {https://link.aps.org/doi/10.1103/PhysRevD.97.124039}
}

@ARTICLE{Kullmann_2022,
       author = {{Kullmann}, I. and {Goriely}, S. and {Just}, O. and {Ardevol-Pulpillo}, R. and {Bauswein}, A. and {Janka}, H. -T.},
        title = "{Dynamical ejecta of neutron star mergers with nucleonic weak processes I: nucleosynthesis}",
      journal = {\mnras},
         year = 2022,
        month = feb,
       volume = {510},
       number = {2},
        pages = {2804-2819},
          doi = {10.1093/mnras/stab3393},
archivePrefix = {arXiv},
       eprint = {2109.02509},
 primaryClass = {astro-ph.HE},
       adsurl = {https://ui.adsabs.harvard.edu/abs/2022MNRAS.510.2804K}
}

@ARTICLE{Sekiguchi_2015,
       author = {{Sekiguchi}, Yuichiro and {Kiuchi}, Kenta and {Kyutoku}, Koutarou and {Shibata}, Masaru},
        title = "{Dynamical mass ejection from binary neutron star mergers: Radiation-hydrodynamics study in general relativity}",
      journal = {\prd},
         year = 2015,
        month = mar,
       volume = {91},
       number = {6},
          eid = {064059},
        pages = {064059},
          doi = {10.1103/PhysRevD.91.064059},
archivePrefix = {arXiv},
       eprint = {1502.06660},
 primaryClass = {astro-ph.HE},
       adsurl = {https://ui.adsabs.harvard.edu/abs/2015PhRvD..91f4059S}
}

@article{Pais_2024,
    author = "Pais, Matteo and Piran, Tsvi and Kiuchi, Kenta and Shibata, Masaru",
    title = "{Simulating Short Gamma-Ray Burst Jets in Realistic Late Binary Neutron Star Merger Environments}",
    eprint = "2407.19002",
    archivePrefix = "arXiv",
    primaryClass = "astro-ph.HE",
    doi = "10.3847/1538-4357/ad7d04",
    journal = "Astrophys. J.",
    volume = "976",
    number = "1",
    pages = "35",
    year = "2024"
}

@article{Kiuchi_2015,
  title = {Efficient magnetic-field amplification due to the Kelvin-Helmholtz instability in binary neutron star mergers},
  author = {Kiuchi, Kenta and Cerd\'a-Dur\'an, Pablo and Kyutoku, Koutarou and Sekiguchi, Yuichiro and Shibata, Masaru},
  journal = {Phys. Rev. D},
  volume = {92},
  issue = {12},
  pages = {124034},
  numpages = {11},
  year = {2015},
  month = {Dec},
  publisher = {American Physical Society},
  doi = {10.1103/PhysRevD.92.124034},
  url = {https://link.aps.org/doi/10.1103/PhysRevD.92.124034}
}

@article{Combi_2023,
doi = {10.3847/1538-4357/acac29},
url = {https://dx.doi.org/10.3847/1538-4357/acac29},
year = {2023},
month = {feb},
publisher = {The American Astronomical Society},
volume = {944},
number = {1},
pages = {28},
author = {Combi, Luciano and Siegel, Daniel M.},
title = {GRMHD Simulations of Neutron-star Mergers with Weak Interactions: r-process Nucleosynthesis and Electromagnetic Signatures of Dynamical Ejecta},
journal = {The Astrophysical Journal}
}

@article{Murguia_Berthier_2014,
   title={NECESSARY CONDITIONS FOR SHORT GAMMA-RAY BURST PRODUCTION IN BINARY NEUTRON STAR MERGERS},
   volume={788},
   ISSN={2041-8213},
   url={http://dx.doi.org/10.1088/2041-8205/788/1/L8},
   DOI={10.1088/2041-8205/788/1/l8},
   number={1},
   journal={The Astrophysical Journal},
   publisher={American Astronomical Society},
   author={Murguia-Berthier, Ariadna and Montes, Gabriela and Ramirez-Ruiz, Enrico and De Colle, Fabio and Lee, William H.},
   year={2014},
   month=may, pages={L8} }

@article{Combi_2023jets,
    author = "Combi, Luciano and Siegel, Daniel M.",
    title = "{Jets from Neutron-Star Merger Remnants and Massive Blue Kilonovae}",
    eprint = "2303.12284",
    archivePrefix = "arXiv",
    primaryClass = "astro-ph.HE",
    doi = "10.1103/PhysRevLett.131.231402",
    journal = "Phys. Rev. Lett.",
    volume = "131",
    number = "23",
    pages = "231402",
    year = "2023"
}

@article{Darbha_2020,
doi = {10.3847/1538-4357/ab9a34},
url = {https://dx.doi.org/10.3847/1538-4357/ab9a34},
year = {2020},
month = {jul},
publisher = {The American Astronomical Society},
volume = {897},
number = {2},
pages = {150},
author = {Darbha, Siva and Kasen, Daniel},
title = {Inclination Dependence of Kilonova Light Curves from Globally Aspherical Geometries},
journal = {The Astrophysical Journal}
}

@article{Duffell_2015,
doi = {10.1088/0004-637X/813/1/64},
url = {https://dx.doi.org/10.1088/0004-637X/813/1/64},
year = {2015},
month = {oct},
publisher = {The American Astronomical Society},
volume = {813},
number = {1},
pages = {64},
author = {Duffell, Paul C. and Quataert, Eliot and MacFadyen, Andrew I.},
title = {A NARROW SHORT-DURATION GRB JET FROM A WIDE CENTRAL ENGINE},
journal = {The Astrophysical Journal}
}

@article{Kumar_2015,
   title={The physics of gamma-ray bursts &amp; relativistic jets},
   volume={561},
   ISSN={0370-1573},
   url={http://dx.doi.org/10.1016/j.physrep.2014.09.008},
   DOI={10.1016/j.physrep.2014.09.008},
   journal={Physics Reports},
   publisher={Elsevier BV},
   author={Kumar, Pawan and Zhang, Bing},
   year={2015},
   month=feb, pages={1–109} }

@ARTICLE{Gottlieb_2025,
       author = {{Gottlieb}, Ore and {Metzger}, Brian D. and {Foucart}, Francois and {Ramirez-Ruiz}, Enrico},
        title = "{A Unified Model of Kilonovae and Gamma-Ray Bursts in Binary Mergers Establishes Neutron Stars as the Central Engines of Short GRBs}",
      journal = {\apj},
         year = 2025,
        month = may,
       volume = {984},
       number = {1},
          eid = {77},
        pages = {77},
          doi = {10.3847/1538-4357/adc577},
archivePrefix = {arXiv},
       eprint = {2411.13657},
 primaryClass = {astro-ph.HE},
       adsurl = {https://ui.adsabs.harvard.edu/abs/2025ApJ...984...77G}
}

@article{Most_2023,
doi = {10.3847/2041-8213/acca84},
url = {https://dx.doi.org/10.3847/2041-8213/acca84},
year = {2023},
month = {apr},
publisher = {The American Astronomical Society},
volume = {947},
number = {1},
pages = {L15},
author = {Most, Elias R. and Quataert, Eliot},
title = {Flares, Jets, and Quasiperiodic Outbursts from Neutron Star Merger Remnants},
journal = {The Astrophysical Journal Letters}
}

@article{Nagakura_2014,
doi = {10.1088/2041-8205/784/2/L28},
url = {https://dx.doi.org/10.1088/2041-8205/784/2/L28},
year = {2014},
month = {mar},
publisher = {The American Astronomical Society},
volume = {784},
number = {2},
pages = {L28},
author = {Nagakura, Hiroki and Hotokezaka, Kenta and Sekiguchi, Yuichiro and Shibata, Masaru and Ioka, Kunihito},
title = {JET COLLIMATION IN THE EJECTA OF DOUBLE NEUTRON STAR MERGERS: A NEW CANONICAL PICTURE OF SHORT GAMMA-RAY BURSTS},
journal = {The Astrophysical Journal Letters}
}

@article{Ruiz_2016,
doi = {10.3847/2041-8205/824/1/L6},
url = {https://dx.doi.org/10.3847/2041-8205/824/1/L6},
year = {2016},
month = {jun},
publisher = {The American Astronomical Society},
volume = {824},
number = {1},
pages = {L6},
author = {Ruiz, Milton and Lang, Ryan N. and Paschalidis, Vasileios and Shapiro, Stuart L.},
title = {BINARY NEUTRON STAR MERGERS: A JET ENGINE FOR SHORT GAMMA-RAY BURSTS},
journal = {The Astrophysical Journal Letters}
}

@article{Villar_2017,
   title={The Combined Ultraviolet, Optical, and Near-infrared Light Curves of the Kilonova Associated with the Binary Neutron Star Merger GW170817: Unified Data Set, Analytic Models, and Physical Implications},
   volume={851},
   ISSN={2041-8213},
   url={http://dx.doi.org/10.3847/2041-8213/aa9c84},
   DOI={10.3847/2041-8213/aa9c84},
   number={1},
   journal={The Astrophysical Journal Letters},
   publisher={American Astronomical Society},
   author={Villar, V. A. and Guillochon, J. and Berger, E. and Metzger, B. D. and Cowperthwaite, P. S. and Nicholl, M. and Alexander, K. D. and Blanchard, P. K. and Chornock, R. and Eftekhari, T. and Fong, W. and Margutti, R. and Williams, P. K. G.},
   year={2017},
   month=dec, pages={L21} }

@article{Thielemann_2017,
   author = "Thielemann, F.-K. and Eichler, M. and Panov, I.V. and Wehmeyer, B.",
   title = "Neutron Star Mergers and Nucleosynthesis of Heavy Elements", 
   journal= "Annual Review of Nuclear and Particle Science",
   year = "2017",
   volume = "67",
   number = "Volume 67, 2017",
   pages = "253-274",
   doi = "https://doi.org/10.1146/annurev-nucl-101916-123246",
   url = "https://www.annualreviews.org/content/journals/10.1146/annurev-nucl-101916-123246",
   publisher = "Annual Reviews",
   issn = "1545-4134",
   type = "Journal Article"
}




\appendix

\section*{APPENDIX}
\renewcommand\thefigure{A\arabic{figure}}
\setcounter{figure}{0}

\subsection{Appendix A}
\label{app:A}
\begin{figure*}
    \centering
    \includegraphics[width=1.0\textwidth]{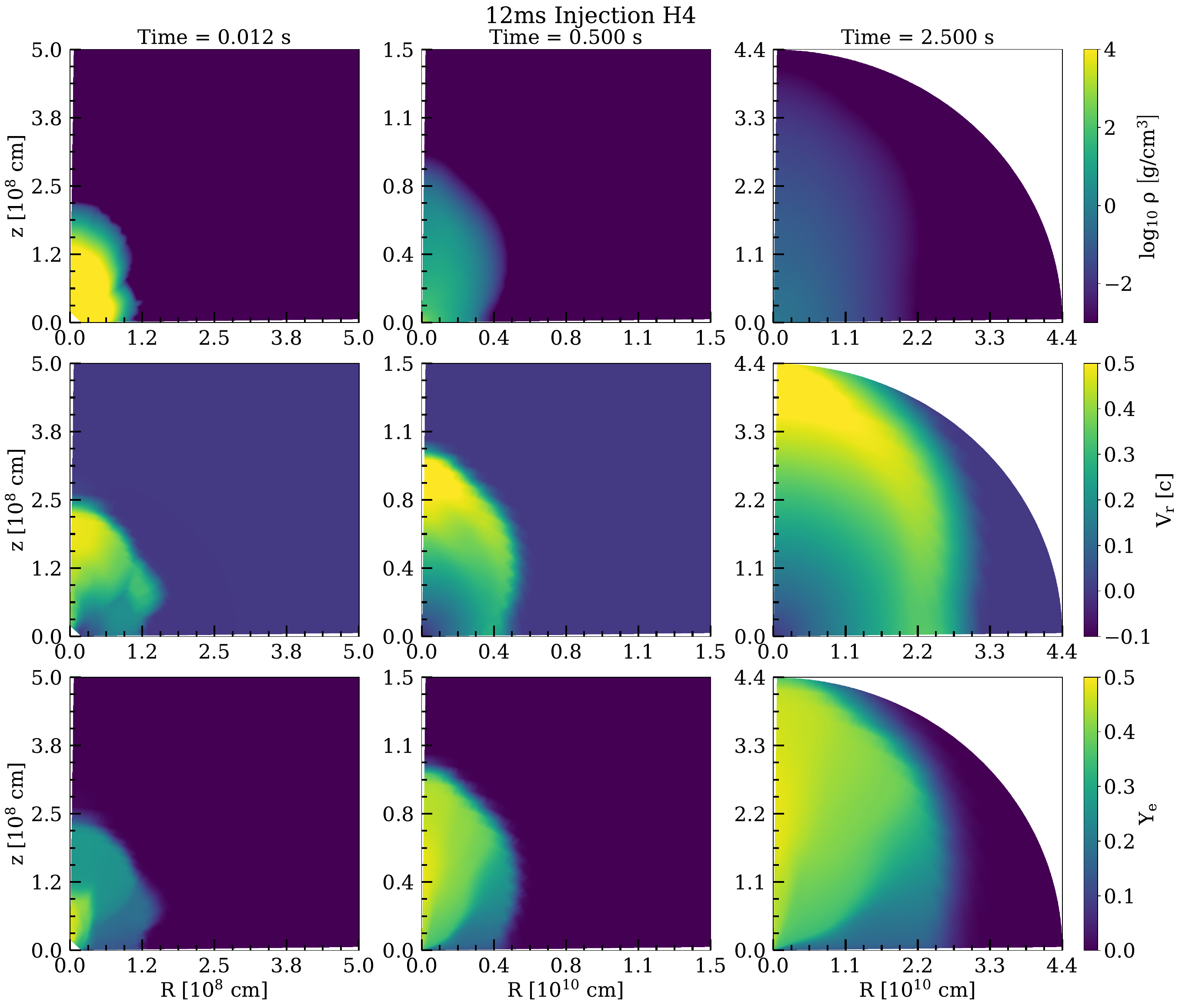}
    \caption{Evolution of the ejecta in the 12ms injection scenario with heating mode 4, shown at three different times: 12ms (left column), 0.5 s (middle), and 2.5 s (right). Each row corresponds to a different quantity: mass density (top), radial velocity (middle), and electron fraction $Y_e$ (bottom). The $v_r$ colorbar ranges from $-0.1c$ to $0.5c$, with negative values indicating fallback, though these are not apparent here because they occur close to the remnant and are obscured due to the large spatial domain used to highlight the ejecta evolution. The spatial domain increases across columns to ensure that key features remain visible, with the final column (2.5 s) covering the full \texttt{FLASH} domain.}
    \label{fig:12ms_H4_3x3}
\end{figure*}

\begin{figure*}
    \centering
    \includegraphics[width=1.0\textwidth]{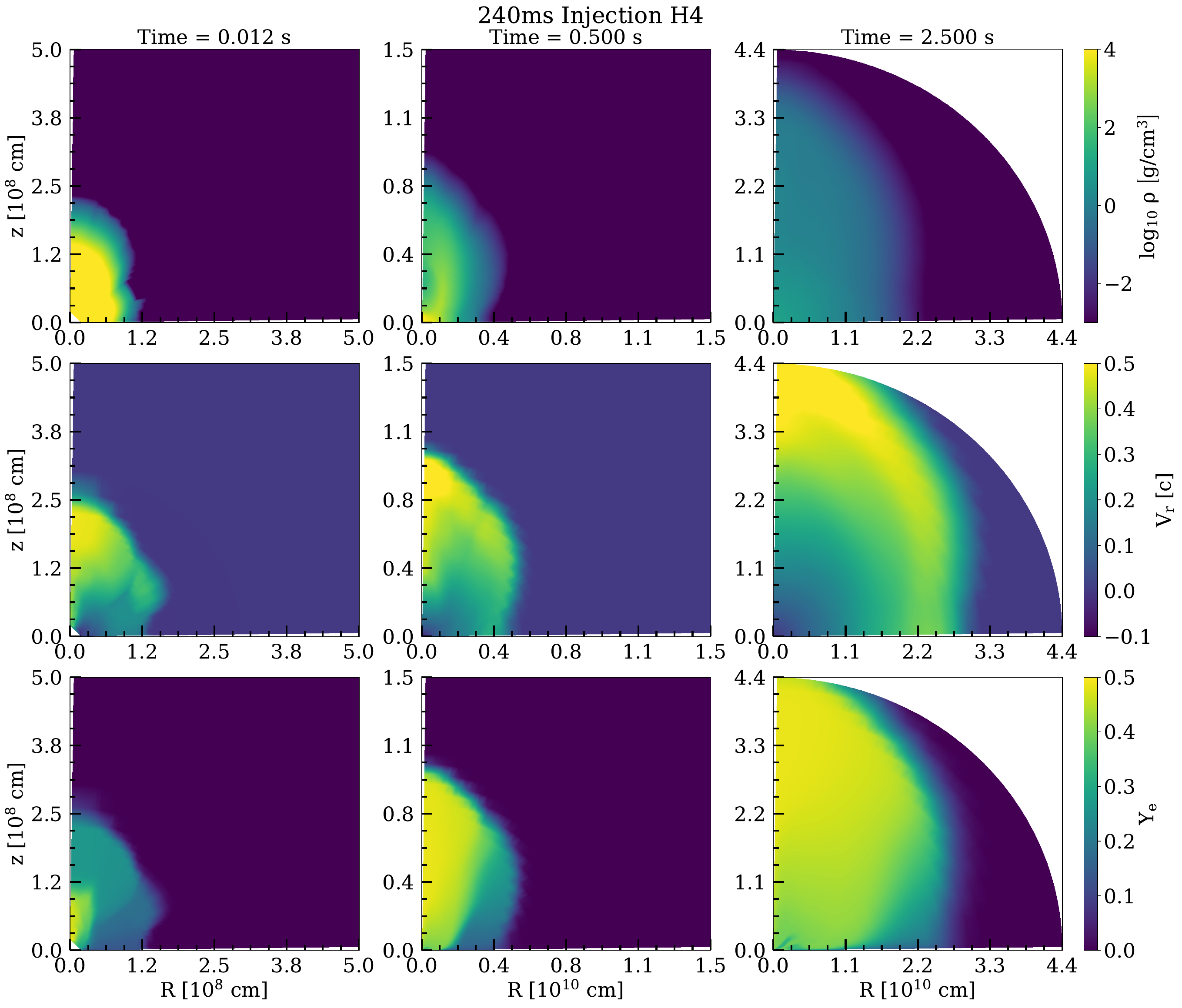}
    \caption{Evolution of the ejecta in the 240ms injection scenario with heating mode 4, shown at three different times: 12ms (left column), 0.5 s (middle), and 2.5 s (right). Each row corresponds to a different quantity: mass density (top), radial velocity (middle), and electron fraction $Y_e$ (bottom). The $v_r$ colorbar ranges from $-0.1c$ to $0.5c$, with negative values indicating fallback, though these are not apparent here because they occur close to the remnant and are obscured due to the large spatial domain used to highlight the ejecta evolution. The spatial domain increases across columns to ensure that key features remain visible, with the final column (2.5 s) covering the full \texttt{FLASH} domain.}
    \label{fig:240ms_H4_3x3}
\end{figure*}

\begin{figure*}
    \centering
    \includegraphics[width=1.0\textwidth]{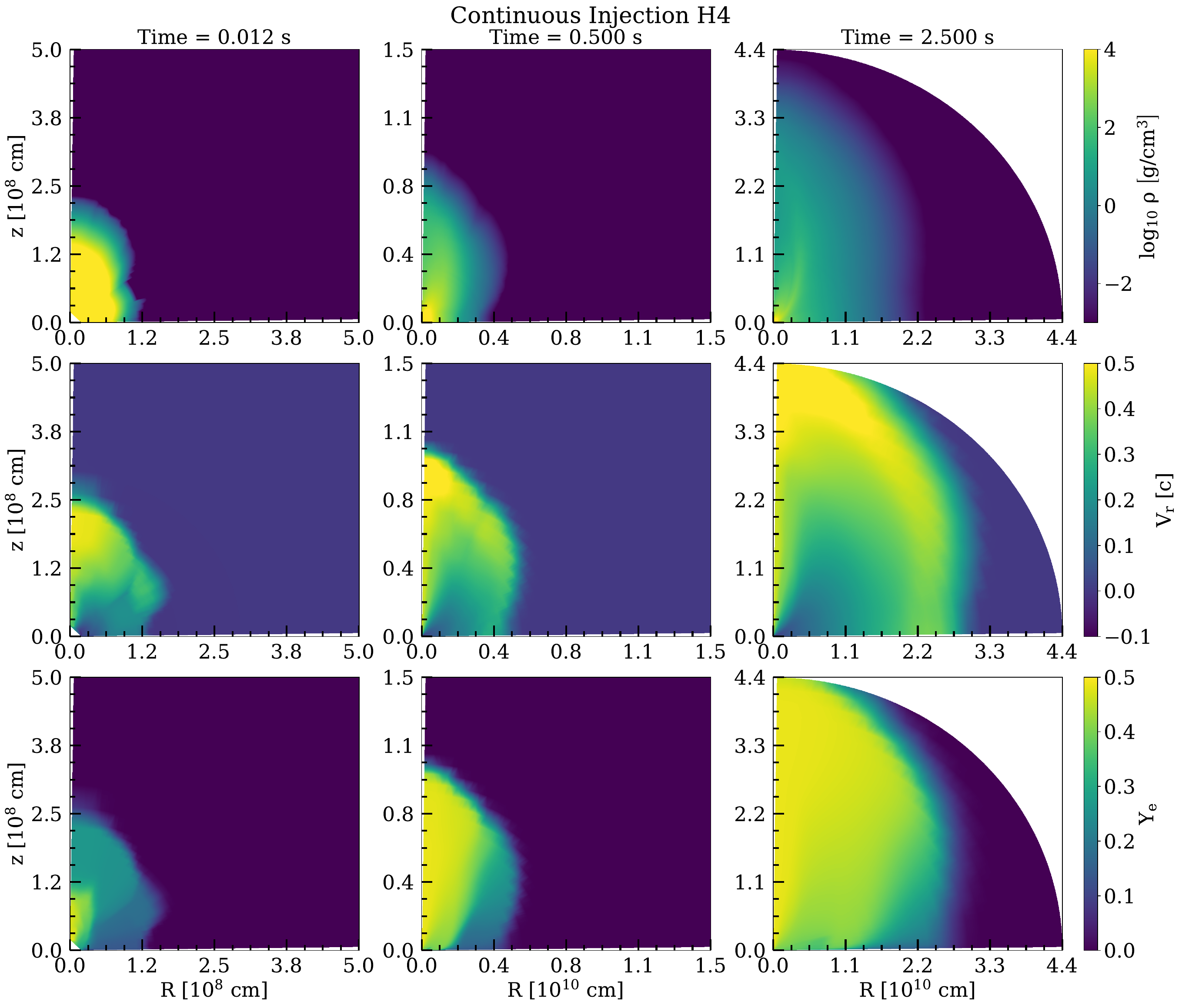}
    \caption{Evolution of the ejecta in the continuous injection scenario with heating mode 4, shown at three different times: 12ms (left column), 0.5 s (middle), and 2.5 s (right). Each row corresponds to a different quantity: mass density (top), radial velocity (middle), and electron fraction $Y_e$ (bottom). The $v_r$ colorbar ranges from $-0.1c$ to $0.5c$, with negative values indicating fallback, though these are not apparent here because they occur close to the remnant and are obscured due to the large spatial domain used to highlight the ejecta evolution. The spatial domain increases across columns to ensure that key features remain visible, with the final column (2.5 s) covering the full \texttt{FLASH} domain.}
    \label{fig:cont_H4_3x3}
\end{figure*}

Figures \crefrange{fig:12ms_H4_3x3}{fig:cont_H4_3x3} show the ejecta evolution for the \texttt{H4} heating rate. In contrast to the \texttt{H3} case, the secondary plume feature becomes more diffuse and essentially disappears, and the outflow adopts a noticeably more spherical morphology. This change can be attributed to the stronger radioactive heating, which accelerates initially slow-moving material by converting thermal energy into kinetic energy. Because equatorial ejecta expand more slowly than polar material, the additional heating has a proportionally greater impact in the lateral direction, enhancing equatorial velocities and promoting a more spherical outflow.

\subsection{Appendix B}
\label{app:B}
\renewcommand\thefigure{B\arabic{figure}}
\setcounter{figure}{0}

\begin{figure*}
    \centering
    \includegraphics[width=0.9\textwidth]{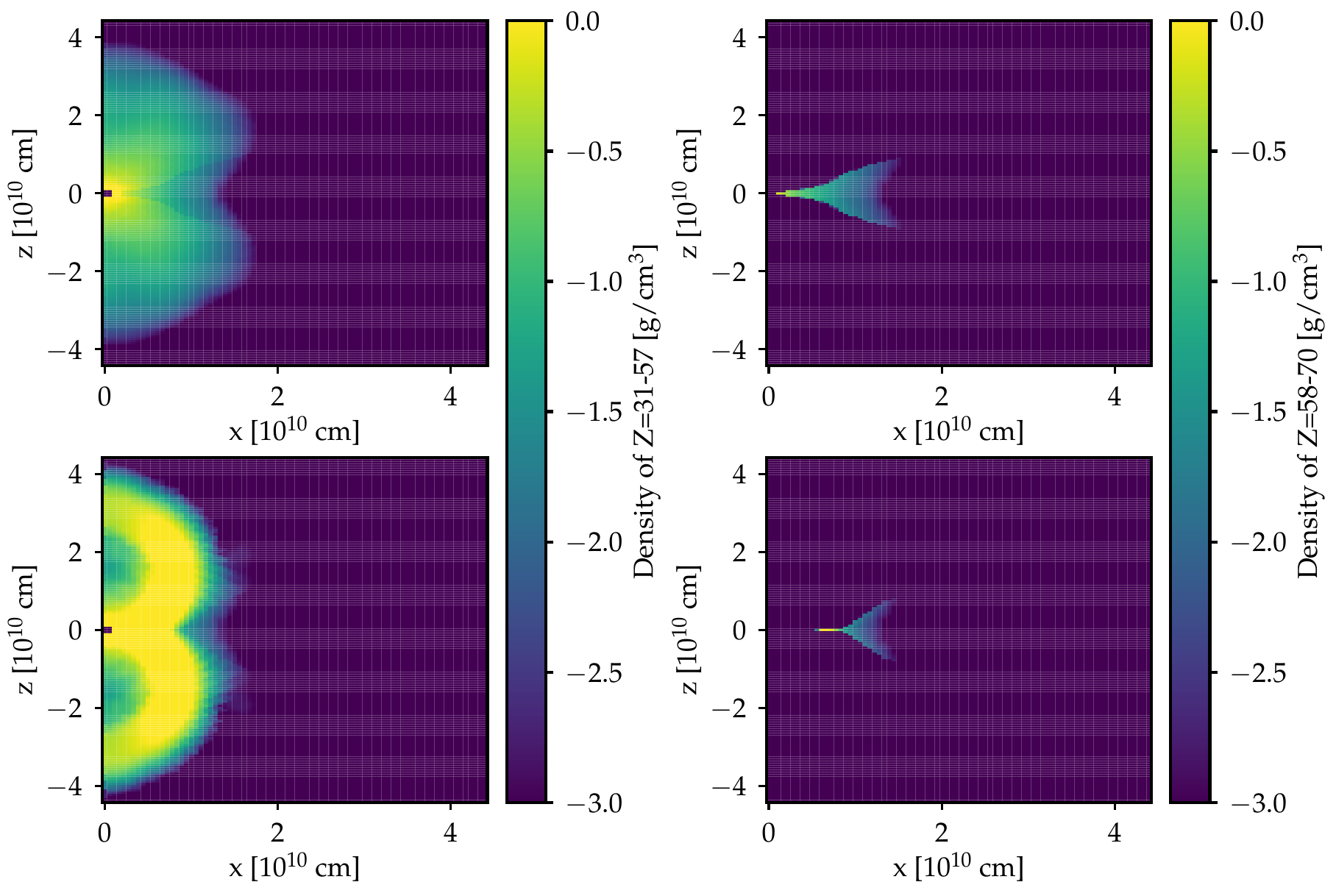}
    \caption{The composition of ejecta used as input for \texttt{Sedona} calculations is plotted for the 12ms scenario (top panels) and the 240ms scenario (bottom panels). The panels on the left show the amount and distribution of non-lathanide isotopes in the ejecta, while the panels on the right reflect the amount and distribution of lanthanides. Given the concentration of high-opacity lanthanide-rich material primarily near the equatorial plane, lanthanide blanketing may be observed at equatorial viewing angles. This has a strong impact on the observed emission for the 12ms scenario, where the relative amount of lanthanides is larger.}
    \label{fig:xla}
\end{figure*}

When constructing the input profiles for \texttt{Sedona}, we assign the composition of various zones based on their $Y_e$. The resulting distribution is shown in Figure \ref{fig:xla}, where we highlight the distribution of lanthanide (Z=58-70) and non-lanthanide elements (Z=31-57) in our ejecta since lanthanide opacities play an important role in shaping the kilonova emission. Note that in order to gauge the relative amounts of (non-)lanthanides in present different zones, we have multiplied the cumulative mass-fraction of (non-)lanthanides in a zone with the zone density. As expected for our high $Y_e$ outflows, most of the ejected material contains little to no lanthanides. The lanthanides that are present appear to be concentrated near the equatorial plane. 

\bsp	
\label{lastpage}
\end{document}